\documentclass[twocolumn,rmp,aps,nofootinbib]{revtex4}
\usepackage{graphics}
\usepackage{epsfig}
\usepackage{amsmath}
\usepackage{longtable}
\usepackage{color}

\def\inbar{\,\vrule height1.5ex width.4pt depth0pt}
\def\IR{\relax{\rm I\kern-.18em R}}
\def\IC{\relax\hbox{$\inbar\kern-.3em{\rm C}$}}

\def\pg{``phonon-glass electron-crystal"}
\def\Ef{E_{\rm F}}
\def\kB{k_{\rm B}}

\def\<{\langle}
\def\>{\rangle}
\def\({\left(}
\def\){\right)}
\def\[{\left[}
\def\]{\right]}
\def\i{\mathrm{i}}
\def\e{\mathrm{e}}
\def \ve {\varepsilon}

\def\sech{\mathop{\mathrm{sech}}\nolimits}

\def\sin{\mathop{\mathrm{sin}}\nolimits}

\def\aa{\textbf{\textit{a}}}
\def\kk{\textbf{\textit{k}}}
\def\pp{\textbf{\textit{p}}}
\def\ww{\textbf{\textit{w}}}
\def\qq{\textbf{\textit{q}}}
\def\rr{\textbf{\textit{r}}}
\def\jj{\textbf{\textit{j}}}
\def\uu{\textbf{\textit{u}}}
\def\dd{\textbf{\textit{d}}}
\def\ee{\textbf{\textit{e}}}

\def\EE{\textbf{\textit{E}}}

\def\DD{\textbf{\textit{D}}}

\def\UU{\textbf{\textit{U}}}
\def\GG{\textbf{\textit{G}}}
\def\HH{\textbf{\textit{H}}}
\def\JJ{\textbf{\textit{J}}}

\def\RR{\textbf{\textit{R}}}

\def\QQ{\textbf{\textit{Q}}}

\begin{document}
\title{Phonon-glass electron-crystal thermoelectric clathrates: Experiments and theory}

\author{Toshiro Takabatake}
\email{takaba@hiroshima-u.ac.jp}
\affiliation{Graduate School of Advanced Sciences of Matter, Hiroshima University,\\
Higashi-Hiroshima 739-8530, Japan}

\author{Koichiro Suekuni}
\affiliation{Graduate School of Advanced Sciences of Matter, Hiroshima University,\\
Higashi-Hiroshima 739-8530, Japan}

\author{Tsuneyoshi Nakayama$^{\dagger~\ddagger}$ and Eiji Kaneshita$^{\star}$}
\email{tnaka@eng.hokudai.ac.jp}
\affiliation{$^\dagger$Max Planck Institute for the Physics of Complex Systems,\\
Dresden 01187, Germany and $^{\ddagger}$Hokkaido University, Sapporo 060-0826, Japan\\
$^{\star}$Sendai National College of Technology, Sendai 989-3128, Japan}

\begin{abstract}
Type-I clathrate compounds have attracted a great deal of interest in connection with the search for efficient thermoelectric materials.
These compounds constitute networked cages consisting of nano-scale 
tetrakaidecahedrons (14 hedrons) and  dodecahedrons (12 hedrons), 
in which the group 1 or 2 elements in the periodic table are encaged as the so-called ``rattling" guest atom.
It is remarkable that, though these compounds have crystalline cubic-structure, 
they exhibit glass-like phonon thermal conductivity over the whole temperature range 
depending on the states of rattling guest atoms in the tetrakaidecahedron.  
In addition, these compounds show unusual glass-like specific heats and THz-frequency phonon dynamics, 
providing a remarkable broad peak almost identical to those observed in topologically disordered 
amorphous materials or structural glasses, the so-called Boson peak.
An efficient thermoelectric effect is realized in compounds showing these glass-like characteristics. 
This decade, a number of experimental works dealing with type-I clathrate compounds have been published. 
These are diffraction experiments, thermal and spectroscopic experiments in addition to those based on heat and electronic transport.
These form the raw materials for this article based on  advances this decade.
The subject of this article involves interesting phenomena from the viewpoint of 
not only physics but also from the view point of the practical problem of elaborating efficient thermoelectric materials.
This review presents a survey of a wide range of experimental investigations of 
type-I clathrate compounds, together with a review of theoretical interpretations of the peculiar thermal and dynamic properties observed in these materials.
\end{abstract}                                                                 

\maketitle
\tableofcontents

\section{INTRODUCTION}
\label{sec:introduction}
\subsection{History and motivation}
\label{subsec:history}
Research on thermoelectricity has a long history.
It started in the 1820's with the discovery of the Seebeck effect, i.e., converting temperature difference to electric current in metals\,\cite{Seebeck:1822a, Seebeck:1826a}.
J.\,T.\,Seebeck discovered in 1821 that an electric current flows in a closed circuit made of two dissimilar metals 
when the two junctions are maintained at different temperatures.
The reverse effect was found 10 years later by \textcite{Peltier:1834a}. 

\begin{figure}[t]
\begin{center}
\includegraphics[width = 0.7\linewidth]{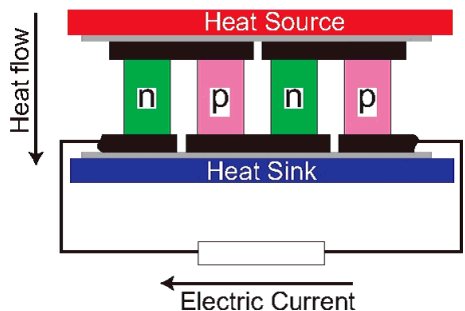}
\caption{(Color online)~Thermoelectric devices contain many couples of $n$-type and $p$-type thermoelectric elements wired electrically in series and thermally in parallel. 
}
\label{fig:takaba_fig01}
\end{center}
\end{figure}

\textcite{Ioffe:1932a} made a breakthrough in the 1930's, noting that
semiconductor thermoelements are more practical than  metals employed at that time.
In fact, modern thermoelectric devices contain many
thermocouples consisting of $n$-type (electron carriers) 
and $p$-type (hole carriers) semiconductors wired electrically in series and thermally in parallel
as shown in Fig.\,\ref{fig:takaba_fig01}.
A.\,F.\,Ioffe is the first to use the quantity $Z$ termed ``material's figure of merit" describing the efficiency of the thermal-to-electrical conversion \cite{Ioffe:1958a, Verdernikov:1998a}. 
How to increase the conversion efficiency can be found from the definition,
\begin{equation}
\label{eq:Fig_merit001}
Z=\frac{S^2\sigma}{\kappa_{\rm tot}}~~
[\mathrm{WK^{-2}m^{-1}}],
\end{equation} 
where $S(T)\,$[V/K] is the Seebeck coefficient, $\sigma(T)\,$[1/($\Omega$m)] the electrical conductivity,
and $\kappa_{\rm tot}=\kappa_{\rm el}+\kappa_{\rm ph}$ is the total thermal conductivity given by 
the sum of the electrical thermal conductivity 
$\kappa_{\rm el}$ and
the phonon thermal conductivity $\kappa_{\rm ph}$.
The physical meaning of Eq.\,(\ref{eq:Fig_merit001})
will be explained in \ref{subsubsec:figure_merit}.

The design concept proposed by A.\,F.\,Ioffe is to reduce the phonon thermal conductivity $\kappa_{\rm ph}(T)$ by forming semiconducting mixed crystals with $narrow$ band gaps composed of heavy elements. 
This approach has led to finding thermoelectric compounds based on Bi-Te alloys and Pb-Te alloys\,\cite{Ioffe:1954a, Ioffe:1956aa, Goldsmid:1954a, Sootsman:2009a, Kanatzidis:2010a, LaLonde:2011a}, which are in practical use at present, though
the elements on the lower right of the periodic table are toxic to humans.
Increasing interest in recent years in the recovery of waste heat has focused attention on earth and human friendly thermoelectric materials with improved efficiency.

Since $Z$ of Eq.\,(\ref{eq:Fig_merit001}) contains the electrical conductivity $\sigma$ in the numerator and the total thermal conductivity $\kappa_{\rm tot}$ in the denominator, 
high performance for thermoelectricity can be achieved for materials with the lowest possible thermal conductivity, the highest possible electrical conductivity and the highest possible Seebeck coefficient.
According to those ideas, efficient thermoelectric effects should be achieved by materials possessing both glass-like phonon thermal conductivity and crystalline electrical conductivity.
In this framework, \textcite{Slack:1995r} has proposed the important concept of ``phonon-glass electron-crystal" for designing efficient thermoelectric materials.
This is really a controversial concept from the aspect  of materials science.
If such materials can be
synthesized, it truly becomes one of the most significant innovations in alternative energy technologies. 
\textcite{Ioffe:1958b} noted in an article entitled ``\textit{The revival of thermoelectricity}" that 
``Evidently we stand on the threshold of a new era in power engineering, heating and refrigeration. 
This prospect flows from the now-rapid advance of thermoelectricity."
We have reached a new stage of thermoelectricity since the concept of phonon-glass electron-crystal introduced \cite{Slack:1995r}.

General phenomena in thermoelectricity have been reviewed in a number of articles 
\cite{Slack:1995r, Rowe:1995a, Mahan:1997a, Mahan:1998r, Sales:1998a, Nolas:2001r, Nolas:2001rr, Tritt:2001r, Dresselhaus:2001a, Sales:2002r, Chen:2003a, Rowe:2003a, Koumoto:2006a, Dresselhaus:2007a, Snyder:2008a,  Goldsmid:2010r, LaLonde:2011a, Gonzalves:2013r}.

\subsection{The concept of ``phonon-glass and electron-crystal'' exploring efficient thermoelectric materials}
\label{subsec:concept}
The concept \pg\ coined by \textcite{Slack:1995r} is actually a summary of ideas discussed by Ioffe in the 1950's\,\cite{Ioffe:1958a, Verdernikov:1998a}. 
In general, glasses and amorphous materials, having no periodic arrangement of atoms, show the lowest thermal conductivities.

The concept has stimulated numerous attempts to search for efficient thermoelectric materials, and the first achievement was for skutterudite compounds\,\cite{Nolas:1995a, Nolas:1996a, Sales:1996a}.
The filled skutterudites are described with a chemical formula $\mathcal{R}_y$T$_{4}$X$_{12}$ ($0 \leq y \leq 1$), 
where  $\mathcal{R}$ is a positive element of the lanthanides or actinides, or a group 1 or 2 element. 
The element T is a transition metal, and X is a pnictogen belonging to the group 15 elements (P, As, Sb).
The crystal structure of skutterudite involves voids (or cages composed of T$_{8}$X$_{12}$) which are filled by $\mathcal{R}$ guests\,\cite{ Takabatake:2006}.
Because the radius of the $\mathcal{R}$ filler is significantly smaller than that of the voids, 
the fillers are trapped in largely anharmonic potentials and  can vibrate with large amplitudes\,\cite{Chakoumakos:1999a}.

\textcite{Morelli:1997a} have reported the significant reduction in phonon thermal conductivity $\kappa_{\rm ph}$ of Ce$_y$Co$_{4}$Sb$_{12}$ relative to unfilled skutterudite CoSb$_{3}$.
It is remarkable that a small amount of Ce filler has a strong influence on the $\kappa_{\rm ph}$.
The similar outcome was observed in La-filled skutterudite La$_y$Co$_{4}$Sb$_{12}$\,\cite{Nolas:1998aaa}.
Furthermore, it has been revealed that all voids need not be occupied 
in order to achieve the maximum reduction in $\kappa_{\rm ph}$\,\cite{Tang:2001a}.
For instance, the value of $\kappa_{\rm ph}$ of Ce$_{y}$Fe$_{4-x}$Co$_{x}$Sb$_{12}$ with $y$=0.7 
has a minimum value lower than 2\,[Wm$^{-1}$K$^{-1}$] above room temperature \,\cite{Uher:2001r}.

The data strongly support that low phonon thermal conductivities $\kappa_{\rm ph}$ are obtained when guest atoms fill voids at random.
At the same time, since electrical conduction is mainly due to the pnictogen orbitals
in the group 15 elements, the carrier mobility remains large enough to maintain its crystalline character.
Thus,  in general, materials like skutterudites with  voids, cavities, cages, and empty sublattices are thought to be favorable candidates for
realizing the \pg\ concept.


Following these works, \textcite{Nolas:1998a, Nolas:1998aa} demonstrated that type-I clathrate compounds
have potential for use as high-efficiency thermoelectric materials. 
In fact, clathrate compounds belong to the category mentioned above.
Highly efficient thermoelectric properties have been exhibited by these compounds encapsulating  ``rattlers".
The terms ``rattler" or ``rattling" were introduced by \textcite{Sales:1997a} for the atoms $\mathcal{R}$=La, Ce, Th in the filled skutterudite structure $\mathcal{R}_{1-y}$Fe$_{4-x}$Co$_x$Sb$_{12}$ and Xe clathrate hydrates by\,\textcite{Tse:1997a} which are weakly bound in an oversized atomic cage.
\textcite{Sales:1997a} have stated that rare earth elements, undergoing incoherent rattling motion in cages, lower the phonon thermal conductivity at
room temperature down to a value comparable to that of vitreous silica.
\textcite{Tse:1997a} have noted that the coupling between   guest modes and acoustic phonons arising from networked cages lead to the glass-like behavior in the thermal conductivity of clathrate hydrates.
 
The type-I clathrate compounds constitute 
networked cages consisting of nano-scale tetrakaidecahedrons with 14 flat faces (14 hedrons) and  dodecahedrons with 12 flat faces (12 hedrons), in which the group 1 or 2 element in the periodic table can fill in the cages as the so-called ``rattling" guest atom.
We employ the term ``guest" in this article, though some use ``filler" instead. 
For $\mathcal{R}_{8}$Ga$_{16}$Ge$_{30}$ with the guest atom $\mathcal{R}$=Ba, Sr and Eu, \textcite{Sales:2001a} have found, by means of neutron diffraction measurement, that the nuclear density 
of guest atoms Sr and Eu are distributed in off-center positions apart from a center of 14 hedrons, while Ba takes on-center position in 14 hedrons.
Glass-like low phonon thermal conductivity has been realized in compounds with off-center guest atoms Sr and Eu, but not in those with on-center guest atom Ba.
It is crucial, though scaffold networks take crystalline cubic-structure, that there emerge glass-like phonon thermal conductivities \textit{depending~on} the states of guest atoms in tetrakaidecahedrons.  
\section{BACKGROUND ON THERMOELECTRICITY}
\subsection{The Wiedemann-Franz law and the Lorenz number}
\label{wiedeman_franz}
\textcite{Wiedemann:1853a} found, when investigating the thermal conductivity $\kappa$\,[W(mK)$^{-1}$] and the electrical conductivity $\sigma$\,[($\Omega$m)$^{-1}$] of several metals, that the two quantities are linearly proportional to each other at room temperature in the form  
 \begin{eqnarray}
\label{eq:Wiedemann_Franz0}
  \kappa=\rm{const.}\times\sigma.
\end{eqnarray}
This relation indicates that electric and thermal conductions are definitely correlated. 
20 years later, L.\,V.\,\textcite{Lorenz:1872a} found, from the data of the temperature dependence of Eq.\,(\ref{eq:Wiedemann_Franz0}), the ratio is also proportional  to  the absolute  temperature $T$ as
\begin{eqnarray}
\label{eq:Wiedeman1}
  \kappa= LT\sigma,
\end{eqnarray}
where $L$ is a proportionality constant termed the Lorenz number.
This relation is  generally called the Wiedemann-Franz law or the Wiedemann-Franz-Lorenz law.

The first theoretical derivation of the Lorenz number was made by \textcite{Sommerfeld:1927a, Sommerfeld:1928a} using the Fermi-Dirac distribution function of degenerate metals, which provides
\begin{eqnarray}
\label{eq:Lorenz1}
L_0=\frac{\pi^2}{3} \left( \frac{k_{\rm B}}{e} \right)^2=2.44 \times 10^{-8}\,[\rm{W\Omega K{^{-2}}}],
\end{eqnarray}
where $k_\mathrm{B}$ is the Boltzmann constant, and $e$ the electron charge.  
The $L_0$ defined via universal constants is called the Sommerfeld value of the Lorenz number. 
This relation can be 
simply derived from the free electron model.
The expression of the conductivity $\sigma$ is 
\begin{eqnarray}
\label{eq:Drude1}
  \sigma=\frac{ne^2\tau_{\rm el}}{m_{\rm e}^*},
\end{eqnarray}
where $\tau_{\rm el}$ is the average collision time of the electrons, $n$ the number of electrons per volume at the Fermi energy, and $m_{\rm e}^*$ is the effective mass of an electron.
The electronic contribution to the thermal conductivity $\kappa_{\rm el}$ is expressed in the same way  by the relation
\begin{eqnarray}
\label{eq:Drude2}
  \kappa_{\rm el}=\frac{1}{3}C_{\rm el} v_{\rm F}\ell_{\rm el}=\frac{\pi^2nk_{\rm B}^2T\tau_{\rm el}}{3m_{\rm e}^*},
\end{eqnarray}
where $C_{\rm el}$ is the  specific heat of the electrons, $v_{\rm F}$ the Fermi velocity, and $\ell_{\rm el}$ the mean free path of the electrons.

The ratio of Eqs.\,(\ref{eq:Drude1}) and (\ref{eq:Drude2}) provides
\begin{eqnarray}
\label{eq:Lorenz00}
\frac{\kappa_{\rm el}}{\sigma}= \frac{\pi^2}{3} \left( \frac{k_{\rm B}}{e} \right)^2 T=L_0T,
\end{eqnarray}
where the collision time $\tau_{\rm el}$ is canceled out.
The mean free path $\ell_{\rm el}$ is reduced by collision with impurities and phonons in impure metals or in disordered alloys or semiconductors.
Note that Eq.\,(\ref{eq:Lorenz00}) has been obtained from the average collision-time approximation for free electrons in addition to the assumption that only electrons carry heat.

The Wiedemann-Franz law is valid for the cases that the elastic scattering of electrons dominates.
The Wiedemann-Franz law holds, to a good approximation, at low temperatures where the elastic scattering by impurities or defects is relevant and at high temperatures where the change in energy of each electron scattered by electron-electron or electron-phonon interactions is negligible compared with the energy scale $\kB T$.
At intermediate temperatures 10-200\,[K], electronic energy-loss becomes of the order of $\kB T$ and Eq.\,(\ref{eq:Lorenz00}) is no longer valid.
See, for example, \textcite{Ashcroft:1976} on the detailed arguments beyond the relaxation-time approximation for electron transport relevant to the Wiedemann-Franz law.

It had been already pointed out by \textcite{Lorenz:1881a, Lorenz:1881aa} that  $L$ is not a  universal factor,  but depends  on the kind of metal.
The actual values of the Lorenz number $L$ are given, for example, in \textcite{Kaye:1966} and \textcite{Kumar:1993a} for elemental metals, semi-metals, metallic alloys and compounds, and degenerate semiconductors.
These provide $L$=2.3$-$2.6$\times$10$^{-8}$\,[W$\Omega$K$^{-2}$] close to the Sommerfeld value $L_{\rm 0}$ for almost all elemental metals and smaller values $L$=1.7$-$2.5$\times 10^{-8}\,[\rm{W\Omega K{^{-2}}}]$ 
in degenerate semiconductors.
In contrast, the Lorenz  numbers of transition metals deviate considerably from the Sommerfeld value\,\cite{Chari:1989a,Chari:1990a}.



The experimental values of the thermal conductivity $\kappa$ in Eq. \,(\ref{eq:Wiedeman1}) is the total thermal conductivity $\kappa_{\rm tot}$ including the phonon contribution to heat transport.
The phonon component of the thermal conductivity $\kappa_{\rm ph}$  can be estimated by subtracting the electric component $\kappa_{\rm el}$ from the measured total thermal conductivity $\kappa_{\rm tot}$.  
Suppression of the electronic contribution to the thermal  conductivity, and hence the separation of the phonon and electronic parts of conductivity can be achieved by applying a transverse magnetic field. 

In the presence of a magnetic field, the Lorenz  number varies with it. 
In this connection, we should mention the Nernst-Ettingshausen effect\,\cite{Ettingshausen:1886a}.
This is a thermomagnetic phenomenon 
where an electric current is generated perpendicular to an applied magnetic field and a temperature gradient.
This was discovered by Etttingshausen and his student Nernst in 1886 
when heating one side of the sample during the investigation of the Hall effect in bismuth.
The reverse effect is called the Ettingshausen effect in which a temperature gradient appears
when applying a current and a magnetic field.
There is a close correspondence between the expressions on cooling power and the coefficient of performance obtained by using Ettingshausen and Peltier effects
\cite{O'Brien:1958a}.
The Ettingshausen refrigerator requires only one material, so the issue of matching $n$-type and $p$-type transport does not arise.
Detailed discussions about the Ettingshausen effect are given, for example, in \textcite{Nolas:2001r}.

\subsection{Heat and electric transport as irreversible processes}
\label{subsec:transport}
\subsubsection{Peltier and Seebeck coefficients}
\label{subsubsec:Peltier_Seebeck}
 The Onsager reciprocal relations express the equality of certain ratios between generalized fluxes and forces based on the principle of detailed balance when a notion of local equilibrium is applicable.
The thermoelectric effects discussed here deal with the relations between two kinds of flux (the heat flux density $\QQ$ and the particle flux density $\JJ$), which are associated with two kinds of generalized forces (the gradient of the inverse of temperature $\nabla (1/T)$ and the gradient of electrochemical potential $\nabla \mu$).
\textcite{Callen:1948a, Callen:1952a} have developed a general argument of the application of Onsager's theory  to thermoelectric phenomena.
The relations given by \textcite{Callen:1948a} are expressed in the form of linear combinations among the generalized fluxes and forces, 
\begin{eqnarray}
\label{eq:Onsager_1}
-\JJ &= L_{11}\frac{1}{T}\nabla{\mu} + L_{12}{\bf \nabla}\left( \frac{1}{T}\right),\\
\label{eq:Onsager_2}
\QQ &= L_{21}\frac{1}{T}\nabla{\mu} + L_{22} {\bf \nabla}\left( \frac{1}{T}\right),
\end{eqnarray}
where  $\mu$ is the electrochemical potential composed of both a chemical potential $\mu_C$ and an electrical one $\mu_E$ with $\mu=\mu_C+\mu_E$.
For the charge $e$ of a particle, $\mu_E=e\phi$, where $\phi$ is the electrostatic potential.
The chemical potential $\mu_C$ is a function of the temperature $T$ and the particle density $n$, and the $\nabla\mu_C$ plays a minor role in electrical conductors.
The Onsager reciprocal relation\,\cite{Onsager:1931a} states that 
\begin{eqnarray}
\label{eq:coefficients}
L_{12}=L_{21}.
\end{eqnarray}
Provided that an applied magnetic field $\HH$ exists, this relation becomes $L_{ij}(\HH)=L_{ji}(-\HH)$.

The kinetic coefficients $L_{ij}$ are closely related
to the electrical $\sigma$\,[($\Omega$m)$^{-1}$] and thermal conductivity $\kappa$\,[W(Km)$^{-1}$] as seen by expressing Eqs.\,(\ref{eq:Onsager_1}) and (\ref{eq:Onsager_2}) in terms of $\sigma$ ad $\kappa$.
Here $\sigma$ is defined as the electrical current density $\jj=e\JJ$ per unit electric field  $\EE=-\nabla\mu_E/e$ in an isothermal system, which provides 
\begin{eqnarray}
\label{eq:sigma}
   \sigma=\frac{\jj}{\EE}=-\frac{\jj}{\nabla\mu_E/e},~~~~\mathrm{for}~~\nabla T=0. 
\end{eqnarray}
Substituting Eq.\,(\ref{eq:sigma}) into Eq.\,(\ref{eq:Onsager_1}), we have the relation
\begin{eqnarray}
\label{eq:sigma11}
   \sigma=\frac{e^2L_{11}}{T}.
\end{eqnarray}
The thermal conductivity  $\kappa$ is the heat current $\QQ$ per unit temperature gradient given by 
\begin{eqnarray}
\label{eq:Thermal_Conduc11}
   \kappa=-\frac{\QQ}{\nabla T},~~~~\mathrm{for}~~\jj=0.
\end{eqnarray}
The above can be related to the kinetic coefficient $L_{ij}$ using Eqs.\,(\ref{eq:Onsager_1}) and (\ref{eq:Onsager_2}), which is written down as
\begin{eqnarray}
\label{eq:Thermal_Conduc22}
   \kappa=\frac{L_{11}L_{22}-L_{12}^2}{T^2L_{11}}.
\end{eqnarray}

The Seebeck effect implies the temperature gradient $\nabla T$ yields the electric field $\EE'$ in a system.
This is expressed as $\EE'=-S\nabla T$, where $S$[V/K] is called the Seebeck coefficient or the thermoelectric power of a medium.
The use of $S$ enables us to rewrite Eqs.\,(\ref{eq:Onsager_1}) and (\ref{eq:Onsager_2}) in terms of the electric field $\EE$\,[V/m] and the temperature gradient $\nabla T$\,[K/m] as
\begin{eqnarray}
\label{eq:Callen_1}
\jj &=& \sigma \left(\EE-S\nabla T\right),\\
\label{eq:Callen_2}
\QQ &=& \sigma ST \EE-\kappa\nabla T,
\end{eqnarray}
where $S=-eL_{12}/(\sigma T^2)$.

Equations (\ref{eq:Callen_1}) and (\ref{eq:Callen_2}) recover the Onsager reciprocal relation $L_{12}=L_{21}=-T^2\sigma S/e$.
From Eqs.\,(\ref{eq:Callen_1}) and (\ref{eq:Callen_2}) with $\nabla T=0$, the Peltier coefficient $\Pi$ is defined by 
\begin{eqnarray}
\label{eq:Peltier_0}
   \QQ=TS\jj~~\longrightarrow~~S=\frac{\Pi}{T}.
\end{eqnarray}
Thus, the Peltier and the Seebeck effects are essentially the same in thermoelectrics. 
The Peltier coefficient has a simple physical meaning
of the energy carried by charged particles per unit charge.

The Seebeck coefficient $S$ becomes negative for electron carriers such as the case of $n$-type semiconductors, and positive for hole carriers in $p$-type semiconductors.

\subsubsection{Figure of merit}
\label{subsubsec:figure_merit}
The direct expression of the heat current $\QQ$ driven by an electrical current $\jj$ and the temperature gradient $\nabla T$ is obtained from Eqs.\,(\ref{eq:Callen_1}) and (\ref{eq:Callen_2}).
This yields 
\begin{eqnarray}
\label{eq:Seebeck_11}
   \QQ=ST\jj-\kappa_{\rm tot}\nabla T,~~\mathrm{with}~\kappa_{\rm tot}\equiv\kappa\left( 1-Z'T\right), 
\end{eqnarray}
where $Z'\equiv\sigma S^2/\kappa$.
Since $\kappa_{\rm tot}$ is a positive quantity, $Z'$ is bounded by the inequality $Z'T<1$ as seen from Eq.\,(\ref{eq:Seebeck_11}).

Instead of $Z'$, let us introduce the quantity called the material's figure of merit defined by\,{\cite{Mahan:1998r}
\begin{eqnarray}
\label{eq:Fig_Merit_0}
   Z=\frac{\sigma S^2}{\kappa_{\rm tot}}=\frac{Z'}{1-Z'T},
\end{eqnarray}
where $Z$ has the dimension of inverse temperature, so the efficiency of thermoelectric conversion is alternatively defined by
the dimension-less parameter $ZT$ termed ``dimensionless figure of merit", which is given by
\begin{eqnarray}
\label{eq:fig_merit1}  
ZT=\frac{S^2\sigma T}{\kappa_{\rm eh}+\kappa_{\rm ph}}.
\end{eqnarray}
Here $\kappa_{\rm tot}=\kappa_{\rm ph}+\kappa_{\rm eh}$ is the total thermal conductivity composed of the phonon part $\kappa_{\rm ph}$ and the electronic part $\kappa_{\rm eh}$.
We use the notation of $\kappa_{\rm eh}$ expressing hole or electron carriers in semiconductors.
When electron-hole pairs are present at temperatures comparable to the gap energy, a bipolar term contributing to the electric thermal conductivity should be taken into account.
See, for example, \textcite{Yang:2004}.

The Carnot limit (also known as the Carnot efficiency) specifies limits on the maximum efficiency that any heat engine can obtain. 
The Carnot limit solely depends on the temperature difference between the hot ($T_{\rm H}$) and cold ($T_{\rm C}$) temperature reservoirs.
It is clear that the hotter the heat source, the higher the possible efficiency.
The maximum efficiency $\eta_{\rm Carnot}$ is given by the dimensionless ratio between the temperature difference $\Delta T=T_{\rm H}-T_{\rm C}$ and $T_{\rm H}$.
The formula for the actual efficiency $\eta$ of a thermoelectric generator is expressed by using  $\eta_{\rm Carnot}$ as \cite{Goldsmid:2010r}
\begin{eqnarray}
\label{eq:Carnot}  
     \eta= \eta_{\rm Carnot}\frac{\sqrt{1+ZT}-1}{\sqrt{1+ZT}+T_{\rm C}/T_{\rm H}}.
\end{eqnarray}

Thus the performance of thermoelectric materials is characterized by $ZT$  determined by three parameters: the Seebeck coefficient $S$\,[V/K], the electrical conductivity $\sigma$\,[($\Omega$m)$^{-1}$], and the thermal conductivity $\kappa$\,[W(Km)$^{-1}$].
Since it is required for efficient $\eta$ to take values above around 10\% at the temperature difference of 300\,[K],
thermoelectric materials satisfying the relation $ZT>$1 are thought to be especially efficient in practical applications.

\subsubsection{The Mott formula for the Seebeck coefficient}
\label{subsubsec:Seebeck_coeff}

In three dimensional systems, the $x$-component of the electric current density $j_x$ and the electric heat flux density $Q_x$ under the electric field  $F_x$ can be expressed in terms of the electron mobility $\mu(E)$ of the charge $-e<0$ as
\begin{eqnarray}
\label{eq:Fritzsche_01}
   j_x &=& eF_x\int\mu(E)f \left( 1-f\right)N(E)dE ,\\
\label{eq:Fritzsche_02}
   Q_x &=& -F_x\int\mu(E)\left( E-\zeta\right) f\left( 1-f\right) N(E)dE,
\end{eqnarray}
where $N(E)$ is the electron density of states, $\zeta$ the chemical potential, and $f(E)$ the Fermi-Dirac distribution function. 
Equation~(\ref{eq:Fritzsche_01}) provides the conductivity expressed by an integral over the single electron states\,\cite{Fritzsche:1971a}  
\begin{eqnarray}
\label{eq:Electric_coduc_0}  
   \sigma =e\int \mu(E)f\left(1-f\right)N(E)dE\equiv\int\sigma(E)dE,
\end {eqnarray}
where $\sigma(E)dE$ is the conductivity in the energy interval between $E$ and $E+dE$.

The product of the Peltier coefficient $\Pi$ and the conductivity $\sigma$ is obtained by differentiating Eq.~(\ref{eq:Fritzsche_02}) with respect to the electric field $F_x$.
The Seebeck coefficient is obtained from Eqs.\, (\ref{eq:Callen_2}) and (\ref{eq:Fritzsche_02}) in comparison,
\begin{eqnarray}
\label{eq:Seebeck_10} 
   S(T)= -\frac{\kB}{e}\int\left(\frac{E-\zeta}{\kB T}\right) 
   \frac{\sigma(E)}{\sigma}dE.
\end{eqnarray}
This yields $S<0$ for electrons, and $S>0$ for holes  taking account of the signs of $-e$.

We can set $f(1-f)=-\kB T \partial f/dE$ when only states near the Fermi energy $E_{\rm F}$ contribute to the conductivity.
By expanding $\sigma(E)$ defined in Eq.\,(\ref{eq:Electric_coduc_0}) around $E=E_{\rm F}$ up to the order $(k_{\rm B}T/\Ef)^2$, we have the convenient expression of
the Seebeck coefficient from Eq. (\ref{eq:Seebeck_10}) of the form
\begin{eqnarray}
\label{eq:Mott_Jones1}    
   S(T)\cong -\frac{\pi^2 k_{\rm B}^2 T}{3e} \left[ \frac{\partial\rm{ln}\sigma(\it{E})}{\partial E} \right]_{E =\Ef}.
\end{eqnarray} 
This is the Mott formula\,\cite{Mott:1936r}, showing that the Seebeck coefficient $S$ is proportional to the logarithmic derivative of $\sigma(E)$ or $\mu(E)N(E)$ with respect to the energy at $E =\Ef$.

The Mott formula can be applied not only to monovalent metals, but also to other metals or degenerate semiconductors. 
It is valid even for alloys at temperatures above the Debye temperature $\theta_{\rm D}$,
and at low temperatures as long as the residual resistivity is large compared with the resistivity due to phonon scattering, in addition to doped semiconductors exhibiting the metal-insulator Anderson transition\,\cite{Cutler:1969a}.
Since the Mott relation is correct only to second order in $k_{\rm B}T/\zeta$, it is not applicable to transition metals with narrow bands and at high temperatures\,\cite{Chari:1989a, Chari:1990a}.
\textcite{Kirchner:2013a} have theoretically discussed the effect of nonlinear thermoelectric response for systems with narrow band
structures.

Equation (\ref{eq:Seebeck_10}) shows that the Seebeck coefficient $|S|$ is proportional to the resistivity $\rho=1/\sigma$ and the energy difference $E-\zeta$ from the chemical potential. 
This indicates that heavily-doped degenerate semiconductors are suitable for increasing $|S|$. 
Equation (\ref{eq:Mott_Jones1}) in the degenerate limit yields, for example, from the definitions $\mu (E)\propto E^\alpha$, with $\alpha$ characterizing the scattering mechanisms, and the energy dependence of the electron density of states $N(E)\propto E^{3/2}$,
\begin{eqnarray}\label{eq:Mott_Jones3}
S \cong -\frac{\pi^2 k_{\rm B}^2 T}{3e\Ef} \left(\alpha +1.5\right),
\end{eqnarray}
where $\alpha$=$-$0.5 for scattering by acoustic phonons and $\alpha$=1.5 for ionized impurity scattering.
The actual evaluation of the above equation gives
$S$\,$\cong-$2.44$\times 10^{-2}T\left(\alpha+1.5\right)/\Ef$[eV] in units of [$\mu$V/K].
This is a measure of the Fermi energy $E_{\rm F}$.

The discussion given above has been focused on degenerate electrons. In semiconductors, the Fermi level lies in the gap and there exist positively charged holes. 
These positive carriers contribute to the transport in addition to negatively charged electrons.  
One needs to assign different time-scales to electrons and holes depending on the band index\,\cite{Nolas:2001rr, Yang:2004}.

\section{CRYSTAL STRUCTURES OF TYPE-I CLATHRATE COMPOUNDS}
\label{sec:type_I_structures}

\subsection{Various types of clathrate compounds}
\label{sec:Various}
\subsubsection{Classification of clathrate compounds}
Clathrates or clathrate compounds consist of regular lattices of cages in which guest atoms or molecules are encapsulated. 
Cages with restricted geometry form the basic framework of a lattice with translational invariance. 
Clathrates are also called host-guest complexes. 
Typical examples of host-guest complexes are inclusion compounds and intercalation compounds. 
Gas hydrate is a special type of clathrate compound consisting of water molecules and gas molecules. 
The history of gas hydrates can be traced back to the work of 200 years ago 
by \textcite{Davy:1811a, Davy:1811aa}, who discovered chlorine hydrate 
in the combination of  water host molecule (H$_{2}$O) and guest chlorine molecule (Cl$_2$) when pressurizing water with chlorine gas below 9\,$^\circ$C. 
Davy and his assistant Michael Faraday continued the research on chlorine hydrate at the Royal Institution of London from 1813. 
\textcite{Faraday:1823a} determined the composition of chlorine clathrate hydrate of nearly 1 part of Cl$_2$ and 10 parts of H$_2$O molecules.

The water molecule  H$_{2}$O forms a variety of cages different in size and in geometrical structure.
Typical examples of hydrate-forming substances include CH, CO, and HS molecules.
The terms ``gas hydrate" and ``clathrate hydrates" have been used for these solids\,\cite{Davidson:1973r}.
Oxygen atoms forming a cage are tetrahedrally connected 
by hydrogen bonds, which make cages with open space for guest atoms/molecules.
In the early stages of this research, the vast majority of clathrate hydrates had been classified into two types, 
which are type-I and type-I$\hspace{-.1em}$I belonging to the cubic space groups $Pm\bar{3}n$ and $Fd\bar{3}m$, respectively
\,\cite{Pauling:1952a, Muller:1952a, Claussen:1951a, Claussen:1951b, Claussen:1951c, von Stackelberg:1951a, von Stackelberg:1951b}. 
\textcite{Jeffrey:1984r} has classified clathrate hydrates into seven types 
by introducing five further types of clathrate hydrates in addition to type-I and -I\hspace{-.1em}I structures.
Crystal structures of type-I\hspace{-.1em}I\hspace{-.1em}I, -V\hspace{-.1em}I\hspace{-.1em}I\hspace{-.1em}I, and 
-I\hspace{-.1em}X 
in addition to type-I and -I\hspace{-.1em}I are shown in Fig.\,\ref{fig:takaba_fig02}.
\begin{figure*}[htb] 
\includegraphics[width=0.8\linewidth]{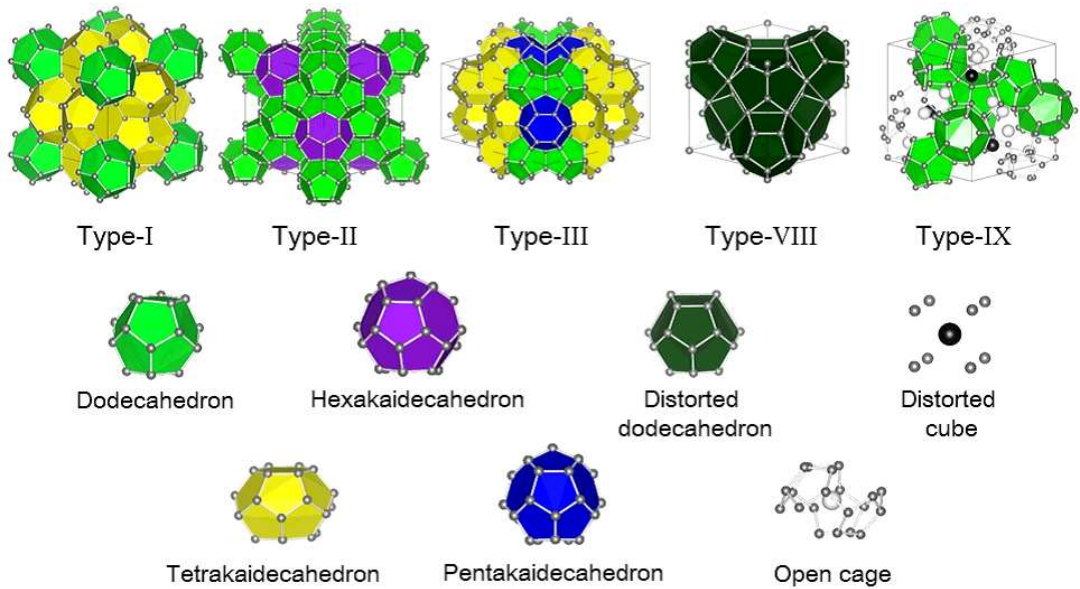}
\caption{\label{fig:takaba_fig02}(Color online) Crystal structure of type-I, type-I\hspace{-.1em}I, type-I\hspace{-.1em}I\hspace{-.1em}I, 
type-V\hspace{-.1em}I\hspace{-.1em}I\hspace{-.1em}I, and type-I\hspace{-.1em}X intermetallic clathrates.} 
\end{figure*}

The unit cell of the type-I structure consists of 46 water molecules which form two types of small and large cages. 
The small cages in the unit cell are two against six large ones. 
The small cage has the shape of a pentagonal dodecahedron (5$^{12}$), 
and the large one that of a tetrakaidecahedron (5$^{12}$6$^{2}$), as given in Fig.\,\ref{fig:takaba_fig02}. 
Typical guests in type-I clathrate hydrates are CO$_2$ in carbon dioxide hydrate and CH$_4$ in methane hydrate.
Methane hydrates have attracted  much interest as a possible new energy source, 
in which large amounts of methane gas are contained both in permafrost formations and under the ocean sea-bed\,\cite{Englezos:1993r, Hester:2009r}.
Hydrogen storage in clathrates is expected to have potentially important energy applications\,\cite{Struzhkin:2007r}.

The unit cell of type-I\hspace{-.1em}I clathrate hydrates consists of 136 water molecules, forming also small and large cages.
The small cages in the unit cell are sixteen against eight large ones. 
The small cage is again a pentagonal dodecahedron (5$^{12}$), but the large one is a hexakaidecahedron (5$^{12}$6$^{4}$). 
Type-I\hspace{-.1em}I hydrates contain larger molecules like CCl$_4$ and SF$_6$ than in type-I hydrates.

The compositions of the type-I and I\hspace{-.1em}I hydrates are expressed as
2D$\cdot$6T$\cdot$46H$_{2}$O and 16D$\cdot$8H$\cdot$136H$_{2}$O, respectively, 
where D, T, and H represent, respectively, the guest atoms/molecules 
in a cage of pentagonal dodecahedron, tetrakaidecahedron, and  hexakaidecahedron composed of water molecules.

\subsubsection{Intermetallic clathrate compounds}

Intermetallic clathrates are inorganic inclusion compounds structurally related to the hydrates\,\cite{Rogl:2006r}.
The clathrates composed of the group 14 elements were discovered as prospective thermoelectric materials which realize the ``phonon-glass electron-crystal" concept \,\cite{Nolas:1998aa, Blake:1999a, Slack:1995r, Iversen:2000a}. 
The group 14 clathrates have been intensively investigated because of their high potential for converting temperature differences to electric energy\,\cite{Kuznetsov:2000a, Saramat:2006a, Kleinke:2010a, Sootsman:2009a, Christensen:2006aaa}.

Frameworks of the group 14 clathrate compounds are constructed from face-sharing polyhedron. 
Five structures given in Fig.\,\ref{fig:takaba_fig02} are known as type-I (2D$\cdot$6T$\cdot$46F), type-I\hspace{-.1em}I (16D$\cdot$8H$\cdot$136F), 
type-I\hspace{-.1em}I\hspace{-.1em}I (10D$\cdot$16T$\cdot$4P$\cdot$172F), type-V\hspace{-.1em}I\hspace{-.1em}I\hspace{-.1em}I (8D$'\cdot$46F), 
and type-I\hspace{-.1em}X (8D$\cdot$12O$\cdot$4C$\cdot$100F)\,\cite{Rogl:2006r, Karttunen:2011a, Mudryk:2002a}.
Here P, D$'$, O, and C express guests in pentakaidecahedron, distorted dodecahedron, 
open cage, and distorted cube, respectively, and F means a framework (cage) atom.
D, T, and H are defined in the previous paragraph.
The type-V\hspace{-.1em}I\hspace{-.1em}I\hspace{-.1em}I and I\hspace{-.1em}X have no representatives among hydrate structures.

A silicon clathrate, for example, is formed by the covalently bonded Si atoms. 
Thereby, Si atoms and $sp^3$ hybridized orbital electrons play roles of O atoms and H atoms in clathrate hydrates
due to their tetrahedral coordination and hydrogen bonding by ``ice rules"\,\cite{Bernal:1933a, Pauling:1948r}. 
\textcite{Kasper:1965a} discovered the first silicon clathrates, type-I Na$_{8}$Si$_{46}$ and type-I\hspace{-.1em}I Na$_{x}$Si$_{136}$,
where cationic Na guests are enclosed in the silicon framework. 
The synthesis of Ge- and Sn-based type-I clathrates with alkali metal atoms as guests was reported by~\textcite{Gallmeier:1969a}.

Ternary type-I clathrates $\mathcal{R}_{8}$M$_{16}$Z$_{30}$, where $\mathcal{R}$ belongs to the group 2, 
M the group 13, and Z the group 14 elements, respectively, were first synthesized by~\textcite{Eisenmann:1986a}.
The $\mathcal{R}$ guest atom is encapsulated by the cage formed by M and Z elements. 
The combination of M and Z elements makes it possible to tune the size of cages.  
When the cage has a large open-space, the encapsulated guest atoms are weakly bounded to the cage.
Such a loosely bound guest is referred to as a ``rattler''\,\cite{Sales:1997a}.
The present article is concerned with type-I $\mathcal{R}_{8}$M$_{16}$Z$_{30}$ clathrate 
with potential application to efficient thermoelectric materials.

Structural and thermoelectric properties for various types of intermetallic clathrates, 
including type-I\hspace{-.1em}I and type-I\hspace{-.1em}I\hspace{-.1em}I clathrates, 
have been treated in the following works: type-I $\mathcal{R}'_{8}$M$_{8}$Z$_{38}$ with alkali metal guest $\mathcal{R}'_{8}$ 
\,\cite{Bobev:2000a, Myles:2001a, Nolas:2000aaa, Tanaka:2010a, Hayashi:2010a};
type-I with transition metal element as cage atom 
\,\cite{Anno:2003a, Akai:2005a, Johnsen:2007a,Christensen:2009a, Christensen:2010a, Melnychenko-Koblyuk:2009a, Nguyen:2010a, Nasir:2010a, Zhang:2011a, Xu:2012a};
type-I mainly constructed from the group 15 elements (As and Sb)
\,\cite{Liu:2009a, He:2012a};
type-I\hspace{-.1em}I\,\cite{Kasper:1965a, Beekman:2008a, Beekman:2010a}, 
type-I\hspace{-.1em}I\hspace{-.1em}I\,\cite{Bobev:2001a, Zaikina:2008a, Zaikina:2010a}, and 
type-I\hspace{-.1em}X clathrates\,\cite{Kroner:1988a, Carrillo-Cabrera:2000a, Fukuoka:2000a, Kim:2000a, Paschen:2002a, Rachi:2005a, Kim:2007a, Kim:2007aa};
``inverse clathrates'' with halogen/tellurium guests\,\cite{Menke:1973a, Kishimoto:2006a, Kishimoto:2007a, Zaikina:2010a, Falmbigl:2012r}.

\subsubsection{Crystal chemistry of type-I clathrate}
\label{sec:cryschem}

\begin{figure} 
\epsfysize=2.3in 
\epsfbox{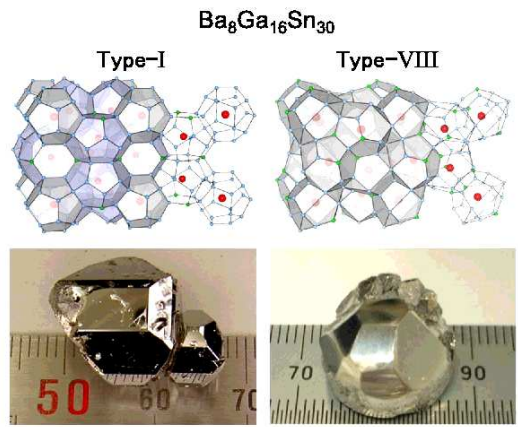}
\caption{\label{fig:takaba_fig03}  (Color online) Crystal structures and single crystals 
of type-I and type-V\hspace{-.1em}I\hspace{-.1em}I\hspace{-.1em}I Ba$_{8}$Ga$_{16}$Sn$_{30}$.
The units of rulers are given in millimeter.
After the work of \textcite{Suekuni:2008a}} 
\end{figure}

Among various types of intermetallic clathrate compounds, the type-I structure of $\mathcal{R}_{8}$M$_{16}$Z$_{30}$ and $\mathcal{R}'_{8}$M$_{8}$Z$_{38}$ stoichiometry has been mostly adopted, where
$\mathcal{R}$ is a guest atom of the group 2 (Ba, Sr) or a divalent rare-earth element (Eu), and $\mathcal{R}'$ a group 1 element (Na, K, Rb, and Cs).
The cage constituents M and Z are the elements of the group 13 (Al, Ga, and In) and the group 14 (Si, Ge, and Sn), respectively\,\cite{Eisenmann:1986a, Bobev:2000a}
It should be noted that only Ba$_{8}$Ga$_{16}$Sn$_{30}$ and Eu$_{8}$Ga$_{16}$Ge$_{30}$
are known to form the type-V\hspace{-.1em}I\hspace{-.1em}I\hspace{-.1em}I structure 
with the same composition as the type-I\,\cite{Eisenmann:1986a, Schnering:1998a, Carrillo-Cabrera:2002a, Paschen:2001a, Leoni:2003a}.
Large single crystals of type-I and type-V\hspace{-.1em}I\hspace{-.1em}I\hspace{-.1em}I Ba$_{8}$Ga$_{16}$Sn$_{30}$, 
which are shown in~Fig.\,\ref{fig:takaba_fig03}, were grown by a flux method using excess amounts of Ga and Sn\,\cite{Suekuni:2008a}.
The type-V\hspace{-.1em}I\hspace{-.1em}I\hspace{-.1em}I structure can be formed in alloyed compounds
Sr$_{8}$Al$_{x}$Ga$_{16-x}$Si$_{30}$ ($8 \leq x \leq 13$) and Sr$_{8}$Al$_{x}$Ga$_{y}$Ge$_{46-x-y}$ 
($6 \leq x \leq 7$ and 10$ \leq y \leq11 $)~\,\cite{Sasaki:2009a, Kishimoto:2008a, Shimizu:2009aa}.
This structure is cubic (space group No.217; $I\bar{4}3m$) 
and the unit cell consists of eight distorted dodecahedra, as shown in Fig.\,\ref{fig:takaba_fig02}.
Aspects of the chemistry and synthesis of intermetallic clathrates have been reported 
in review articles\,\cite{Shevelkov:2011a, Kovnir:2004a}.

Most of the type-I and the type-V\hspace{-.1em}I\hspace{-.1em}I\hspace{-.1em}I clathrates follow the so-called Zintl concept\,\cite{Paschen:2003a, Shevelkov:2011a, Rodriguez:2010a, Toberer:2010a}.
In a Zintl compound, each constituent attains a closed valence shell by combining a formal charge transfer with covalent bonds. 
In intermetallic clathrate compounds, the cage atom Z is partially substituted by the acceptor atoms M 
for charge compensation between the guest and the cage.
Consequently, the cage atoms are tetrahedrally bonded by a $sp^3$ hybridized orbital as shown in Fig.\,\ref{fig:takaba_fig04}.
\begin{figure}[t]
\epsfysize=3.0in 
\epsfbox{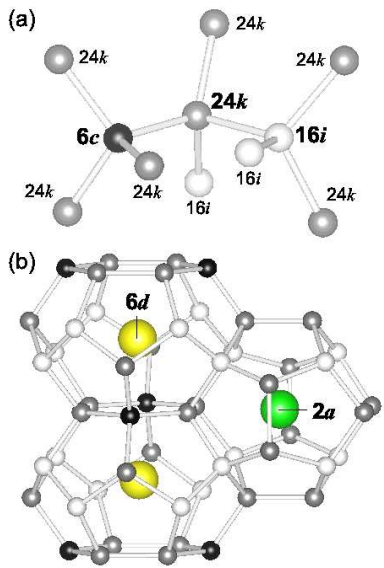}
\caption{\label{fig:takaba_fig04} (Color online) (a) Tetrahedral bonds of framework atoms, and (b) guests in dodecahedral and tetrakaidecahedral cages for type-I clathrates.} 
\end{figure}

In the type-I structure the guests occupy two sites as shown in~Fig.\,\ref{fig:takaba_fig04}(b): 
the 2$a$ site in the dodecahedron and the 6$d$ site in the tetrakaidecahedron.
The cage atoms M and Z in type-I $\mathcal{R}_{8}$M$_{16}$Z$_{30}$ and $\mathcal{R}'_{8}$M$_{8}$Z$_{38}$ occupy three crystallographic sites 6$c$, 16$i$, and 24$k$ sites. 
The 6$c$ site is bound to four 24$k$ sites; the 16$i$ site, to three 24$k$ sites and one 16$i$ site; 
the 24$k$ sites, to one 6$c$ site, two 16$i$ sites, and one 24$k$ site.  
Note that there are no equivalent positions around the 6$c$ site.
The M and Z atoms do not order in the host framework, but are distributed by a certain rule (see below).
First-principles calculations by \textcite{Blake:2001a, Blake:1999a} and \textcite{Gatti:2003a} have suggested that the M atoms in $\mathcal{R}_{8}$M$_{16}$Z$_{30}$ preferentially occupy the 6$c$ site 
which does not have adjoining equivalent sites.
This is because the bonds between M atoms are energetically unfavorable.  
In fact, as described in the next subsection, the preferential occupation of the M atom at 6$c$ sites 
was experimentally proved from analysis of x-ray and neutron diffraction experiments\,\cite{Eisenmann:1986a, Zhang:2002a, Bentien:2005a, Christensen:2006a, Christensen:2007a, Christensen:2010a}. 
For type-I Ba$_{8}$Ga$_{16}$Sn$_{30}$, extended x-ray absorption fine-structure\,(EXAFS) analyses have shown that the Ga-Ga bond is unfavorable;
only 15\% of the Ga nearest neighbors are Ga\,\cite{Kozina:2009a}.

First-principles calculations have pointed out that type-I clathrates 
with $\mathcal{R}_{8}$M$_{16}$Z$_{30}$ stoichiometry have a semiconducting electronic structure\,\cite{Blake:2001a, Blake:1999a, Madsen:2003a, Nenghabi:2008a, Uemura:2008a, Kono:2010a, Akai:2009a}.
Despite the balanced electron count [$\mathcal{R}^{2+}$]$_{8}$[M$^{1-}$]$_{16}$[Z$^{0}$]$_{30}$, 
these compounds often behave like a metal or a doped semiconductor in their electrical conduction.
This is attributed to a minor off-stoichiometry: [$\mathcal{R}^{2+}$]$_{8}$[M$^{1-}$]$_{16+x}$[Z$^{0}$]$_{30-x}$ and/or vacancy formation on the cage sites.
It is crucial for efficient thermoelectric materials that 
the carrier type and the carrier density can be tuned by adjusting $x$ in $\mathcal{R}_{8}$M$_{16+x}$Z$_{30-x}$. 
For instance, in type-I Ba$_{8}$Ga$_{16+x}$Ge$_{30-x}$, a positive value of $x$ yields $p$-type conduction and negative $x$ leads to $n$-type\,
\cite{Anno:2002a, Anno:2003a, Avila:2006a, Avila:2006aa, Tang:2010aa}.
A tunable carrier type in one compound has the advantage of the possibility of constructing a thermoelectric module using $p$- and $n$-type legs.

\subsection{On-center and off-center positions of guest atoms in type-I clathrate compounds}
\label{subsec:states_rattling}
\subsubsection{X-ray and neutron diffractions and EXAFS}
\label{subsubsec:xray_diffraction}
This subsection  describes the atomic configuration in type-I clathrate compounds determined 
by means of x-ray diffraction\,\cite{Nolas:2000a, Bentien:2005a, Christensen:2006a, Christensen:2010a}, 
neutron diffraction\,\cite{Sales:2001a, Chakoumakos:2000a, Chakoumakos:2001a, Christensen:2006a, Christensen:2010a},
and extended x-ray absorption fine-structure\,(EXAFS)\,\cite{Baumbach:2005a, Jiang:2008a, Kozina:2009a, Mansour:2012}.

In type-I clathrates, the guest sites in the dodecahedron and tetrakaidecahedron are denoted by guest\,(1) and (2) sites, respectively.
\textcite{Chakoumakos:2000a} have investigated the vibrational characteristics of Sr(2) guest-atom encaged in type-I Sr$_{8}$Ga$_{16}$Ge$_{30}$ in terms of neutron diffraction at room temperature, claiming that the isotropic atomic displacement parameter (ADP) for Sr(2) is enormous. 
Taking into account the anisotropic displacement, the vibrations of Sr(2) show a large ADP in the plane parallel to the six-member ring cage.
These results suggest that the site of Sr(2) could be equally described by a fractionally occupied four-fold split-site.
In fact, the differential Fourier map around the cage center 6$d$ site shows
a residual nuclear density with lobes in the directions of the split positions 24$k$.
Following the above work, \textcite{Sales:2001a} and \textcite{Chakoumakos:2001a} have performed a neutron diffraction study 
on single crystals of type-I $\mathcal{R}_{8}$Ga$_{16}$Ge$_{30}$ ($\mathcal{R}$=Ba, Sr, Eu). 
Difference Fourier maps suggest that the nuclear density of Ba(2) is centered at the cage center, 
while those of Sr(2) and Eu(2) move off to one of four off-center sites 24$k$, 0.3 \AA~and 0.4 \AA~away from the cage center, respectively.
Refinements result in similar residual factors for two models with a four split site on 24$k$ and 24$j$, as shown in~Fig.\,\ref{fig:takaba_fig05}. 
Therefore, the disorder of the guest position is more complex than the case of just four sites.
In fact, \textcite{Fujiwara:2012a} have reported that the Sr(2) are widely distributed to 24$k$ and 24$j$ sites.

\begin{figure} 
\epsfysize=3.0in 
\epsfbox{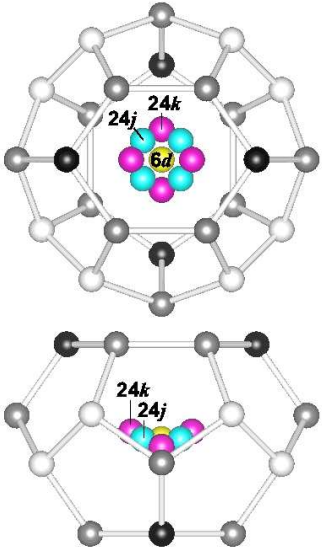}
\caption{\label{fig:takaba_fig05} (Color online) Split 24$j$ and 24$k$ sites for guest (2) atoms around 6$d$ sites in tetrakaidecahedral cages for type-I clathrate.} 
\end{figure}

By combining x-ray and neutron diffraction methods, \textcite{Christensen:2006a,Christensen:2010a} have revealed that
the Ba(2) guests in $n$- and $p$-type Ba$_{8}$Ga$_{16}$Ge$_{30}$ also occupy slightly off-center sites 24$k$ or 24$j$.
A more definite off-center 24$k$ site for Ba(2) situated 0.43 \AA~away from the center 
was found in type-I Ba$_{8}$Ga$_{16}$Sn$_{30}$ ($\beta$-BGS), where the cage size is larger by 8 \% than that of Ba$_{8}$Ga$_{16}$Ge$_{30}$.
The off-center displacement of 0.43 \AA~is comparable to that of Eu(2)\,\cite{Avila:2008a, Suekuni:2008a}.

EXAFS has also been employed to investigate the rattling sites of the guests 
in $\mathcal{R}_{8}$Ga$_{16}$Ge$_{30}$ ($\mathcal{R}$=Ba, Sr, Eu)\,\cite{Baumbach:2005a, Jiang:2008a, Mansour:2012}.
The analysis has confirmed that Eu(1) is located at the center 2$a$ sites, 
but Eu(2) is located an off-center site 0.445 \AA~away from the center of the cage.
For $\mathcal{R}$=Sr, $\sim$75\% of Sr(2) atoms take off-center sites  0.40 \AA~away from the center 
and the rest reside at the center.
The results of EXAFS were in reasonable agreement with those of x-ray and neutron diffraction.
Furthermore, the analysis for $\mathcal{R}$=Ba indicates a somewhat off-center site for Ba(2).

To summarize, it has been revealed that the degree of off-center displacement of the guest(2) atoms primarily  
depends on the relative size of the cage to that of the guest atom.
Actually, neutron diffraction experiments have shown that  
the guest(2) atoms occupy four off-center positions away from the center of the cages
when the size of the cage is large enough, or equivalently, the ionic radius of the guest atom is small enough\,\cite{Sales:2001a}.

Table \ref{tab:tab1} gives the sets of the data obtained from various types of experimental methods on characteristic properties of type-I and type-V\hspace{-.1em}I\hspace{-.1em}I\hspace{-.1em}I $\mathcal{R}_{8}$M$_{16}$Z$_{30}$.
This is useful for the purpose of overviewing the characteristics of actual clathrate compounds.

\begingroup
\begin{longtable*}[b]{@{\extracolsep{\fill}}llllllll}
\multicolumn{8}{c}{\parbox{7in}{\label{tab:tab1}Table 1: Key properties of type-I and type-V\hspace{-.1em}I\hspace{-.1em}I\hspace{-.1em}I clathrates $\mathcal{R}_{8}$M$_{16}$Z$_{30}$.
The samples are in the forms of poly-crystallines, powders, or single crystals. 
Listed are carrier types $n$ or $p$, $\kappa_{\rm ph}(T)$, on- or off-center state of the guest atom $\mathcal{R}$(2), and two characteristic temperatures  $\theta$ of $\mathcal{R}$(2).
Features of $\kappa_{\rm ph}(T)$ are denoted by C: crystalline or G: glass-like.
The exponents of the temperature dependence of $\kappa_{\rm ph}(T)$ below 1\,[K] are given for single crystals.
The states of guest atom $\mathcal{R}$(2), on- or off-center, have been judged by various experiments; x-ray diffraction\,(XRD), neutron diffraction\,(ND), resonant x-ray diffraction\,(RXD), extended x-ray absorption fine structure spectroscopy\,(EXAFS), resonant ultrasound spectroscopy\,(RUS), M$\ddot{\rm o}$ssbauer spectroscopy\,(M$\ddot{\rm o}$ssbauer), 
microwave absorption spectroscopy\,(MW), specific heat\,(SH), electrical resistivity\,(ER), Raman scattering\,(Raman), THz spectroscopy\,(THz), 
inelastic neutron diffraction\,(INS), and nuclear inelastic scattering\,( NIS).
Two characteristic temperatures $\theta$ on $\mathcal{R}$(2) correspond to vibrational modes in out-of-the plane/on-the-plane perpendicular to the four-fold axis of the tetrakaidecahedron.
Single or non-separated values denote the cases not distinguished or assumed off-center 24$k$ or 24$j$ sites. 
The values in parentheses may include the components of $\mathcal{R}$(1) in the dodecahedron in addition to $\mathcal{R}$(2).
For type-V\hspace{-.1em}I\hspace{-.1em}I\hspace{-.1em}I clathrates, the characteristic temperature $\theta$ is for the guest atom in a distorted dodecahedron.}}\\


\hline
\hline
Compounds&Sample&Carrier &$\kappa_{\rm ph}(T)$&Experiments&Rattling states&$\theta$ [K]&References\\
Type of $\mathcal{R}_{8}$M$_{16}$Z$_{30}$&form&types&~&(Temperature)&of $\mathcal{R}$(2)&for $\mathcal{R}$(2)\\
\hline
\endfirsthead

\multicolumn{8}{l}{continued from previous page}\\
\hline
Compounds&Sample&Carrier &$\kappa_{\rm ph}(T)$&Experiments&Rattling states&$\theta$ [K]&References\\
Type of $\mathcal{R}_{8}$M$_{16}$Z$_{30}$&form&types&~&(Temperature)&of $\mathcal{R}$(2)&for $\mathcal{R}$(2)\\
\hline
\endhead

\hline
\multicolumn{6}{l}{$^a$E$_{\rm g}$(1)/E$_{\rm g}$(A),T$_{\rm 2g}$(1) rattling modes in the tetrakaidecahadron}&\multicolumn{2}{r}{table continued on next page}\\
\endfoot

\hline
\hline
\multicolumn{8}{l}{$^a$E$_{\rm g}$(1)/E$_{\rm g}$(A),T$_{\rm 2g}$(1) rattling modes in the tetrakaidecahadron}\\
\endlastfoot

I-Ba$_{8}$Al$_{16}$Si$_{30}$&Poly&n&C&~&~&~&\textcite{Mudryk:2002a}\\
~&Single&~&~&XRD&on-center&92,100/68,74&\textcite{Christensen:2010a}\\
(Ba$_{8}$Al$_{14}$Si$_{31}$)&Single&n&C&ND(298\,[K])&on&~&\textcite{Condron:2006a}\\
\hline\\
I-Sr$_{8}$Al$_{10}$Si$_{36}$&Single&~&~&XRD(90\, [K])&on&~&\textcite{Roudebush:2012a}\\
I-Sr$_{8}$Al$_{11}$Si$_{35}$&Single&n&C&SH&~&125/67&unpublished our data\\
\hline\\
I-Ba$_{8}$Ga$_{16}$Si$_{30}$&Single&~&~&ND&on&98/69&\textcite{Bentien:2002a}\\
~&Powder&n&C&ND&off-center&85&\textcite{Qiu:2004}\\
~&Single&~&~&ND&off&~&\textcite{Bentien:2005a}\\
~&Powder&~&~&XRD&on&101/77&\textcite{Bentien:2005a}\\
~&Single&~&~&XRD&on&92-95/66-69&\textcite{Christensen:2010a}\\
(Ba$_{8}$Ga$_{12}$Si$_{33}$)&Powder&~&~&XRD(R.T.)&on&~&\textcite{Mudryk:2002a}\\
(Ba$_{8}$Ga$_{15-16}$Si$_{31-30}$)&Powder&~&~&XRD(R.T.)&on&88-96/53-57&\textcite{Anno:2012a}\\
~&Single&~&~&Raman(4\,[K])&off&-/59&\textcite{Takasu:2008a}\\
\hline\\
I-Sr$_{8}$Ga$_{16}$Si$_{30}$&Single&n&C&SH&~&120/59&\textcite{Suekuni:2007a}\\
~&Single&n&C&Raman(50\,[K])&off&-/56&\textcite{Takasu:2008a}\\
\hline
I-Ba$_{8}$Al$_{16}$Ge$_{30}$&Single&~&~&XRD&on&81-85/61-69&\textcite{Christensen:2007a}\\
~&Single&~&~&XRD&off&81-84&\textcite{Christensen:2007a}\\
~&Single&~&~&ND(20\,[K])&off&~&\textcite{Christensen:2007a}\\
~&Poly&n&C&~&~&~&\textcite{Christensen:2007a}\\
\hline\\
I-Ba$_{8}$Ga$_{16}$Ge$_{30}$&~&~&~&XRD(R.T.?)&on&(51)&\textcite{Eisenmann:1986a}\\
~&~&~&~&~&~&~&\textcite{Sales:1999a}\\
~&Single&~&C&ND(15\,[K])&?&~&\textcite{Keppens:2000a}\\
~&Single&~&~&ND&on&~&\textcite{Chakoumakos:2001a}\\
~&Single&n&~&ND&on&64&\textcite{Sales:2001a}\\
~&~&n&~&XRD(293\,[K])&off&72&\textcite{Paschen:2001a}\\
~&Powder&~&~&XRD&on&101/73&\textcite{Bentien:2005a}\\
~&Single&p&~&ND&on&90/62&\textcite{Christensen:2006a}\\
~&Single&p&~&XRD&on&87/60&\textcite{Christensen:2006a}\\
~&Single &n&~&ND&on&89/59&\textcite{Christensen:2006a}\\
~&Single&n&~&XRD&on&84/60&\textcite{Christensen:2006a}\\
~&Single&p&~&ND&off&88&\textcite{Christensen:2006a}\\
~&Single&p&~&XRD&off&82&\textcite{Christensen:2006a}\\
~&Single &n&~&ND&off&88&\textcite{Christensen:2006a}\\
~&Single&n&~&XRD&off&81&\textcite{Christensen:2006a}\\
~&Single&p&~&XRD(20\,[K])&off:65\%,on:33\%&~&\textcite{Fujiwara:2012a}\\
~&Single&n&~&XRD(20\,[K])&on&~&\textcite{Fujiwara:2012a}\\
~&Powder&n,p&~&EXAFS&off&86&\textcite{Jiang:2008a}\\
~&Poly&n&C&~&~&~&\textcite{Uher:1999a}\\
~&Single&n&C&SH&~&(60)&\textcite{Sales:2001a}\\
~&~&n&~&SH&~&(80)&\textcite{Paschen:2001a}\\
~&Single&~&G: $T^{\sim 1.8}$&~&~&~&\textcite{Paschen:2003a}\\
~&Single&p&G: $T^{1.5}$&SH&~&38&\textcite{Bentien:2004a}\\
~&~&~&~&SH&~&80&\textcite{Bentien:2005a}\\
~&Single&p&G: $T^{2}$&SH&~&42&\textcite{Umeo:2005a}\\
~&Single&p&G: $T^{2}$&SH&~&87/49&\textcite{Avila:2006a, Avila:2006aa}\\
~&Single&n&C: $T^{2}$&SH&~&87/49&\textcite{Avila:2006a, Avila:2006aa}\\
~&Single&n&~&SH&~&74&\textcite{Xu:2011a}\\
~&Single&n, p&~&Raman(2\,[K])&off&92/47,45-47$^a$&\textcite{Takasu:2006a, Takasu:2010a}\\
~&Single&p&~&THz(6.6\,[K])&~&86/55&\textcite{Mori:2009a}\\
~&Poly&~&~&INS(295\,[K])&~&55&\textcite{Hermann:2005a}\\
~&Powder&p&~&INS&~&-/50&\textcite{Christensen:2006aa}\\
~&Powder&n&~&INS&~&65/52&\textcite{Christensen:2006aa}\\
~&Single&n&~&INS(R.T.)&~&-/52&\textcite{Lee:2007a}\\
~&Single&n&~&INS&~&87/53&\textcite{Christensen:2008a}\\
\hline\\
I-Ba$_{8}$In$_{16}$Ge$_{30}$&Powder&~&~&XRD&on&65/65&\textcite{Bentien:2005a}\\
~&Single&n&C&~&~&~&\textcite{Bentien:2006a}\\
(with vacancy)&Poly&n&G&~&~&~&\textcite{Bentien:2006a}\\
~&Single&n&C&SH&~&99/38&\textcite{Suekuni:2008aa}\\
\hline\\
I-Sr$_{8}$Ga$_{16}$Ge$_{30}$&Powder&~&~&XRD(R.T.)&on&85/39&\textcite{Schujman:2000a}\\
~&Pow.,Sin.&~&~&ND&off&~&\textcite{Chakoumakos:2000a}\\
~&Pow.,Sin.&~&~&ND&off or on&85&\textcite{Chakoumakos:2000a}\\
~&Single&~&~&XRD(R.T.)&on&74&\textcite{Nolas:2000a}\\
~&Single&~&G&ND(15\,[K])&off&~&\textcite{Keppens:2000a}\\
~&Powder&~&~&XRD&~&90/29&\textcite{Iversen:2000a}\\
~&Single&n&G&ND&off&~&\textcite{Sales:2001a}\\
~&Single&n&G&ND&on&80&\textcite{Sales:2001a}\\
~&Powder&~&~&RXD(R.T.)&off&~&\textcite{Zhang:2002a}\\
~&Powder&n&G&ND&off&76&\textcite{Qiu:2004}\\
~&Powder&~&~&XRD&on&104/163&\textcite{Bentien:2005a}\\
~&Single&n&~&XRD(20\,[K])&off:98\%,on:2\%&~&\textcite{Fujiwara:2012a}\\
~&Poly&n&G: $T^{2}$&~&~&~&\textcite{Nolas:1998a}\\
~&~&~&~&~&~&~&\textcite{Cohn:1999a}\\
~&Poly&n&G/C&~&~&~&\textcite{Uher:1999a}\\
~&Poly&n&G&~&~&~&\textcite{Nolas:1998aa}\\
~&~&~&G&~&~&~&\textcite{Nolas:2000a}\\
~&Powder&n&~&EXAFS&off&126-128&\textcite{Baumbach:2005a}\\
~&Powder&n&~&EXAFS&off:75\%,on:25\%&147-156&\textcite{Baumbach:2005a}\\
~&~&~&~&SH&~&(50)&\textcite{Chakoumakos:2000a}\\
~&Single&n&G&SH&~&(53)&\textcite{Sales:2001a}\\
~&~&~&~&SH&~&(79)&\textcite{Paschen:2001a}\\
~&~&~&~&SH&~&80&\textcite{Bentien:2005a}\\
~&Single&n&G&SH&~&34&\textcite{Umeo:2005a}\\
~&Single&n&G&SH&~&90/35&\textcite{Suekuni:2007a}\\
~&Single&n&~&SH&~&80&\textcite{Xu:2011a}\\
~&Single&~&~&RUS&~&45&\textcite{Keppens:2002a}\\
~&Poly&~&~&Raman(300\,[K])&~&-/46&\textcite{Nolas:2000aa}\\
~&Single&n&~&Raman(2\,[K])&off&95/42,42$^a$&\textcite{Takasu:2006a}\\
~&Poly&~&~&INS(295\,[K])&~&49&\textcite{Hermann:2005a}\\
~&Powder&n&~&INS&~&62/48&\textcite{Christensen:2006aa}\\
~&Single&n&~&INS&~&-/46&\textcite{Lee:2008a}\\
\hline\\
I-Eu$_{8}$Ga$_{16}$Ge$_{30}$&Single&~&~&XRD(R.T.)&on&53&\textcite{Nolas:2000a}\\
~&Single&~&~&ND&off&~&\textcite{Chakoumakos:2001a}\\
~&Single&n&G&ND&off&~&\textcite{Sales:2001a}\\
~&Single&n&~&XRD(293\,[K])&off&45&\textcite{Paschen:2001a}\\
~&Powder&n&~&EXAFS&off&93-96&\textcite{Baumbach:2005a}\\
~&~&~&G&~&~&~&\textcite{Nolas:2000a}\\
~&Single&n&G&SH&~&(30)&\textcite{Sales:2001a}\\
~&Poly&n&G&~&~&~&\textcite{Paschen:2001a}\\
~&Poly&n&G&~&~&~&\textcite{Bentien:2005aa}\\
~&Single&~&~&RUS&~&22&\textcite{Zerec:2004a}\\
~&Poly&~&~&M$\ddot{\rm o}$ssbauer, MW&off&-/31&\textcite{Hermann:2006a}\\
~&Poly&~&~&Raman(300\,[K])&~&-/33&\textcite{Nolas:2000aa}\\
~&Single&n&~&Raman(2\,[K])&off&69/26,26$^a$&\textcite{Takasu:2006a}\\
~&Powder&~&~&NIS(25\,[K])&~&87,57,35&\textcite{Hermann:2005a}\\
\hline\\
V\hspace{-.1em}I\hspace{-.1em}I\hspace{-.1em}I-Eu$_{8}$Ga$_{16}$Ge$_{30}$&Single&n&~&XRD(293\,[K])&~&45&\textcite{Paschen:2001a}\\
~&Poly&n&C&SH&~&72&\textcite{Paschen:2001a}\\
\hline\\
I-Ba$_{8}$Ga$_{16}$Sn$_{30}$&Single&p&G: $T^{\sim 2}$&XRD&off&~&\textcite{Suekuni:2008a}\\
~&Single&n&G: $T^{\sim 2}$&XRD&off&78/60&\textcite{Suekuni:2008a}\\
~&Single&n&G: $T^{\sim 2}$&SH&off&-/20&\textcite{Avila:2008a}\\
~&Single&p&G: $T^{\sim 2}$&SH&off&-/20&\textcite{Suekuni:2008a}\\
~&~&p?&~&SH&~&55,14&\textcite{Zheng:2012a}\\
~&~&p?&~&ER&~&54&\textcite{Zheng:2012a}\\
~&Single&n&~&Raman(4\,[K])&off&68/27,20$^a$&\textcite{Suekuni:2010a}\\
~&Single&p&~&Raman(4\,[K])&off&-/27,21$^a$&\textcite{Suekuni:2010a}\\
~&Single&n&~&THz(7\,[K])&~&64/34&\textcite{Mori:2011a}\\
~&~&~&~&NMR&off&-/20&\textcite{Zheng:2011a}\\
\hline\\
V\hspace{-.1em}I\hspace{-.1em}I\hspace{-.1em}I-Ba$_{8}$Ga$_{16}$Sn$_{30}$&Sin.,Pow.&n&~&XRD&~&64&\textcite{Huo:2005a}\\
~&Single&n&C&SH&~&50&\textcite{Huo:2005a}\\
~&~&~&~&~&~&~&\textcite{Avila:2006aa}\\
~&Single&p&G&SH&~&50&\textcite{Avila:2006aa}\\
\end{longtable*}
\endgroup

\section{SPECIFIC HEATS OF TYPE-I CLATHRATE COMPOUNDS}
\label{sec:thermal_experiments}
It is common to use an appropriate energy scale in each experiment.
These are Joule (J) or Kelvin (K) for thermal measurements, 
cm$^{-1}$ or THz for optical spectroscopies, and eV for inelastic neutron or x-ray scatterings.
The conversion relation is 1\,[THz] = 47.99\,[K] = 33.36\,[cm$^{-1}$] = 4.136\,[meV].
This relation will be helpful for readers in connection with the following descriptions on the dynamical behaviors of type-I clathrate compounds.

\subsection{Observed characteristics of specific heats}
\label{subsec:specific_heats_exper}

Heat capacity provides fundamental information on the thermal properties of materials. 
It represents the amount of energy needed to raise the temperature of a given material by one degree. 
The heat capacity per unit mass is called specific heat. 
Molar specific heat is referred to as specific heat $C_{\rm V}$ in the present article.
Temperature dependences of specific heats reflect the phonon density of states, the electronic one at $E_{\rm F}$, the magnetic degree of freedom, superconducting properties, and so on.

Three characteristic temperature-regions are manifested in the observed specific heats of type-I clathrate compounds by subtracting the electronic contribution.
These are temperature regions below 1\,[K], around several [K], and above several 10\,[K].
We will describe in details these characteristics in the following subsections.

\subsubsection{Low temperature specific heats below 1\,K}
\label{subsubsec:low_temperatures}
A series of specific heat experiments on type-I  clathrate compounds have made clear that the characteristics critically depend on the states of guest atoms in cages\,\cite{Nolas:1995a,Nolas:1996a,Nolas:1998a,Nolas:1998aa,
Nolas:1998aaa,Nolas:1998aa,Nolas:2000a,Nolas:2001r,
Cohn:1999a,Sales:1996a,Sales:1998a,Sales:2001a,
Bentien:2004a,Bentien:2005a,Avila:2006a,Avila:2006aa,Suekuni:2007a,
Suekuni:2008a,Suekuni:2008aa,Suekuni:2010a, Xu:2010a}.
The most intriguing findings in the regime below about 1\,[K] are that the temperature dependences and the magnitude are almost identical to those of glasses
for type-I clathrate compounds with \textit{off-center} guest atoms.
 
The specific heats of type-I clathrates with off-center guest atoms are scaled in the following form
\begin{equation}
\label{Two-level_Specific}
C_\mathrm{V}(T)\cong \alpha T^{1+\delta}+ \beta T^3.
\end{equation}
The first term proportional to $T^{1+\delta}$ is not  exactly linear in $T$
at temperatures below 1\,[K] with the positive small factor $\delta$, a dependence that is identical to that of structural glasses\,\cite{Zeller:1971a, Freeman:1986a, Stephens:1930a}. 
This term has no connection with conduction electron specific heat proportional 
to $\gamma T$.

From careful measurements of specific heats of type-I Ba$_{8}$Ga$_{16}$Sn$_{30}$~($\beta$-BGS), it has been estimated that $\alpha$\,$\cong$\,30\,$\left[ \mathrm{mJ\,mol^{-1}\,K^{-2}}\right]$
and $\beta$\,$\cong$\,50\,$\left[\mathrm{mJ\,mol^{-1}\,K^{-4}}\right]$ with $\delta$\,=\,0.2\,\cite{Suekuni:2008a, Suekuni:2008aa}.
In silica glass\,\cite{Zeller:1971a}, $\alpha$\,$\cong$\,0.072\,$\left[ \mathrm{mJ\,mol^{-1}\,K^{-2}}\right]$ and $\beta$\,$\cong$\,0.108\,$\left[ \mathrm{mJ\,mol^{-1}\,K^{-4}}\right]$ with the positive small factor $\delta$\,$\cong$\,0.2.
Most carefully measured silica glass\,\cite{Lasjaunias:1975a} gives $\delta=0.22-0.30$ between $T$=25\,[mK] and 0.9\,[K].

The coefficient of the second term $\beta$ should be separated into $\beta_{\rm D}$ and $\beta_{\rm B}$, where $\beta_{\rm D}$ is the coefficient from the Debye phonon specific heat and $\beta_{\rm B}$ is that of the excess specific heat originating from the tail of the hump observed at around $T\simeq$4\,[K] in $\beta$-BGS.

\subsubsection{Excess specific heats over Debye phonons at temperatures of several Kelvin}
\label{subsubsec:excess_density}
Atomic vibrations of type-I clathrates containing off-center guest atoms display glass-like low-lying modes.
Actually, 
the excess densities of states in type-I clathrate compounds have been observed as a hump in specific heat experiments in the temperature range around several K, i.e., broad peaks observed in the temperature range at 10$-$100
times smaller than the Debye temperature $\theta_{\rm D}=\hbar\omega_{\rm D}/k_{\rm B}$.
The characteristics are almost identical to those of structural glasses observed by a variety of measurements such as optical spectroscopies, inelastic neutron scattering, and specific heats in the THz frequency
range\,(see, for example, a review by \textcite{Nakayama:2002r}). 
The Debye frequency $\omega_{\rm D}$ is related to the velocities of transverse $v_{\rm t}$ and longitudinal $v_{\rm \ell}$ acoustic phonons by
\begin{eqnarray}
\label{eq:Specific_1c}   
\omega_D^3=\frac{18\pi^2N}{V}\left(\frac{2}{v_t^3}+\frac{1}{v_\ell^3}\right)^{-1}.
\end{eqnarray}
Here $N$ is the number of atoms in the volume $V$. 
The experimental values of velocities of silica glass are $v_{\rm t}=3.767\times 10^5$\,[cm s$^{-1}$] and $v_{\rm \ell}=5.970\times 10^5$\,[cm s$^{-1}$].
These give a Debye frequency $\omega_{\rm D}/2\pi$ of 10.40\,[THz] corresponding to the Debye temperature $\theta_{\rm D}$=500\,[K] from the relation $\theta_{\rm D}=\hbar\omega_{\rm D}/k_{\rm B}$.
The specific heat of the Debye phonons is expressed by
\begin{equation}
\label{Debye:T3}
   C_{\rm ph}(T) \cong \frac{12\pi^4Nk_{\rm B}}{5V}\left(\frac{T}{\theta_{\rm D}}\right)^3.
\end{equation}
This formula for the phonon specific heat holds in the temperature regime $T\ll\theta_{\rm D}$.
However, observed specific heats of structural glasses do not show the temperature dependence of Eq.\,(\ref{Debye:T3}) at low temperatures but
manifest the excess humps in the temperature range around several [K].

\textcite{Flubacher:1959a} and \textcite{Anderson:1959a} have observed the excess  specific heat of silica glass
at around 10\,[K] over the Debye contribution of Eq.\,(\ref{Debye:T3}). 
This is clearly realized by plotting $C_{\rm V}(T)/T^3$\,vs.\,$T$ \cite{Zeller:1971a, Pohl:1981a, Buchenau:1986a}. 
These observations have been ascribed to the presence of excess vibrational states over the Debye density of states $D(\omega)\propto \omega^2$.
The term ``Boson peak" refers to this excess contribution to the conventional Debye density of states.
The issue in structural glasses has been revitalized according to the advent of various experimental techniques.
However, the nature of these vibrational states in glasses has been the subject of a very intensive and controversial debate for decades.

The humps in $C_{\rm V}/T^3$ of type-I Ba$_{8}$Ga$_{16}$Sn$_{30}$ ($\beta$-BGS) and type-I Sr$_{8}$Ga$_{16}$Ge$_{30}$
appear around $T$\,$\simeq$5\,[K]\,\cite{Avila:2008a, Umeo:2005a, Suekuni:2007a, Suekuni:2008a, Suekuni:2008aa, Xu:2011a}.
See Fig.\,\ref{fig:takaba_fig07}.
This excess hump possess almost identical characteristics to the so-called Boson peak.

\subsubsection{Characteristic temperatures of excess specific heats}
\label{Analysis* specific heats}
The data on specific heats have been used to characterize vibrational properties of type-I clathrate compounds. 
Analysis based on the Einstein model has been carried out on type-I EGG, SGG, and BGG single crystals by~\textcite{Sales:2001a}, where the cages and the guests were assumed to behave as a Debye lattice and Einstein oscillators, respectively. 
At temperatures above 80\,[K], the specific heat data were the same among three samples to within the experimental error, 
and well described by the Debye specific heat with a $\theta_{\rm D}$ of about 300\,[K]. 
At low temperatures, however, the Einstein oscillators predicted significantly different behaviors to the three samples.
The Einstein temperatures $\theta_{\rm E}$ corresponding to the vibrations of Ba, Sr, and Eu guest atoms were estimated to be 60, 53, and 30\,[K], respectively.

\textcite{Paschen:2001a} have reported different analysis for these three compounds.
The data of $C_{\rm V}$\,vs.\,$T$ were fitted by including lattice and electronic contributions given by $C_{\rm ph}$ and $C_{\rm e}$.
The fitting parameters are $\theta_{\rm D}$, $\theta_{\rm E}$, 
and numbers of atoms $N_{\rm D}$ and $N_{\rm E}$ contributing to the Debye and Einstein specific heats, respectively, where
their sum $N_{\rm D}$+$N_{\rm E}$ is taken as 54.
For BGG, the obtained parameters were 
$\theta_{\rm D}$=\,355\,[K], $\theta_{\rm E}$=80\,[K], $N_{\rm D}$=39, and $N_{\rm E}$=15.
Similar results were obtained for SGG.
The fits to the data for BGG and SGG yield $N_{\rm E}$=15 and 16, respectively, instead of the number of guests equal to 8.
For $\beta$-EGG, $\theta_{\rm E}$ was not evaluated  
due to the large magnetic contribution of Eu$^{2+}$ at temperatures up to 50\,[K].
The value of $\theta_{\rm D}$ was estimated to be 302\,[K] using $\theta_{\rm E}$ obtained from room-temperature isotropic ADP
and $C_{\rm V}(T)$ data at temperatures above the Curie temperature of 36\,[K].

\textcite{Bentien:2004a} have analyzed $C_{\rm V}$ for BGG using two kinds of the Einstein oscillators. 
The obtained characteristic temperatures are $\theta_{\rm D}$=324(4)\,[K], $\theta_{\rm E1}$=78(2)\,[K], and $\theta_{\rm E2}$=38(1)\,[K] 
with $N_{\rm D}$=2.8(6), $N_{\rm E1}$=9.7(5), and $N_{\rm E2}$=1.5(5), respectively.
A linear fit to the data of $C_{\rm V}/T$\,{\rm~vs}.\,$T^2$ between 1.5\,[K$^{2}$] and 4\,[K$^{2}$] gave $\theta_{\rm D}$=311(10)\,[K], 
in agreement with the result from the fitting of the data 
$C_{\rm V}$\,{\rm~vs}.\,$T$ above 5\,[K]. 
The Sommerfeld coefficient $\gamma$ was estimated to be 14\,[mJ/K$^{2}$mol].

In the above analysis on BGG, $N_{\rm D}$ and $N_{\rm E}$ are free fitting parameters.
\textcite{Bentien:2005a} then imposed the constraints $N_{\rm D}$=46 and  $N_{\rm E1} + N_{\rm E2}$=8.
By using $\theta_{\rm D}$=312\,[K] obtained from the temperature dependence of the ADP of cage atoms, 
the parameters were evaluated to be $\gamma$=35\,[mJ/K$^{2}$mol],
 $\theta_{\rm E1}$=80\,[K] and  $N_{\rm E1}$=6.5, and $\theta_{\rm E2}$=42\,[K] and $N_{\rm E2}$1.5, respectively. 
For  type-I SGG, the evaluated parameters were $\gamma =$~40\,[mJ/K$^{2}$mol], $\theta_{\rm E1}$=80\,[K], and $\theta_{\rm E2}$=33\,[K]. 
They have argued that $N_{\rm E1}$ and $N_{\rm E2}$  almost correspond to the numbers of the large and small cages.
However, the result disagreed with the fact that the guest in a small cage has a higher vibrational energy\,\cite{Christensen:2010a}. 

\begin{figure} 
\epsfysize=3.8in 
\epsfbox{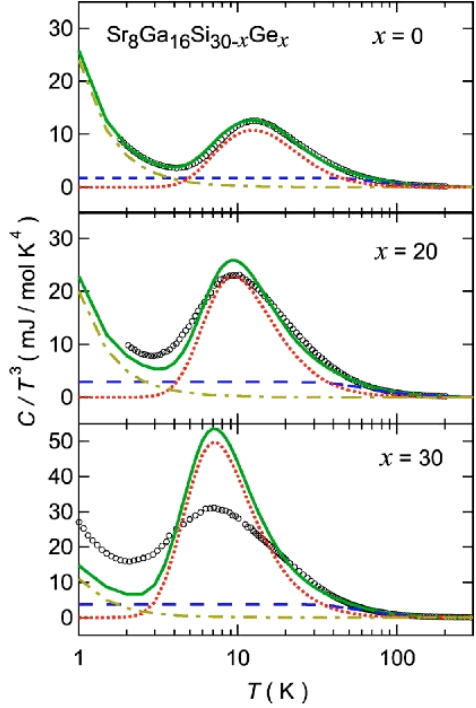}
\caption{\label{fig:takaba_fig06} (Color online) The specific heat data plotted as $C_{\rm V}(T)/T^3$\,{\rm~vs}.\,$T$  (open circles) for type-I Sr$_{8}$Ga$_{16}$Si$_{30-x}$Ge$_{x}$. 
The dotted, dashed, and dashed-dotted lines are from the Einstein model, the Debye model, and from electronic components, respectively.
After the work of \textcite{Suekuni:2007a}.} 
\end{figure}

\textcite{Avila:2006aa} have analyzed the specific heat data of BGG by plotting $C_{\rm V}/T^{3}$\,{\rm~vs}.\,$T$. 
The contributions of the guest atoms from these systems appear as broad peaks 
over electronic and Debye {\it backgrounds} originating from the stiff cage.
This treatment has reproduced the specific heat data
for type-V\hspace{-.1em}I\hspace{-.1em}I\hspace{-.1em}I Ba$_{8}$Ga$_{16}$Sn$_{30}$\,\cite{Huo:2005a}.
It should be recalled that six Ba(2) atoms at the 6$d$ site in the tetrakaidecahedra show 
strongly anisotropic vibration with larger amplitudes within the plane perpendicular to the four-fold axis.
So, at least two characteristic temperatures
are required to describe the vibrations of Ba(2) when applying the Einstein model: in plane $\theta_{\rm E2}^{\parallel}$ and out of plane $\theta_{\rm E2}^{\perp}$.
Furthermore, the characteristic temperature of a Ba(1) atom at a 2$a$ site in the dodecahedra was assumed to be 
the single parameter $\theta_{\rm E1}$ because of the isotropic shape of the cage.
The dimensionality and the numbers of oscillators were predefined: 
$D \times N_{\rm E1}$=3$\times$2, $D\times N_{\rm E2}^{\parallel}$=2$\times$6, and $D\times N_{\rm E2}^{\perp}$=1$\times$6.
Additional constraints $\theta_{\rm E2}^{\parallel}<\theta_{\rm E2}^{\perp}$ and $\theta_{\rm E2}^{\parallel}<\theta_{\rm E1}$ were
imposed.
The values of $\gamma$ and $\theta_{\rm D}$ determined from the plot of $C/T$\,{\rm~vs}.\,$T^{2}$ were fixed together with the number $N_{\rm E}$.
Therefore, only three fitting parameters were left, which characterize the vibrational energies of the Ba guests.
The best fits give the values $\theta_{\rm E1}$=87.2\,[K], $\theta_{\rm E2}^{\perp}$=87.1\,[K], and $\theta_{\rm E2}^{\parallel}$=49.4\,[K].

\begin{figure}[t]
\epsfysize=2.5in 
\epsfbox{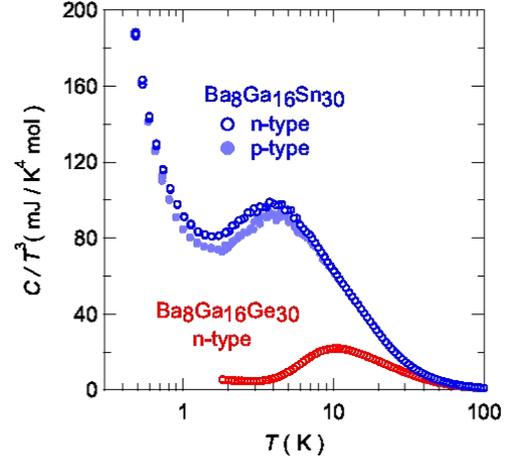}
\caption{\label{fig:takaba_fig07} (Color online) The specific heat data plotted as $C_{\rm V}/T^3$\,{\rm~vs}.\,$T$ on $n$-type type-I Ba$_{8}$Ga$_{16}$Ge$_{30}$ 
and $n$- and $p$-type type-I Ba$_{8}$Ga$_{16}$Sn$_{30}$ single crystals. Adopted from \textcite{Avila:2006aa, Avila:2008a, Suekuni:2008a}.} 
\end{figure}

Specific heat experiments on cage-size tuned Sr$_{8}$Ga$_{16}$Si$_{30-x}$Ge$_{x}$ have been performed
by~\textcite{Suekuni:2007a}.
With increasing $x$ from 0 to 30, the cage size was expanded by 8\%.
The data in Fig.\,\ref{fig:takaba_fig06} plotted as 
$C_{\rm V}/T^{3}$\,{\rm~vs}.\,$T$ show that the peak temperatures of the Boson-peak like hump shift lower
from 10.5 to 7\,[K] with increasing hump heights.
These observations do not depend on the carrier density, whose contribution becomes vanishingly small at temperatures above 4\,[K].
In the analysis based on the Einstein model, the constraints $\theta_{\rm E2}^{\parallel}$=$\theta_{\rm EL}$ 
and $\theta_{\rm E2}^{\perp}$=$\theta_{\rm E1}$=$\theta_{\rm EH}$ were imposed, and the rest of the fitting parameters were taken as the lower and higher characteristic temperatures $\theta_{\rm EL}$ and $\theta_{\rm EH}$.
For $x$=0, a good fit was obtained with $\theta_{\rm EL}$=59\,[K] and $\theta_{\rm EH}$=120\,[K], 
indicating the validity of the Einstein model for Sr guests in Sr$_{8}$Ga$_{16}$Si$_{30}$. 
However, for $x$=20, 30, the Einstein model apparently did not yield an adequate fitting to the observed data, as seen in Fig.\,\ref{fig:takaba_fig06}.
These discrepancies indicate the relevance of the anharmonic potential for the Sr(2) sites on increasing cage size for $x$=20, 30.

\textcite{Avila:2008a} and \textcite{Suekuni:2008a} have reported specific heat measurements on type-I BGS, 
which has a larger cage and a larger off-center displacement of Ba(2) than that of BGG.
The $C/T^{3}$\,{\rm~vs}.\,$T$ plot for $\beta$-BGS shows larger Boson-peak like hump around 4\,[K] than that around 10\,[K]\,(see Fig~\ref{fig:takaba_fig07}). 
The temperature of 4\,[K] is the lowest among type-I clathrates, and corresponds to a vibrational energy equivalent to 20\,[K] for Ba(2) in the plane perpendicular to the four-fold axis.
The height of the hump for $\beta$-BGS was not reproduced by the Einstein model unlike BGG, 
that indicates a strong anharmonicity of Ba(2) vibrations. 
The low-temperature $T$-linear coefficient for $n$-type $\beta$-BGS was estimated to be 29\,[mJ/K$^{2}$mol], 
which is about 5 times larger than 6\,[mJ/K$^{2}$mol] for $n$-type BGG, regardless of a lower carrier density in $\beta$-BGS.
It was argued that the larger coefficient in $\beta$-BGS is attributed to a tunneling contribution of Ba(2) among the off-center sites.
The Boson-peak like hump at low temperature and the anomalously large temperature-linear coefficient in $\beta$-BGS 
are almost identical to that in vitreous silica.

\begin{figure} 
\epsfysize=2.2in 
\epsfbox{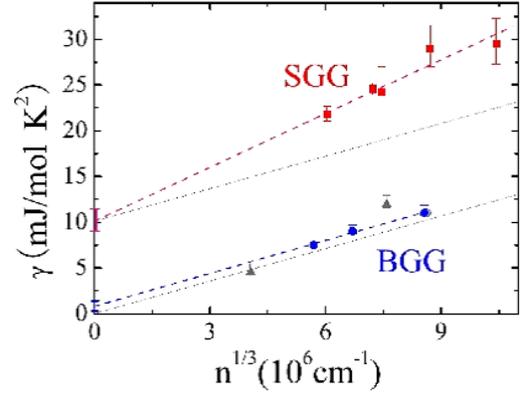}
\caption{\label{fig:takaba_fig08} (Color online) The temperature-linear factor $\gamma$ of low-temperature specific heat for type-I $\mathcal{R}_{8}$Ga$_{16}$Ge$_{30}$ ($\mathcal{R}$=Sr and Ba) 
plotted as a function of the carrier concentration $n^{1/3}$.
Adopted from \textcite{Xu:2010a}.} 
\end{figure}

\textcite{Xu:2010a} have made an analysis of the data of low-temperature specific heats of type-I BGG and SGG.
By using the relation $\gamma=\alpha+\gamma_{e}=\alpha+cm_{e}^{*}n^{1/3}$ 
($c$ is a known constant) together with the measured values of $\gamma$ and the carrier density $n$, 
they have obtained $\alpha$ and the carrier effective mass $m_{e}^{*}$.
The relationships between $\gamma$ and $n^{1/3}$ are shown in~Fig.\,\ref{fig:takaba_fig08}.
The $\alpha$ value of 10.1$\pm$1.5\,[mJ/K$^{2}$mol] for  type-I SGG is larger than that of 0.8$\pm$0.4\,[mJ/K$^{2}$mol] for BGG. 
It has been suggested that the finite $\alpha$ value in these clathrates 
presumably results from the random displacement of Ga and Ge in the host framework.
The larger $\alpha$ value in type-I SGG indicates large disorder.
The $m_{e}^{*}$ was obtained from the slopes of the $\gamma$\,{\rm~vs}.\,$n^{1/3}$ plot to be 1.68$m_{0}$ for type-I SGG and 1.01$m_{0}$ for BGG,
where $m_{0}$ is the free-electron mass.
The enhancement of $m_{e}^{*}$ was interpreted in terms of electron-phonon interactions. 
The value of $m_{e}^{*}$ is described as $m_{e}^{*} = m_{0} (1+\lambda_{\rm e-ph}+\lambda_{\rm s})$, 
where $\lambda_{\rm e-ph}$ is the electron-phonon term and $\lambda_{\rm s}$ is the spin fluctuation one. 
The last term is not important in SGG and BGG, and then values of the $\lambda_{\rm e-ph}$ were estimated to be 0.68 for SGG and 0.01 for BGG, respectively. 
For BGG, the low value of $\lambda_{\rm e-ph}$ indicates the weak interaction between electrons and vibrations of guest atoms. 
In contrast,  $\lambda_{\rm e-ph}$ is enhanced for SGG. 
This enhancement is attributed to the anharmonic phonon of the Sr(2).

\section{TRANSPORTS PROPERTIES OF TYPE-I CLATHRATE COMPOUNDS}
\label{sec:ele_experiments}
\subsection{Electrical resistivity and Seebeck coefficient}
\label{subsec:rho_S}

The type-I clathrates exhibit semiconducting or metallic behaviors in the electrical transport and thermoelectric properties 
depending on the magnitude of gap energy and doping level.
Semiconducting behaviors are characterized by the exponential decrease of $\rho$ and the $1/T$ dependence of $|S|$ with increasing temperature, 
which can be attributed to the thermal excitations of electrons over the intrinsic energy gap between the valence and conduction bands, or hopping excitations between impurity states and the band edge.
Metallic behaviors are characterized by the increase of the $\rho$ and $|S|$ with increasing temperature, 
which are caused by heavily-doped electrons or holes.
The self-doping is due to the off-stoichiometry and/or vacancy in the cage structure.
Binary and ternary type-I clathrates have gained much attention because of high thermoelectric performance.

Among type-I Si clathrates, Na$_{2}$Ba$_{6}$Si$_{46}$ and Ba$_{8}$Si$_{46}$ are superconductors
with the transition temperatures $T_{\rm sc}$ of 4\,[K] and 8\,[K], respectively\,\cite{Yamanaka:1995, Yamanaka:2000}. 
The superconductivity in such a compound based on the covalent $sp^{3}$ network has attracted great attention.
It was followed by the discovery of the superconductivity in diamond\,\cite{Ekimov:2004a} and diamond-structured Si\,\cite{Bustarret:2006a}.
False superconductivity in BGG was observed for a polycrystalline sample of Ba$_{7.62}$Ga$_{17.30}$Ge$_{28.70}$\,\cite{Bryan:1999a}.
Magnetic measurement for the finely grounded sample showed a diamagnetic signal below $T_{\rm sc}$=7.5\,[K].
The diamagnetic signal was approximately 40\%~of the shielding expected for a perfect diamagnet.
The $ac$ resistance measurement on the pelletized sample showed a resistive transition below 4.8\,[K].
The higher $T_{\rm sc}$ in the diamagnetic signal was attributed to 
the surface oxidation and the non-stoichiometric surface states in the polycrystalline grains.

\begin{figure} 
\epsfysize=2.2in 
\epsfbox{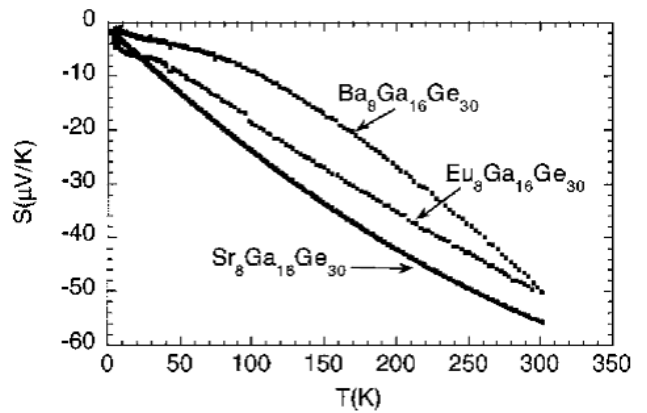}
\caption{\label{fig:takaba_fig09} Temperature dependences of the thermopower $S$\,[$\mu$V/K] for 
$\mathcal{R}_{8}$Ga$_{16}$Ge$_{30}$ ($\mathcal{R}$= Ba, Sr, Eu) single crystals.
After \textcite{Sales:2001a}.} 
\end{figure}

\textcite{Sales:2001a} have carefully examined the superconducting properties by using single crystals.
The crystal of BGG was grown by slowly cooling molten mixtures of the stoichiometric elements. 
The crystal with a carrier density $n\simeq$10$^{21}$\,[cm$^{-3}$] provided no evidence of superconductivity down to 2\,[K] in either $C_{\rm V}$, $\rho$, or $ac$ susceptibility.
Thus, the superconductivity observed by~\textcite{Bryan:1999a} may be due to small amounts of impurity phases 
such as Ga and Ba-Ge alloys with $T_{\rm sc}$=1$-$10\,[K]\,\cite{Cohen:1967a, Li:2003a, Li:2004a}.

Good electrical transport and thermoelectric properties 
have been observed in the type-I clathrates $\mathcal{R}_{8}$Ga$_{16}$Ge$_{30}$ ($\mathcal{R}$=\,Ba, Sr, and Eu). 
The replacement of Ga for Ge produces a charge compensation for the divalent alkali-earth ion $\mathcal{R}^{2+}$.
Adjusting the Ga/Ge ratio can tune the carrier density and improve the thermoelectric power factor 
$PF=S^{2}$/$\rho$ and the dimensionless figure of merit $ZT$.
Electrical transport and thermoelectric properties for single crystals of SGG and $\beta$-EGG were examined by~\textcite{Sales:2001a}.
Both $\rho$ and $S$ showed metallic behavior because of the heavily doped level of electrons at $n\simeq$10$^{21}$\,[cm$^{-3}$].
The sharp drop in $\rho$ for $\beta$-EGG below 35\,[K] is characteristic of the loss of spin disorder scattering
due to the long-range magnetic order of the Eu$^{2+}$ magnetic moments.
The room-temperature $S$ of $-$50\,[$\mu$V/K] is typical for heavily doped $n$-type semiconductors. See Fig.\,\ref{fig:takaba_fig09}.
The Hall mobility $\mu_{\rm H}$ of the carriers increases with decreasing temperature,
which suggests that the scattering of carriers is dominated by acoustic phonons.

The thermoelectric properties for hot-pressed polycrystals of SGG have been reported by~\textcite{Nolas:1998aa, Nolas:2003r}.
The values of the $n$ in the article in 1998 were corrected in the review in 2003.
The sample with $n\simeq$10$^{19}$\,[cm$^{-3}$] allowed the measurement of the semiconducting $\rho$ contribution below 100\,[K], as shown in Fig.\,\ref{fig:takaba_fig10}.
The negative $S$ decreases with increasing temperature and reaches a value of $-$320\,[$\mu$V/K] at 300\,[K], as shown in Fig.\,\ref{fig:takaba_fig11}. 
The sample with $n\simeq$10$^{20}$\,[cm$^{-3}$] exhibited metallic behaviors in both $\rho$ and $S$ in the temperature range from 5 to 300\,[K].
The thermal conductivity $\kappa$ is approximately 1\,[W/(mK)] and the value of $ZT$ reaches 0.25 at 300\,[K]. 

\begin{figure} 
\epsfysize=2.7in 
\epsfbox{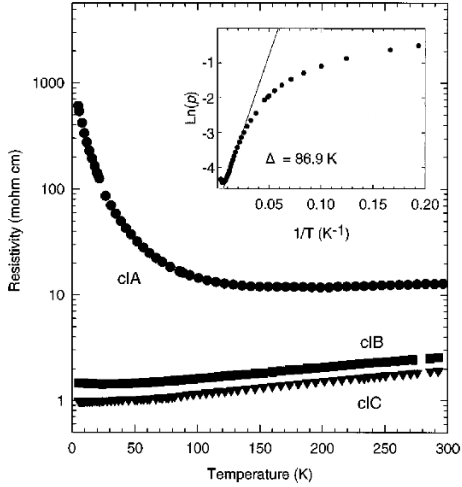}
\caption{\label{fig:takaba_fig10} Temperature dependences of electrical resistivity $\rho$ for 
Sr$_{8}$Ga$_{16}$Ge$_{30}$ polycrystalline samples with various carrier density (See text).
After the work of \textcite{Nolas:1998aa}.} 
\end{figure}
\begin{figure} 
\epsfysize=2.4in 
\epsfbox{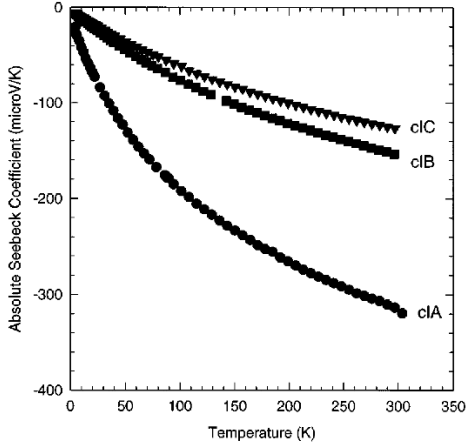}
\caption{\label{fig:takaba_fig11} Temperature dependences of the thermopower $S$\,[$\mu$V/K] for type-I
Sr$_{8}$Ga$_{16}$Ge$_{30}$ polycrystalline samples with various carrier density (see text).
After the work of \textcite{Nolas:1998aa}.} 
\end{figure}

The value of the carrier density $n$ in type-I SGG was tuned by adjusting the nominal composition in polycrystalline samples. 
\textcite{Fujita:2006a} prepared the samples by arc melting the elements with the nominal ratio of Sr\,:\,Ga\,:\,Ge$=$8\,:\,$x$\,:\,46$-x$ ($x$=13$-$20).
The ingot was powdered and sintered by the spark-plasma-sintering (SPS) technique.
With increasing $x$, the actual Ga composition tended to increase, and the carrier density $n$ decreased from 11.1 to 1.8$\times$10$^{20}$\,[cm$^{-3}$].
The effective mass $m_{\rm e}^{*}$ was almost constant at 3.1$m_{0}$, and $\mu_{\rm H}$ was 8$-$13\,[cm$^{2}$/Vs], at room temperature. 
The electrical conductivity $\sigma$ and the negative $S$ exhibited metallic behavior from 300\,[K] up to 700\,[K];
the metallic behavior turned into an intrinsic semidonducting one above 700$-$800\,[K].
The tuning of the carrier density $n$ to 5.6$\times$10$^{20}$\,[cm$^{-3}$] maximized $ZT$ of 0.62 at 800\,[K]. 

\textcite{Wang:2009a} have started with the nominal composition of Sr\,:\,Ga\,:\,Ge$=$9.6\,:\,$x$\,:\,30 ($x \,$=$\,$18$-$23).
The grounded polycrystalline sample was ultrasonically washed with diluted hydrochloric acid to remove the excess Ga. 
Dense samples were obtained by sintering the ground samples.
The carrier density $n$ decreased from 2.81$\times$10$^{20}$ to 0.33$\times$10$^{20}$\,[cm$^{-3}$] as the Ga/Ge ratio increased from 0.481 to 0.517.
The samples with $n<10^{20}$\,[cm$^{-3}$] showed a high $S$ of 350$-$400\,[$\mu$V/K] at 450\,[K] and intrinsic semiconducting conduction above 450 K
while those with higher $n$ exhibited a metallic behavior up to 700\,[K]. 
The value of $ZT$ reached 0.85 at 650\,[K] for the sample with the lowest Ga/Ge ratio of 0.481.

For $\beta$-Eu$_{8}$Ga$_{16+x}$Ge$_{30-x}$ (EGG), the carrier density $n$ was controlled by changing the annealing temperature between 823 and 970\,[K]\,\cite{Pacheco:2005a, Bentien:2005aa}.
The samples for $-$1.01\,$<x<$\,$-$0.47 had $n$ between 0.43 and 1.53 e$^{-}$ per unit cell.
All samples showed a metallic behavior in $\rho$ with a hump around 35\,[K] attributable to the ferromagnetic transition. 
$S$ exhibited metallic behavior and reached $-$(75$-$160)\,[$\mu$V/K] at 400\,[K].

\begin{figure} 
\epsfysize=2.5in 
\epsfbox{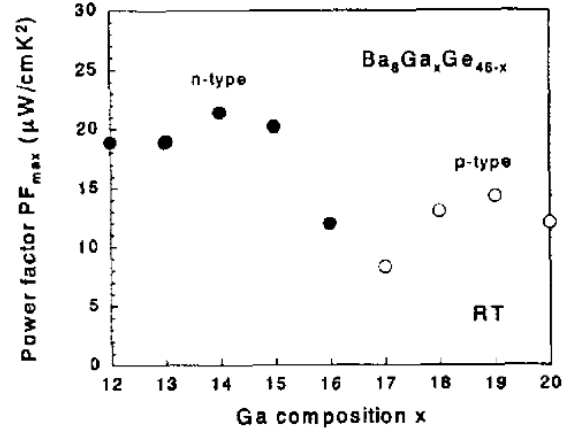}
\caption{\label{fig:takaba_fig12} Thermoelectric power factor at room temperature 
as a function of Ga composition $x$ for $n$- and $p$-type Ba$_{8}$Ga$_{x}$Ge$_{46-x}$ polycrystalline samples.
Adopted from \textcite{Anno:2002a}.} 
\end{figure}

For SGG and $\beta$-EGG, no $p$-type conduction has been observed yet.
In contrast, the carrier types and density have been
tuned in BGG\,\cite{Anno:2002a, Anno:2003a, Avila:2006a, Avila:2006aa, Cederkrantz:2009a,Tang:2010aa}.
\textcite{Anno:2002a} have tuned the carrier type by adjusting the initial composition of Ba\,:\,Ga\,:\,Ge=8\,:\,$x$\,:\,46$-x$ ($x=$12$-$20).  
Polycrystalline samples were synthesized by the arc-melting and SPS.
The carrier type changed from $n$-type ($x$=12$-$16) to $p$-type ($x$=17$-$20), which was confirmed by the change in the signs of $S$ and $R_{\rm H}$.
The $PF$ at high temperatures had a maximum of
22\,[$\mu$W/K$^2$cm] at $x$=14 for $n$-type 
and that of 15\,[$\mu$W/K$^2$cm] at $x$=19 for $p$-type. See Fig.\,\ref{fig:takaba_fig12}.
$\kappa$ was less than 2.5\,[W/(mK)].
The estimated $ZT$ value for $n$-type ($x$=15) and $p$-type ($x$=18) reached approximately 1.0 at 900\,[K]. 
The tunable carrier type in BGG is advantageous for constructing a thermoelectric module using $p$- and $n$-type legs.

\textcite{Martin:2006, Martin:2007} have performed transport experiments on  polycrystalline Si-Ge type-I clathrates with the nominal
composition Ba$_{8}$Ga$_{16}$Si$_{x}$Ge$_{30-x}$ such that a constant Ga-to-group-IV element ratio is maintained but with increasing Si
substitution 4$<x<$14.
Electrical transport measurements on $n$-type specimens show
a modest increase in the absolute Seebeck coefficient and a decrease in electrical resistivity with increasing Si content. 
Substitution of 20 at.\% Si within the Ga-Ge lattice framework of the type I clathrate BGG results in thermoelectric performance enhancement.
They have interpreted these observations as a modification in the band structure with Si substitution. 
The band modification has been confirmed by the XAFS study by \textcite{Mansour:2012}.

\textcite{Martin:2008a} have identified optimal carrier density $n$ for $n$-type BGG.
A polycrystalline sample with $n$=9.86$\times$10$^{20}$\,[cm$^{-3}$] exhibited high $ZT$=0.8 at 953\,[K].
The $ZT$ value was comparable to the values for $n$-type BGG reported 
by \textcite{Kuznetsov:2000a} ($ZT$=0.7 for polycrystalline sample at 700\,[K]),
by \textcite{Christensen:2006aaa} ($ZT$=0.9 for a Czochralski-grown single crystal at 1000\,[K])
by \textcite{Toberer:2008a} ($ZT$=0.8 for a flux-grown single crystal and $ZT$=0.74 for a polycrystalline sample at 1000\,[K]),
and by \textcite{Hou:2009a} ($ZT\sim$0.8 for a Czochralski-grown crystal at 850\,[K]). 
The highest $ZT$ of 1.35 at 900\,[K] was reported for an $n$-type crystal synthesized by the Czochralski method\,\cite{Saramat:2006a},
but it has not been confirmed.

We mention the band calculations and x-ray photoemission spectroscopy studies 
on $\mathcal{R}_{8}$M$_{16}$Z$_{30}$ ($\mathcal{R}$\,=\,Sr, Ba)\,\cite{Blake:2001a, Madsen:2003a, Nenghabi:2008a, Nenghabi:2008aa, Tang:2009a, Kono:2010a, Tang:2011a}.
\textcite{Madsen:2003a} have pointed out that undoped SGG and BGG are semiconductors 
with band gaps of 0.71 and 0.89\,[eV], respectively.
\textcite{Tang:2009a} have revealed that the valence band is constructed mainly from the Ga/Ge 4$s$ and 4$p$ wave functions with little contribution of the Sr/Ba atomic orbitals. 
However, the conduction band hybridizes with the unoccupied $d$ states of the guest atom.
These results indicate that electrical transport and thermoelectric properties of the $n$-type sample are expected to depend strongly on the guest atom
while those of $p$-type clathrate compounds are relatively unaffected.
\textcite{Kono:2010a} have calculated the electronic structures and the thermoelectric properties for 
type-I and type-V\hspace{-.1em}I\hspace{-.1em}I\hspace{-.1em}I Ba$_{8}$Ga$_{16}$Sn$_{30}$ ($\beta$- and $\alpha$-BGS). 
The calculation has shown that these compounds are indirect semiconductors with band gaps of 0.51\,[eV] and 0.32\,[eV], respectively.
The Seebeck coefficient of $n$-type $\beta$-BGS is higher than that of $n$-type $\alpha$-BGS.

\subsection{Phonon thermal conductivity}
\label{subsec:kappa}
As mentioned in Sec.\,\ref{sec:type_I_structures}, ternary type-I clathrates have received much attention as potential high-performance thermoelectric (TE) materials because of not only the good electronic properties but also the low $\kappa_{\rm ph}$ of $\simeq$\,1\,[W/(mK)].
\textcite{Nolas:1998a,Nolas:1998aa} have reported glass-like temperature dependences of $\kappa_{\rm ph}$ for SGG polycrystalline samples.
It was then regarded to satisfy the true ``phonon-glass electron-crystal" (PGEC) concept. 
The nature of the THz-frequency vibrational states in structural glasses has been the subject of intensive and controversial debate for decades.
This is due to the difficulty in identifying the microscopic structure of glasses.
The microscopic structure of clathrate compounds are well defined, and this makes it possible to interpret the origin of glass-like thermal and dynamic properties generally observed in type-I clathrates containing off-center guest atoms.

\textcite{Nolas:1998a} and \textcite{Cohn:1999a} have  investigated the temperature dependence of the $\kappa_{\rm ph}$ down to 60\,[mK] for a SGG polycrystalline sample.
$\kappa_{\rm ph}$ varies as $T^{2}$ below 1\,[K] and has a plateau around 10\,[K], 
which is quite similar to the behavior of vitreous SiO$_{2}$ in the temperature dependence in addition to the magnitude, as shown in~Fig.\,\ref{fig:takaba_fig13}.
The glass-like plateau in $\kappa_{\rm ph}$($T$) has been also observed for a single crystal of SGG. 
However, BGG showed a crystalline peak due to Umklapp process\,\cite{Keppens:2000a}, 
which indicates the existence of translational symmetry in the sample. 
We should mention that the plateau in $\kappa_{\rm ph}$($T$) has been observed in gas hydrates
in the same temperature region\,\cite{Krivchikov:2005a, English:2010r}.

\begin{figure} 
\epsfysize=2.3in 
\epsfbox{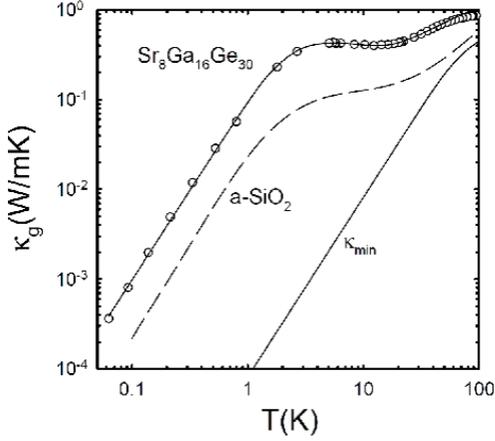}
\caption{\label{fig:takaba_fig13} Temperature dependences of phonon thermal conductivity $\kappa_{\rm ph}$ 
for polycrystalline samples of type-I Sr$_{8}$Ga$_{16}$Ge$_{30}$. The dashed line represents $\kappa_{\rm ph}$ for vitreous SiO$_{2}$.
After \textcite{Cohn:1999a}.} 
\end{figure}

\textcite{Sales:2001a} have reported a temperature dependence of heat transport for BGG, SGG, and $\beta$-EGG using single crystals with $n$-type carriers.
The crystals were grown by slowly cooling molten stoichiometric mixtures of the elements.
For SGG and $\beta$-EGG crystals, $\kappa_{\rm ph}$ exhibits a glass-like plateau in the temperature range 10-20\,[K]. See Fig.\,\ref{fig:takaba_fig14}.
The room-temperature values of $\kappa_{\rm ph}$ for SGG and $\beta$-EGG are 1.0 and 0.6\,[W/(mK)], respectively.
In contrast, $\kappa_{\rm ph}$ for BGG shows a typical crystalline Umklapp peak at around 10\,[K]. 
Nevertheless, the room-temperature $\kappa_{\rm ph}$ value is no more than 1.3\,[W/(mK)]. 

\begin{figure} 
\epsfysize=2.1in 
\epsfbox{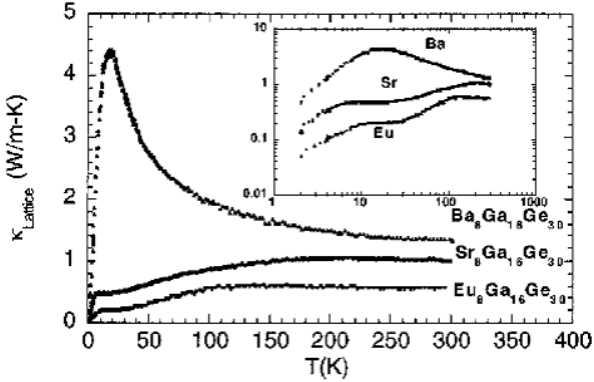}
\caption{\label{fig:takaba_fig14} Temperature dependences of phonon thermal conductivity $\kappa_{\rm ph}(T)$
for type-I $\mathcal{R}_{8}$Ga$_{16}$Ge$_{30}$ ($\mathcal{R}$ = Ba, Sr, Eu) single crystals.
Adopted from \textcite{Sales:2001a}.} 
\end{figure}
\begin{figure} 
\epsfysize=2.3in 
\epsfbox{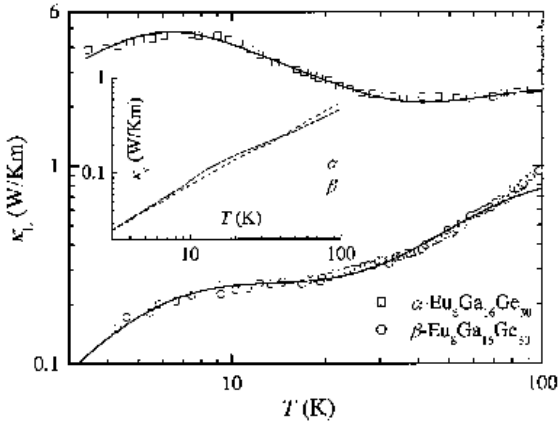}
\caption{\label{fig:takaba_fig15} Temperature dependences of phonon thermal conductivity $\kappa_{\rm ph}$ 
for type-I and type-V\hspace{-.1em}I\hspace{-.1em}I\hspace{-.1em}I Eu$_{8}$Ga$_{16}$Ge$_{30}$ polycrystalline samples.
After \textcite{Paschen:2001a}.} 
\end{figure}

The data of $\kappa_{\rm ph}$ on type-I ($\beta$-phase) and type-V\hspace{-.1em}I\hspace{-.1em}I\hspace{-.1em}I ($\alpha$-phase) EGG 
by~\textcite{Paschen:2001a} are given in Fig. {\ref{fig:takaba_fig15}}.
In contrast to the plateau for the $\beta$-EGG sample, 
the crystalline Umklapp peak was observed around 5\,[K] for a type-V\hspace{-.1em}I\hspace{-.1em}I\hspace{-.1em}I  sample.
Type-I and V\hspace{-.1em}I\hspace{-.1em}I\hspace{-.1em}I samples of EGG possess magnetic moments realizing ferromagnetic orderings at 36 and 10.5\,[K], respectively.

\begin{figure} 
\epsfysize=2.4in 
\epsfbox{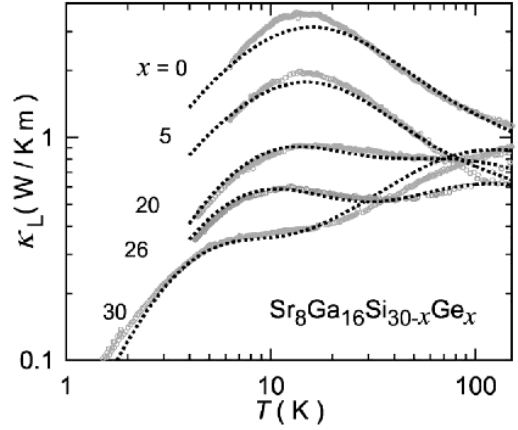}
\caption{\label{fig:takaba_fig16} Temperature dependences of phonon thermal conductivity $\kappa_{\rm ph}(T)$\,{\rm~vs}.\,$T$ on a log-log scale
for type-I Sr$_{8}$Ga$_{16}$Si$_{30-x}$Ge$_{x}$ single crystals.
After \textcite{Suekuni:2007a}.} 
\end{figure}

\textcite{Martin:2006, Martin:2007} have reported on the temperature dependence of $\kappa_{\rm ph}(T)$ of polycrystalline Si-Ge type-I clathrates with nominal
composition of Ba$_{8}$Ga$_{16}$Si$_{x}$Ge$_{30-x}$ in the temperature range 10-400\,[K] with increasing Si substitution 4$<x<$14.

\textcite{Suekuni:2007a} have reported the variations of $\kappa_{\rm ph}$ on type-I Sr$_{8}$Ga$_{16}$Si$_{30-x}$Ge$_{x}$ (0 $\leq x \leq$ 30)\,(SGSG) of single crystals.
In these alloy series, the cage size is enlarged by the substitution of Ge for Si.
The cage size expands by 8\% and the guest vibrational energy decreases as $x$ increases from 0 to 30.
Thereby, the height of the peak in $\kappa_{\rm ph}$ at around 10$-$20\,[K] was reduced 
and the temperature dependence of $\kappa_{\rm ph}$($T$) changed from crystal-like to glass-like as illustrated in~Fig.\,\ref{fig:takaba_fig16}.
The drastic reduction in $\kappa_{\rm ph}$ suggests a strong interaction between acoustic phonons of cages and vibrational modes of Sr guests.
A detailed theoretical interpretation on the above will be given in Sec.\,\ref{Dispersion_1}.
$\kappa_{\rm ph}(T)$ for $x$ = 30 varies as $T^{2-\delta}$ with a small factor $\delta<$1 below 1 K\,\cite{Umeo:2005a}, 
which confirms the data for the polycrystalline samples\,\cite{Cohn:1999a}.

Some data for the thermal conductivity in polycrystalline SGG samples have displayed crystal-like peaks at 4-10 K\,\cite{Uher:1999a}.
These peaks may be due to metallic impurities.
The samples were synthesized in a quartz tube or in a pyrolytic boron nitride (pBN) crucible placed inside a quartz tube,
which were referred to as as-grown samples.
The sample synthesized in the pBN was then crushed and sintered in a carbon die.
For the as-grown samples, $\kappa_{\rm ph}$ shows crystalline behavior: $\kappa_{\rm ph}\propto1/T$ and a large peak at around 4$-$10\,[K]. 
On the other hand, for the sintered sample, $\kappa_{\rm ph}$ shows a glass-like temperature dependence.
Despite the similar value of the carrier density $n$ = 10$^{21}$\,[cm$^{-3}$], the behaviors of $\rho$ and $S$ seem to strongly depend on the crucible material.
For samples synthesized in the quartz tube, $\rho$ decreases to zero at helium temperatures. 
This is totally different from the data for single crystals with a similar carrier density of 10$^{21}$\,[cm$^{-3}$]\,\cite{Sales:2001a}.
The quite low $\rho$ can be ascribed to the presence of highly-conducting metallic impurities, e.g., gallium.  
In fact, the $S$ goes to zero at low temperatures, which is a result of a short circuit. 
Therefore, the crystal-like behavior in $\kappa$ reported for some SGG polycrystalline samples is not intrinsic but is attributable to the metallic impurities.

\begin{figure} 
\epsfysize=2.4in 
\epsfbox{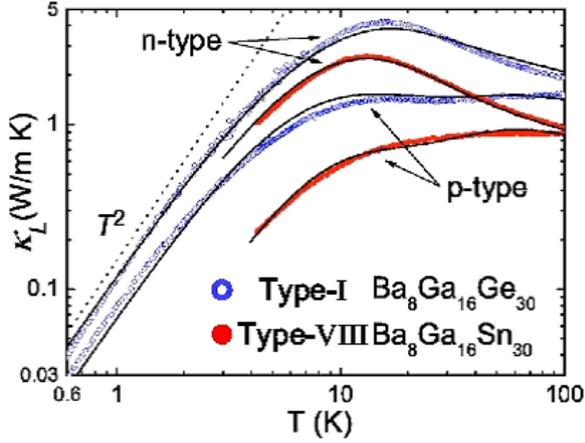}
\caption{\label{fig:takaba_fig17} (Color online) Temperature dependences of phonon thermal conductivity $\kappa_{\rm ph}$ 
for $n$- and $p$-type Ba$_{8}$Ga$_{16}$Ge$_{30}$
and type-V\hspace{-.1em}I\hspace{-.1em}I\hspace{-.1em}I
Ba$_{8}$Ga$_{16}$Sn$_{30}$ single crystals on a log-log scale.
The data are adopted from the works of  \textcite{Avila:2006aa}.} 
\end{figure}
\begin{figure} 
\epsfysize=2.4in 
\epsfbox{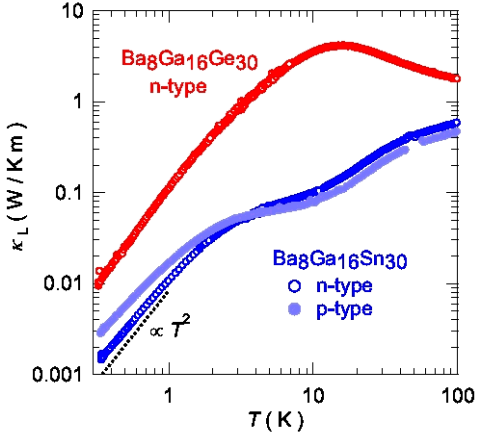}
\caption{\label{fig:takaba_fig18} (Color online) Temperature dependences of phonon thermal conductivity $\kappa_{\rm ph}(T)$ 
of $n$-type and $p$-type type-I Ba$_{8}$Ga$_{16}$Ge$_{30}$ and Ba$_{8}$Ga$_{16}$Sn$_{30}$ single crystals on a log-log scale.
After \textcite{Avila:2006aa, Avila:2008a, Suekuni:2008a}.} 
\end{figure}

For type-I BGG, a large difference in $\kappa_{\rm ph}$($T$) between the $n$- and $p$-type samples
has been reported by different groups\,\cite{Bentien:2004a,Avila:2006a,Avila:2006aa, Tang:2009a}.
The single crystals were synthesized by the flux method using excess Ga.
\textcite{Avila:2006a, Avila:2006aa} grew the crystals with both $p$- and $n$-type carriers from nominal compositions of 
Ba$:$Ga$:$Ge$=$8$:$38$:${\it x}, where $x$=30 for $p$-type and $x$=34 for $n$-type.
$\kappa_{\rm ph}$ for the $n$-type samples showed a typical crystalline peak at 15\,[K], 
which is totally suppressed for $p$-type samples, as shown in Fig.\,\ref{fig:takaba_fig17}. 
$\kappa_{\rm ph}$ below 1\,[K] for the $p$-type single crystal follows
$\sim T^{2}$\cite{Avila:2006aa} or $\sim T^{1.5}$\cite{Bentien:2006a};
that for $n$-type follows $\sim T^{2}$\,\cite{Avila:2006aa}.
However, no difference was observed in either the crystal structure nor specific heat $C_{\rm V}$($T\gtrsim 4$\,[K]) for these compounds.
Similar difference in $\kappa_{\rm ph}$ between $p$- and $n$-type samples was observed also in type-I Ba$_{8}$Ni$_{6-x}$Ge$_{40-x}$, Ba$_{8}$Au$_{x}$Si$_{46-x}$, and type-V\hspace{-.1em}I\hspace{-.1em}I\hspace{-.1em}I Ba$_{8}$Ga$_{16}$Sn$_{30}$\,\cite{Bentien:2006a, Avila:2006aa, Aydemir:2011a}.

The Ba guest in type-I Ba$_{8}$Ga$_{16}$Sn$_{30}$ ($\beta$-BGS) has larger off-center displacement 
and lower guest vibration energy than that of BGG\,\cite{Avila:2008a,Suekuni:2008a}.
The carrier types of $\beta$-BGS are tunable as in BGG.
$\kappa_{\rm ph}$ varies as $T^{2-\delta}$ below 
1\,[K] and exhibits a plateau around 4\,[K] irrespective of the carrier types.
See Fig.\,\ref{fig:takaba_fig18}.
The value of $\kappa_{\rm ph}$ below 10\,[K] is the lowest among intermetallic clathrates.
Because of the glass-like $\kappa_{\rm ph}$ and the  Boson-peak like structure in the specific heat $C_{\rm V}$, 
$\beta$-BGS is regarded as a true phonon-glass material.

\begin{figure*}
\epsfysize=2.3in 
\epsfbox{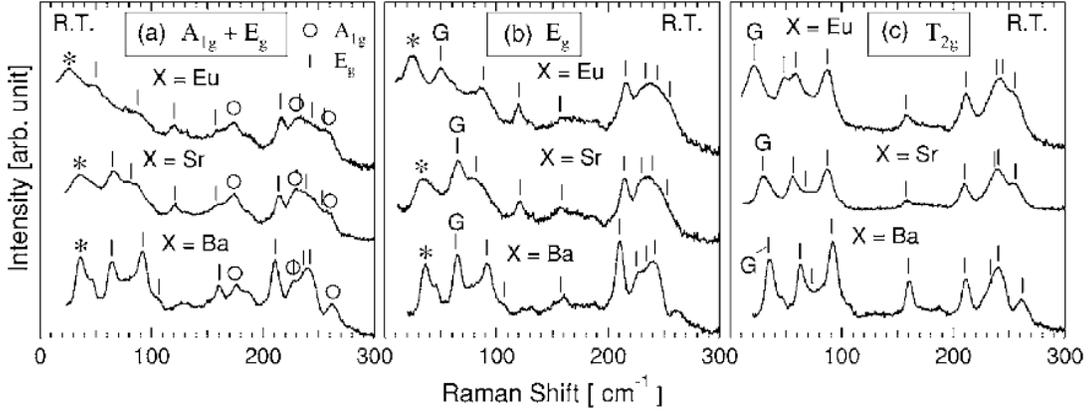}
\caption{\label{fig:takaba_fig19} Polarization dependence of Raman spectra for type-I $\mathcal{R}_{8}$Ga$_{16}$Ge$_{30}$ ($\mathcal{R}$ = Ba, Sr, Eu) at room temperature.
``G'' are the symmetry-allowed Raman active phonons, and bars and open circles denote the assigned peak. The peaks marked by $\ast$ are additional modes.
After \textcite{Takasu:2006a}.} 
\end{figure*}

\section{SPECTROSCOPIES IN THz FREQUENCY REGION}
\label{sec:THz}
The excess densities of states realized as a hump in specific heats at around 10\,[K] for type-I clathrate compounds have also been observed by various types of spectroscopies such as Raman scattering, infra-red absorption, and inelastic neutron scattering measurements.
This Section describes in detail these observations.
\subsection{Optical spectroscopies}
\label{optic_spectroscopy}
\subsubsection{Raman spectra and those assignments}
\label{sec:raman}
The reflected light from crystals yields small shifts in initial frequency.
This shift is called the Raman shift or scattering, which enables us investigate characteristic frequencies relevant to optical phonons in the vicinity of $\Gamma$-point ($\qq$=0) in the Brillouin zone.
This is due to the fact that the wave vector $\kk$ of photons is negligibly small compared to that of $\qq$ of optical phonon.
It has been known that the Raman spectroscopy is a powerful method to access the dynamics relevant to the excess density of states.

\textcite{Krishnan:1953a} found during
his Raman scattering experiments on silica glass a broad band in the vicinity of 30$-$120 \,[cm$^{-1}$], which markedly differs from the behavior of crystal quartz.
Note that 1\,[THz]=33.35\,[cm$^{-1}$].  
This band has its maximum intensity at the low frequency end and the intensity falls off continuously. 
The broad and intense band  has been also observed in various types of topologically disordered amorphous materials or structural glasses by Raman scattering, inelastic neutron scattering and infrared absorption. 
It should be noted that \textcite{Helen:2000a} have performed hyper-Raman scattering experiments
to investigate the broad band in the THz region in silica glass.
The relevant mode
obtained by hyper-Raman scattering principally involves rotational motion of SiO$_4$ tetrahedra.
This mode corresponds to the mode suggested by \textcite{Buchenau:1984a, Buchenau:1986a} from the analysis of their inelastic neutron scattering data.

Type-I clathrates have three distinct cage sites 6$c$, 16$i$, and 24$k$. 
The guest(1) occupies the 2$a$ sites in the dodecahedra, and the guest(2) does the 6$d$ sites in the tetrakaidecahedra. 
There are 54 atoms per unit cell and thus 162\,(3$\times$54) phonon modes at the $\Gamma$-point. 
Group theory of lattice vibrations for type-I clathrates with a cubic primitive structure $Pm\bar{3}n$ 
indicates the number of Raman-active vibrational modes 3$A_{\rm 1g}$+7$E_{\rm g}$+8$T_{\rm 2g}$ relevant to the cage atoms. 
The group theory also ensures the existence of two guest(2) modes $E_{\rm g}$ and $T_{\rm 2g}$ 
under the assumption that the guest(2) occupies the on-center 6$d$ site.
Because the vibrations of the guest(1) at the 2$a$ site are $not$ Raman active, 
all observed guest modes are attributed to the guest(2) atoms.

Raman scattering from a polycrystalline sample provides all active modes.
The study of $\beta$-$\mathcal{R}_{8}$Ga$_{16}$Ge$_{30}$ ($\mathcal{R}$= Sr and Eu) 
has provided assignments of the guest(2) and cage modes\,\cite{Nolas:2000aa}.
However, they were unable to identify all of the modes due to the numerous  Raman active modes.  

\textcite{Takasu:2006a} have given full assignments to the Raman active modes observed for type-I $\mathcal{R}_{8}$Ga$_{16}$Ge$_{30}$ ($\mathcal{R}$=Ba, Sr and Eu) by combining polarized Raman scattering measurements on single crystals with first principles calculations. 
Vibrational modes with each irreducible representation have been determined by polarization dependence measurements.
The polarization geometry is represented by the notation ($\alpha$, $\beta$), 
where $\alpha$ and $\beta$ denote the polarization directions of incident and scattered light, respectively. 
Three geometries of ($x$, $x$), ($x$,$y$), and ($x$+$y$,$x$-$y$) were employed, 
where $x$ and $y$ correspond to the [100] and [010] crystal axes, respectively. 
The phonon with $A_{1g}$ symmetry appears in ($x$,$x$), 
$E_{\rm g}$ in both ($x$,$x$) and ($x$+$y$,$x$-$y$), and $T_{\rm 2g}$ in ($x$,$y$).
The $A_{\rm 1g}$ spectra have been obtained by subtracting the $E_{\rm g}$ spectra from the $A_{1g}$+$E_{\rm g}$.

The Raman spectra of the $A_{\rm 1g}$+$E_{\rm g}$, $E_{\rm g}$, and $T_{\rm 2g}$ modes are shown in Fig.\,\ref{fig:takaba_fig19}
for the three compounds at room temperature\,\cite{Takasu:2006a}.  
The modes above 70\,[cm$^{-1}$] are fully assigned as cage modes 
in comparison with the energies obtained by experiments and calculations.
For $\mathcal{R}$=Ba, the $E_{\rm g}$(1) mode at 64\,[cm$^{-1}$] and $T_{\rm 2g}$(1) at 34\,[cm$^{-1}$], 
marked by “G” in Fig.\,\ref{fig:takaba_fig19}, is assigned to the modes for Ba(2).
The $E_{\rm g}$(1) mode concerns guest vibrations along the four-fold axis; 
and the $T_{\rm 2g}$(1) concerns the vibrations in the plane perpendicular to the fourfold axis and along the [100] axis.
The modes for Sr(2) and Eu(2) have been assigned by comparing the analogous spectral shape between $\mathcal{R}$=Ba, Sr, and Eu.
The energy of the $T_{\rm 2g}$(1) mode at room temperature decreased on going from Ba(2) (34\,[cm$^{-1}$]), Sr(2) (30\,[cm$^{-1}$]), and to Eu(2) (21\,[cm$^{-1}$]).
The energies for Ba(2) and Sr(2) agreed with those obtained from the analysis of specific heat data\,\cite{Avila:2006aa, Suekuni:2007a}.

Besides the assigned peaks, there are two additional modes, marked by asterisks in Fig.\,\ref{fig:takaba_fig19}, 
in the low-energy region of the $A_{\rm 1g}$+$E_{\rm g}$ and $E_{\rm g}$ spectra\,\cite{Takasu:2006a}. 
Particularly for $\mathcal{R}$=Eu, the good subtraction of the $E_{\rm g}$ from $A_{\rm 1g}$+$E_{\rm g}$ has allowed us to assign
the additional modes to one $A_{\rm 1g}$ and one $E_{\rm g}$ mode.
These modes were identified as the vibrations of the guest(2). 

Neutron diffraction experiments have provided evidence that the nuclear density of the guest atom Eu(2) departs from the 6$d$ site of cage-center to one of four off-center 24$k$ sites along the [1,0,0] direction\,\cite{Chakoumakos:2001a, Sales:2001a}.
This deviation lowers the site symmetry from 4$\bar{2}m$ (6$d$) to $m$ (24$k$).
The lowered symmetry does not affect the $E_{\rm g}$(1) and $T_{\rm 2g}$(1) modes,
but the displacement along the [1,0,0] direction, corresponding to the $A_{\rm 1g}$(A) mode, 
becomes Raman active, as shown in Fig.\,\ref{fig:takaba_fig20}\,\cite{Takasu:2008a}.
The additional $E_{\rm g}$(A) mode has been identified as the tangential fluctuation to [1,1,0]. 
Thus, the vibrations of the guest(2) have been described as rotational or librational modes,together with an additional $A_{\rm 1g}$ mode.

\begin{figure} 
\epsfysize=1.2in 
\epsfbox{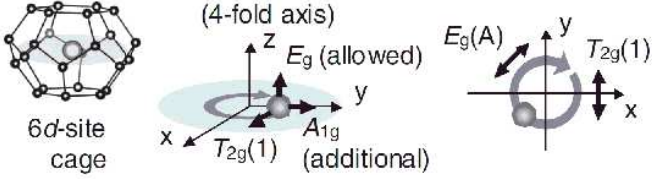}
\caption{\label{fig:takaba_fig20} c) Illustration of librational modes of the guest atom $\mathcal{R}$(2). See the text for the notations of additional $A_{\rm 1g}$ , $E_{\rm g}$(A), and $T_{\rm 2g}$(1).
Adopted from \textcite{Takasu:2008a}.} 
\end{figure}

\begin{figure} 
\epsfysize=1.8in 
\epsfbox{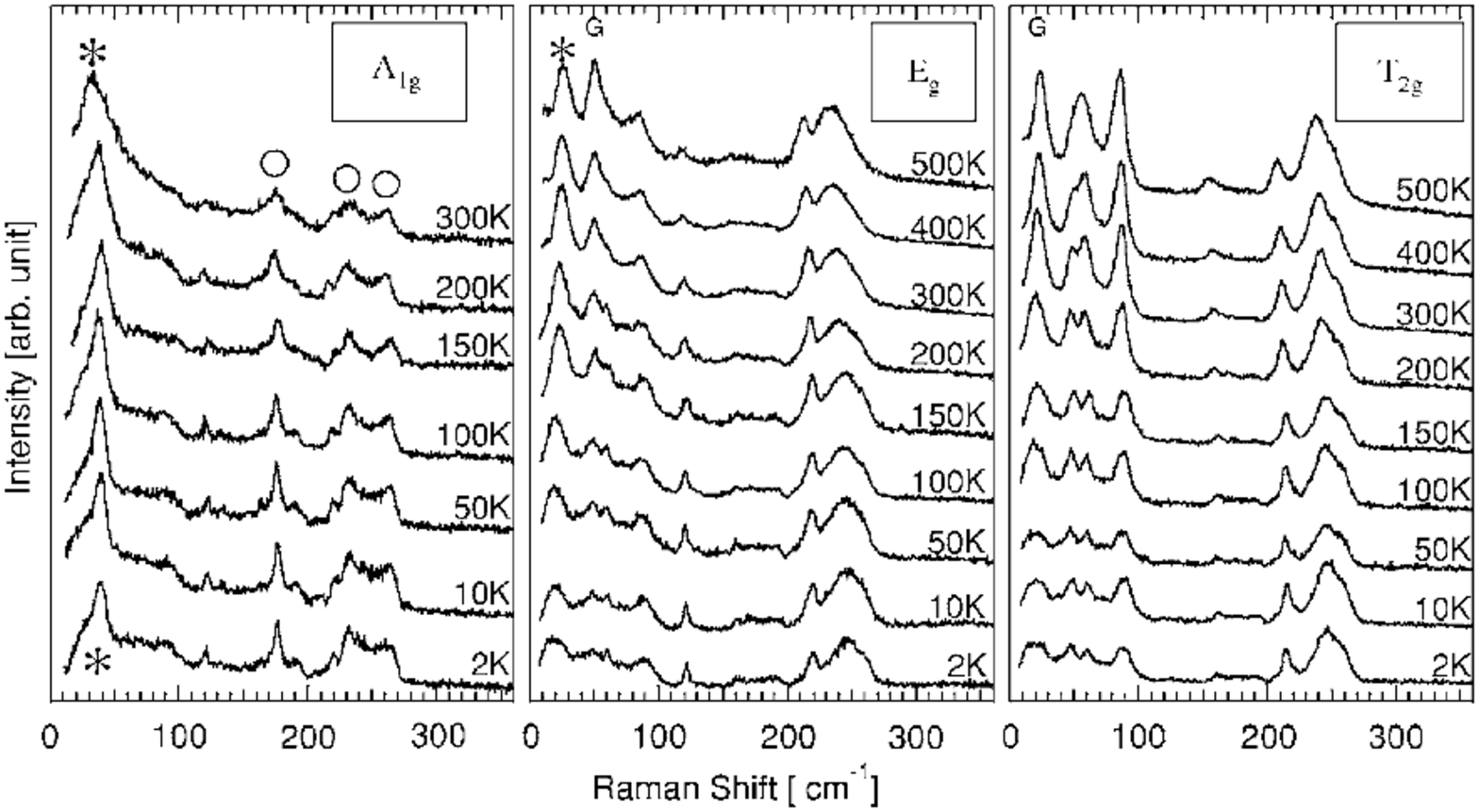}
\caption{\label{fig:takaba_fig21} Temperature dependences of Raman spectra for type-I  Eu$_{8}$Ga$_{16}$Ge$_{30}$.
After \textcite{Takasu:2006a}.} 
\end{figure}

\subsubsection{Temperature, carrier-type, and pressure dependences of Raman spectra}
\label{sec:Temperature_Raman}
Temperature dependences of the Raman spectra for the guest atom $\mathcal{R}$ $=$ Eu are shown in Figs.\,\ref{fig:takaba_fig21} and~\ref{fig:takaba_fig22}
\,\cite{Takasu:2006a, Takasu:2007a}. 
The modes related to cages show no anomalous dependence on cooling from 500\,[K] to 2\,[K],
in which temperature range their energy increases due to the thermal shrinkage effect. 
This is the same tendency for $\mathcal{R}$=Ba and Sr.
However, it should be emphasized that the vibrational energies of the $T_{\rm 2g}$(1) and $E_{\rm g}$(A) on the guest(2) \textit{decrease} on cooling. 
This anomalous softening can be primarily explained by the relevance of the anharmonic potential including a quartic term; $U(u)=k_{0}+k_{2}u^{2}+k_{4}u^{4}$, from which 
the mode frequency $\omega$ is expressed by $\omega^2 = \omega_0^2 + k_4\langle u^2 \rangle$ under a self-consistent phonon approximation where $\omega_0$ is the eigenfrequency of the harmonic term, $k_4$ the coefficient of the quartic anharmonic potential, 
and $\langle u^2 \rangle\propto T$ is a mean-square atomic displacement of the guest(2), respectively. 
The quantity $\langle u^2 \rangle$ explains the temperature dependence of $\omega^2$ and Raman spectra relevant to the guest(2) on cooling.

For $\mathcal{R}$=Ba, the energy of $T_{\rm 2g}$(1) coincides with that of $E_{\rm g}$(A) 
at temperatures between 300\,[K] and 2\,[K], as shown in Fig.\,\ref{fig:takaba_fig22}\,\cite{Takasu:2006a}.
For the $\mathcal{R}$=Sr and Eu, the energy of $T_{\rm 2g}$(1) is lower than that of $E_{\rm g}$(A) at 300\,[K].
This means that the potential minima are located toward the [100] direction 
which points from the high symmetry 6$d$ site to the off-center 24$k$ sites. 

\begin{figure} 
\epsfysize=3.5in 
\epsfbox{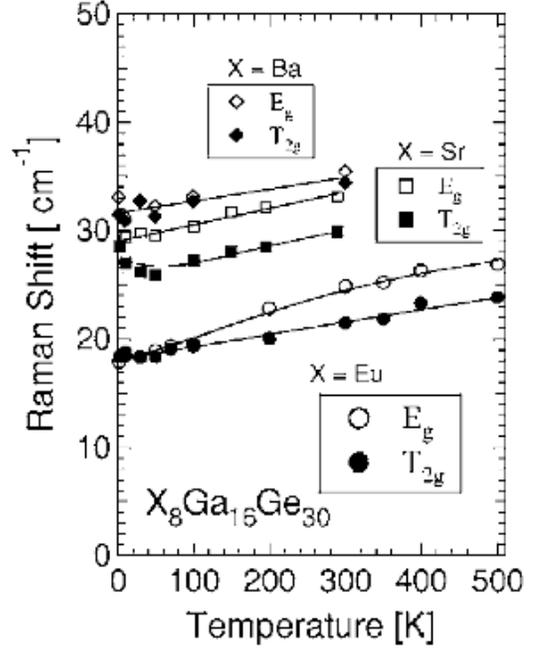}
\caption{\label{fig:takaba_fig22} Temperature dependences of mode energies of guest atoms, 
$T_{\rm 2g}$(1) and $E_{\rm g}$(A), for type-I $\mathcal{R}_{8}$Ga$_{16}$Ge$_{30}$ ($\mathcal{R}$ = Ba, Sr, Eu).
After \textcite{Takasu:2006a}.} 
\end{figure}

Vibrational properties of pseudo-ternary systems $\mathcal{R}_{8}$Ga$_{16}$Si$_{30-x}$Ge$_{x}$ ($\mathcal{R}$= Ba and Sr) 
have also been investigated in terms of the Raman spectroscopy\,\cite{Takasu:2008a}.
The vibrational energies of the guest(2) have decreased clearly with increasing Ge content $x$ and cage size.
For Ba samples of the end compositions $x$ = 0 and $x$ = 30 and the Sr compound with $x$ = 0, 
the energy difference between $T_{\rm 2g}$(1) and $E_{\rm g}$(A) was small, 
indicating an isotropic potential at the guest(2) site for the these samples.
For the Sr samples with $x$=20 and $x$=30, the difference becomes larger with increasing cage size, 
and thereby the energy of the $T_{\rm 2g}$(1) was lower than that of the $E_{\rm g}$(A).
The results indicate that the potential anisotropy for Sr(2) between
the [100] and [110] directions emerges with increasing the cage size. 

Compared with Ba$_{8}$Ga$_{16}$Ge$_{30}$ (BGG), type-I Ba$_{8}$Ga$_{16}$Sn$_{30}$ has a larger cage, which results in a larger off-center deviation of Ba(2).
The energy of $T_{\rm 2g}$(1) is lower than that of the $E_{\rm g}$(A) at temperatures between 300\,[K] and 
4\,[K]\,\cite{Suekuni:2010a},
indicating the strong anisotropy of the potential at the Ba(2) site.
The energy of $T_{\rm 2g}$(1) at 4\,[K] in~Fig~\ref{fig:takaba_fig23} is 14\,[cm$^{-1}$], which is lower than 
that in $\mathcal{R}_{8}$Ga$_{16}$Ge$_{30}$ for $\mathcal{R}$ $=$ Ba (31\,[cm$^{-1}$]), Sr (29\,[cm$^{-1}$]), and Eu (18\,[cm$^{-1}$]).
\begin{figure} 
\epsfysize=4.0in 
\epsfbox{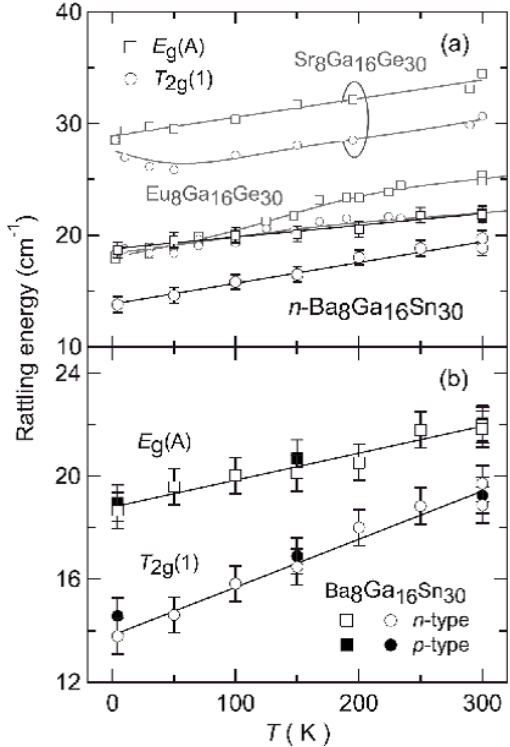}
\caption{\label{fig:takaba_fig23} Temperature dependences of the mode energies, 
$T_{\rm 2g}$(1) and $E_{\rm g}$(A), for $n$- and $p$-type of type-I Ba$_{8}$Ga$_{16}$Sn$_{30}$ and $\mathcal{R}_{8}$Ga$_{16}$Ge$_{30}$ ($\mathcal{R}$=Sr, Eu).
Adopted from \textcite{Takasu:2006a} and \textcite{Suekuni:2010a}.} 
\end{figure}

For BGG, carrier-type ($n$- and $p$-type) dependences of Raman spectra have been studied by \textcite{Takasu:2010a}.
They have observed the additional $E_{\rm g}$(A) mode independent of carrier types.
For both $n$- and $p$-types, the decrease in the energies of $T_{\rm 2g}$(1) and  $E_{\rm g}$(A) modes upon cooling provides information concerning the anharmonic potential for the guest Ba(2).
The energy difference between $T_{\rm 2g}$(1) and  $E_{\rm g}$(A) was larger in the $p$-type,
and the energy of the $T_{\rm 2g}$(1) was clearly lower than that of the $E_{\rm g}$(A).
Therefore, the potential anisotropy at the Ba(2) site is stronger in $p$-type.
Furthermore, it has been clarified that the potential energy difference between the [100] and [110] directions 
is proportional to the off-center deviation of the guest(2) from the cage center in a series of type-I clathrates.
The Raman intensity of cage vibrations at the 6$c$ site for $p$-type 
is weaker than that for $n$-type, indicating that the vibration amplitude at the 6$c$ site is smaller in the case of $p$-type. 
These experimental findings suggest that the 6$c$-site in $p$-type  provides the guests with a larger space.

Raman scattering experiments under high pressure have provided important information on the structural stability of type-I clathrate compounds\,\cite{Shimizu:2009a, Shimizu:2009aa, Kume:2010a}.
The phase transition and the vibrational properties of BGG 
have been investigated at pressures up to 40\,[GPa] at room temperature by \textcite{Kume:2010a}. 
They have demonstrated the occurrence of a first-order phase transition at 33\,[GPa] by combining Raman scattering 
and synchrotron x-ray powder diffraction (XRD) experiments at high pressures. 
The Raman spectra showed anomalies at around 17\,[GPa]. 
They have investigated the shift of vibrational frequency of the guest Ba as a function of the cage size. 
The results have indicated a linear relation between the vibrational frequency and the cage size.

The deviation of nearest-neighboring Ba(2) atoms create electric dipole fields\,\cite{Takasu:2010a}.
The $T_{\rm 1u}$ and $T_{\rm 2g}$ modes are correlated with the motion of the neighboring Ba(2) guests, and
the corresponding dipole moment lies along its direction.
The dipole fields of the $T_{\rm 1u}$ and $T_{\rm 2g}$ modes are directed opposite.
In fact, the dipole interaction yields the energy shift: 
the energy of $T_{\rm 1u}$ is larger by about 6\,[cm$^{-1}$] than that of $T_{2g}$,
indicating the existence of interacting dipoles between the neighboring terakaidecahedra.

\subsubsection{Infrared spectroscopy in the terahertz (THz) frequency range}
\label{sec:terahertz}

\begin{figure} 
\epsfysize=3.7in 
\epsfbox{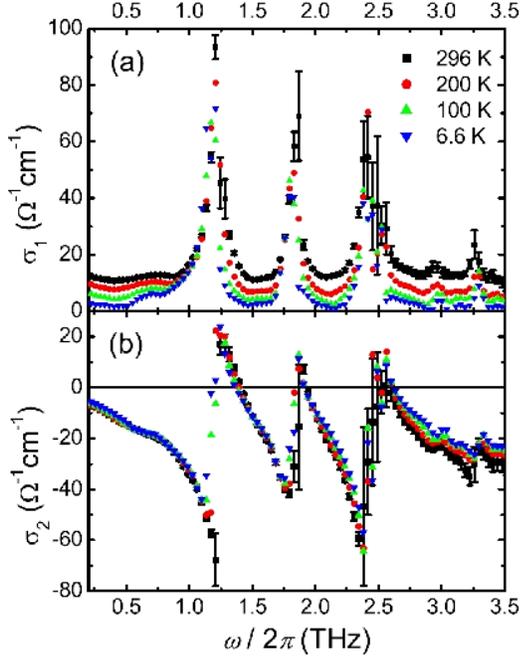}
\caption{\label{fig:takaba_fig24} (Color online) Temperature dependences of conductivity
$\sigma(\omega)$=$\sigma_{1}(\omega) + i\sigma_{2}(\omega)$ for type-I Ba$_{8}$Ga$_{16}$Ge$_{30}$.
After \textcite{Mori:2009a}.} 
\end{figure}

\begin{figure} 
\epsfysize=6.5in 
\epsfbox{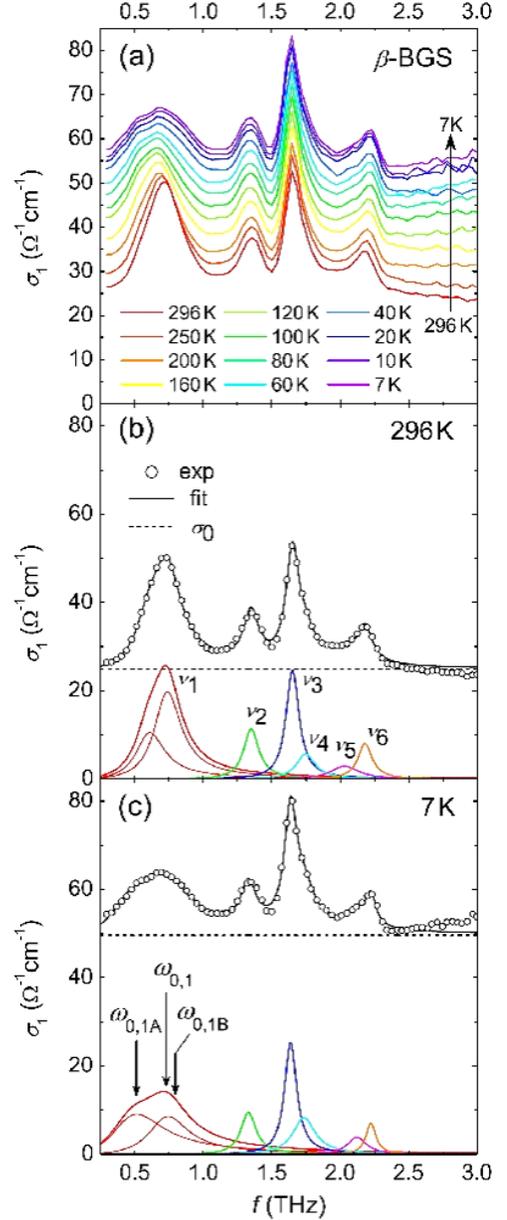}
\caption{\label{fig:takaba_fig25} (a) (Color online) Temperature dependences of real part
of conductivities $\sigma_{1}$ for type-I Ba$_{8}$Ga$_{16}$Sn$_{30}$. 
(b) and (c) represent experimental data (open circles) and fitting results (solid lines) at 296\,[K] and 7\,[K], respectively (see text).
After \textcite{Mori:2011a}.} 
\end{figure}

Terahertz (THz) frequency spectroscopy plays an important role for the investigation of the  characteristics of vibrational modes not accessible by other optical spectroscopies such as Raman scattering.
In this method, input and propagated electromagnetic pulse-shapes are measured.
Fourier analysis of both pulses enables us to obtain the frequency-dependent absorption and dispersion of the observed spectra.
Infrared-active phonon modes can be investigated in terms of this spectroscopy. 
THz time-domain spectroscopy (THz-TDS) has been successfully adopted for investigating vibrational properties of various kinds of clathrates.  
Ultrafast laser pulses generate short pulses of broadband THz radiation, which  provide both high sensitivity and time-resolved phase information for the transmitted THz electric fields. 
Group theory of the lattice vibrations of type-I clathrates with cubic primitive structure $Pm\bar{3}n$ 
gives the infrared-active vibrational modes of ten $T_{\rm 1u}$ for the cage atoms, one $T_{\rm 1u}$ for the guest(1), 
and two $T_{\rm 1u}$ for the guest(2) under the situation that guest(2) occupies the on-center 6$d$ site.

THz-TDS studies on $p$-type Ba$_{8}$Ga$_{16}$Ge$_{30}$ single crystal with on-center guest atoms have been  performed in the frequency range between 0.2\,[THz] and 3.5\,[THz]\,\cite{Mori:2009a}.
As shown in Fig.\,\ref{fig:takaba_fig24}, peaks in conductivity spectra at $\nu_1$=1.15\,[THz] and $\nu_2$=1.80\,[THz] are identified with the guest(2) modes.
They have assigned these modes as vibrational modes of Ba(2) in the plane 
perpendicular to the fourfold axis and that along the [100] axis, respectively.
The broad peak around 2.4\,[THz] ($\nu_3$ and $\nu_4$) results from a hybridization between the cage mode and Ba(1) vibrations. 
Two peaks at higher frequencies are the cage modes.

The energies of $\nu_1$ and $\nu_2$  decrease by 4.2\% with decreasing temperature from 300\,[K] to 6.6\,[K]. 
The decrease is attributed to the anharmonic potential at the Ba(2) site, as will be theoretically discussed in \ref{Anharmonic_potential_Relative}.
For other modes, however, the mode energies slightly increased with decreasing temperature as a result of the effect of thermal shrinkage.
The temperature dependence of each mode was consistent with the results of Raman scattering\,\cite{Takasu:2006a}.

The vibrational modes for type-I Ba$_{8}$Ga$_{16}$Sn$_{30}$ single crystal with off-center guest atoms have been investigated 
in the frequency range between 0.3 and 3.0\,[THz]\,\cite{Mori:2011a}.
As shown in Fig.\,\ref{fig:takaba_fig25}, the mode with the lowest energy $\nu_1$=0.72\,[THz] at 296\,[K] 
is assigned as the vibrations relevant to the off-center Ba(2) guest in the plane perpendicular to the four-fold axis.
The observed peak was fitted by superposing two Lorentzian curves.
The peak energy, defined as the peak position of the superposed curves,
decreases by 7\% with decreasing temperature down to 100\,[K], and then \textit{increases} with further decreasing temperature. 
Such anomalous peak-shape and temperature-dependence were not observed for type-I Ba$_{8}$Ga$_{16}$Ge$_{30}$
containing on-center guest atoms. 
They have tried to explain the peak splitting at low temperatures assuming
a one-dimensional double-well potential.
The energy levels with quantum numbers $n$=0 and 1 are quasi-degenerate, 
whereas the $n$=2 and 3 levels are well separated due to the moderately deep potential well. 
At sufficiently low temperatures, the optical transitions occur mainly from $n$ = 0 to 3 and from $n$=1 to 2 owing to a parity requirement of the wave functions. 
As a result, these two transitions result in double peaks.
However, significant broadening of the spectra for the guest(2) towards low temperature 
was not reproduced by assuming this type of one-dimensional potential.
The mechanism of this broadening at low temperatures will be discussed in \ref{subsubsec:off_center}. 

NMR study provides important information on the motion of encaged guest atoms depending on temperature. 
\textcite{Gou:2005a} have performed NMR studies for Sr$_{8}$Ga$_{16}$Ge$_{30}$ belonging to the category of an off-center system.
They have measured $^{71}$Ga NMR down to 1.9\,[K].
The values of Knight shift and $T_1$ showed low density metallic behavior. 
There is a significant increase in linewidth above 4\,[K] attributing to atomic motion, as shown in Fig.\,\ref{fig:takaba_fig26}.
They have estimated the relevant time scale of roughly $\tau\,\sim$10$^{-5}$\,[s] from the inverse of the linewidth.
From the time scale of the narrowing of the  NMR line, they have concluded that the reduced high-temperature width is due to motional narrowing. 
The point is that this is very slow
compared with relevant vibrational frequencies of $\omega_{\rm ph}/2\pi\simeq$\,1\,[THz], and the relation  $\omega_{\rm ph}\tau\gg1$ holds.
This indicates very slow hopping rates of Sr atoms between sites within the same cage.
Another point is that the motion of guest atoms can be treated as vibrational.

\textcite{Arcon:2013a} have performed $^{71}$Ga NMR experiments on type-I SGG and BGG in order to extract the contributions from different Ga sites at randomly occupied Ga/Ge sites.
They have claimed that the effect of weak covalent bonding should be taken
into account in addition to ionic interactions.
\begin{figure}[t]
\begin{center}
\includegraphics[width = 0.9\linewidth]{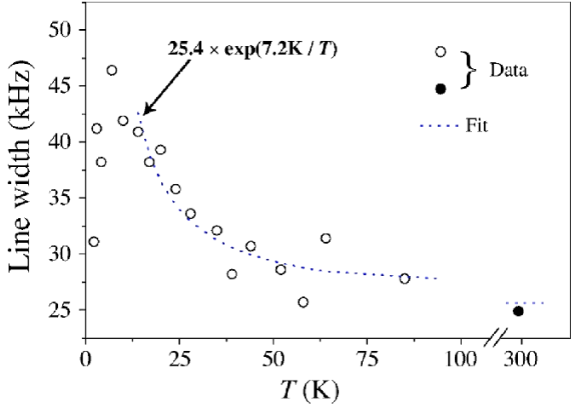}
\caption{$^{71}$Ga NMR linewidth (square root of second moment) ${\rm vs.}$ temperature for type-I Sr$_{8}$Ga$_{16}$Ge$_{30}$.
Adopted from \textcite{Gou:2005a}.
}
\label{fig:takaba_fig26}
\end{center}
\end{figure}

\subsection{Inelastic neutron scattering}
\label{sec:inelastic}

Time-of-flight inelastic neutron scattering (INS) is a powerful technique for the investigation of vibrational states in clathrate compounds. 
In INS experiments, the energy $\hbar\omega$ and wavelength $\lambda$ of incident neutrons can be taken as comparable to  those of excited phonon energies and the length scale of lattice spacing, respectively.  
From this wavelength-energy relation, INS experiments have been proven to be unique for the study of the dynamic properties of materials on the atomic scale\,\cite{Lovesey:1984b, Squires:1978}. 

The quantity obtained by INS experiments is the dynamic structure
factor $S(\QQ, \omega)$, which provides information projected onto plane-wave states since this is defined through the spatio-temporal
Fourier transformation. 
When one projects exact eigenstates onto plane-wave states, a lifetime is obtained in frequency or wave vector space, where the width arises from that a plane-wave would experience.
Thus, it is necessary to take into account this point for the interpretation of the energy width of local modes associated with $\mathcal{R}$(2) guest atoms in cages.

\subsubsection{Dynamic structure factor and the flat dispersion relation
of phonons}
\label{INS_On_center}
Vibrational dynamics in the terahertz (THz) frequency range is the most important aspect for clarifying the glass-like phonon thermal conductivity $\kappa_{\rm ph}(T)$ in type-I clathrate  compounds,
because the upper limits of frequencies of acoustic branches appear in this frequency range.
The THz frequency dynamics concerns the lowest two bands: acoustic and the lowest optic modes originating from networked cages and $\mathcal{R}$(2) guest atoms. 

The deformed vibrations of cages themselves manifest higher energy states than those in the THz frequency region.
For this reason, it is sufficient to treat the rigid cage as having a mass $M$ and an effective charge $-e_C^*$, and the $\mathcal{R}$(2) guest atom with mass $m$ and charge $e_G^*$.
We define the position vector of the $\ell$-th cage at time $t$ as $\RR_\ell+\rr_\ell(t)$,
where $\RR_\ell$ is the equilibrium position of the $\ell$-th cage center.
The  vector $\rr_\ell(t)$ represents a small deviation from  $\RR_\ell$.
The position of $\mathcal{R}$(2) guest atom from the center of $\ell$-th cage $\RR_\ell$ is defined by the
vector $\UU_\ell+\uu_\ell(t)$, where $\UU_\ell$ is
the equilibrium position of the $\mathcal{R}$(2) guest atom with respect to $\RR_\ell$,
and $\uu_\ell(t)$ is a small deviation from  $\UU_\ell$ at time $t$.
Note that $\UU_\ell\neq 0$ corresponds to the case of an \textit{off-center} guest atom,
while the \textit{on-center} case becomes $\UU_\ell=0$.
Figure \ref{fig:takaba_fig27} gives the definitions of the vectors $\RR,\,\rr_\ell(t),\, \UU_\ell$ and $\uu_\ell(t)$.
\begin{figure}[t]
\begin{center}
\includegraphics[width = 0.9\linewidth]{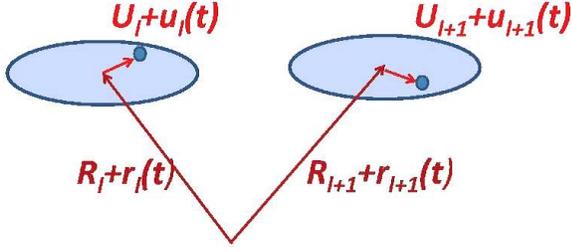}
\caption{(Color online)~
The definition of the position vector of the $\ell$-th cage at time $t$ as $\RR_\ell+\rr_\ell(t)$,
where $\RR_\ell$ is the equilibrium position of the $\ell$-th cage center.
The vector $\rr_\ell(t)$ represents a small deviation from  $\RR_\ell$.
The position of $\mathcal{R}$(2) guest atom from the center of $\ell$-th cage $\RR_\ell$ is defined by the
vector $\UU_\ell+\uu_\ell(t)$, where $\UU_\ell$ is
the equilibrium position of $\mathcal{R}$(2) guest atom from $\RR_\ell$,
and $\uu_\ell(t)$ is a small deviation from  $\UU_\ell$ at time $t$.
}
\label{fig:takaba_fig27}
\end{center}
\end{figure}

The dynamic structure factor $S(\QQ, \omega)$ is proportional to the spatio-temporal Fourier transform
of the density-density correlation function defined by
\begin{equation}
\label{eq:Correltion1}
   G(|\rr-\rr'|, t) = \left\langle \rho(\rr, t)\rho(\rr', 0)\right\rangle,
\end{equation}
where
$\rho(\rr, t)$ is the atomic number density at the time $t$, and the angular brackets denote an  ensemble
average at an equilibrium state. 
The variation of the atomic number density, e.g., induced by vibrations of $\mathcal{R}$(2) guest atom , is defined as
\begin{equation}
\label{Atomic_density}
   \rho(\rr,t)=
   \sum_\ell\delta(\DD_\ell+\dd_\ell(t)-\rr),
\end{equation}
where the definitions $\DD_\ell=\RR_\ell+\UU_\ell$ and $\dd_\ell(t)=\rr_\ell(t)+\uu_\ell(t)$ apply.
The $\QQ$-component of the spatial
Fourier transform of Eq.\,(\ref{Atomic_density}) becomes
\begin{equation}\label{Fourier_density}
   \rho_{\QQ}(t)=\sum_\ell\e^{-i\QQ\cdot\left(\DD_\ell+\dd_\ell(t)\right)}.
\end{equation}
The dynamicstructure factor $S(\QQ, \omega)$ is
proportional to the differential cross section for incident neutrons through the formula,
\begin{equation}
\label{Cross_Section}
   \frac{\mathrm{d}^2\sigma}{\mathrm{d\Omega}\mathrm{d}E}=\frac{k_\mathrm{f}}{k_\mathrm{i}}\frac{\bar{\sigma}}{4\pi}\e^{-\beta\hbar\omega/2}S\left( \QQ,\omega\right) ,
\end{equation}
where $k_\mathrm{i}$ and $k_\mathrm{f}$ are the strength of the initial and final momenta of the neutrons.
The change of momentum is given by
 $\hbar\QQ=\hbar(\kk_\mathrm{f}-\kk_\mathrm{i})$,
and $\hbar\omega=(\hbar^2/2m)(k_\mathrm{f}^2-k_\mathrm{i}^2)$ is the change of energy of the incident neutron.
$\bar{\sigma}$ is an arbitrarily chosen microscopic scattering cross section. 
The dynamic structure factor $S(\QQ, \omega)$ is expressed by the Fourier transform of Eq. (\ref{eq:Correltion1}) as
\begin{align}
\nonumber
   S(\QQ,\omega)&=\frac{1}{2\pi\hbar N}\int_{-\infty}^\infty\mathrm{d}t\e^{-i\omega t}\\
   \label{S_Q_omega}
   &\times\sum_{\ell,\ell'}^{N}\bar{b}_\ell\bar{b}_{\ell'}
   \left\langle \e^{i\QQ\cdot(\DD_\ell+\dd_\ell(t))}\e^{-i\QQ\cdot(\DD_{\ell'}+\dd_{\ell'}(0))}\right\rangle,
\end{align}
where $\bar{b}_\ell$ is the neutron scattering length of atom $\ell$, and $N$ is its total number.
The sum is taken over all atoms $\ell,\ell'~(\ell\neq\ell')$ of the system.  
The expansion of Eq.\,(\ref{Fourier_density}) in terms of the small deviation $\dd_\ell(t)$ yields the first-order density fluctuation,
\begin{equation}
\label{eq:First_order1}
   \Delta\rho_\QQ(t)=\sum_\lambda\Delta
   \rho_\lambda(\QQ, t)+O(d^2),
\end{equation}
where the first term is expressed by 
\begin{align}
\nonumber
\Delta\rho_\lambda(\QQ,t)&=-i\hbar\sum_{\lambda,\ell}\frac{\QQ\cdot\ee_\lambda}{\sqrt{2M_\ell\omega_\lambda}}\\
\label{Delta1}
&
\times\e^{-i\QQ\cdot\DD_\ell}\left( \phi_\lambda\left( \DD_\ell\right)  a_\lambda^\dagger(t)+\mathrm{h.c.}\right).
\end{align}
Here $\mathrm{h.c.}$ means the Hermitian conjugate, $\ee_\lambda$ the polarization vector of the $\lambda$-mode, and
$\phi_\lambda$ the associated eigenfunction, respectively. 
The $a_\lambda^\dagger(a_\lambda)$ is the creation (annihilation)
operator of the phonon mode $\lambda$.
$M_\ell$ means the mass of atoms in the $\ell$-th cell.
The substitution of Eq.\,(\ref{Delta1}) into Eq.\,(\ref{S_Q_omega}) expanded up to the first order in $\dd_\ell(t)$ yields the coherent inelastic term for the dynamic structure factor given by
\begin{align}
\nonumber
   S(\QQ,\omega)&=\frac{n(\beta\omega+1)}{N\bar{\sigma}}\sum_{\lambda}
   4\pi\delta(\omega-\omega_\lambda)\\
   \label{S_Q_omega33}
   &\times \Big\vert\sum_\ell \bar{b}_\ell\frac{\left(\QQ\cdot\ee_\lambda\right) \phi_\lambda\left(\DD_\ell\right)}{\sqrt{2M_\ell\omega_\lambda}}\e^{-i\QQ\cdot\DD_\ell}\Big\vert^2,
\end{align}
where $n(\beta\omega+1)$ is the Bose-Einstein (BE) distribution function with the definition $\beta=1/k_{\rm B}T$.

Equation (\ref{S_Q_omega33}) indicates that $S(\QQ, \omega)$ becomes a flat dispersion relation independent of $\QQ$ for a spatially localized mode $\phi_\lambda$. 
This point is realized in the results of INS experiments on type-I clathrae compounds as given in \ref{subsub_INS_On} and \ref{subsub_INS_Off}.
The phonon densities of states (PDOS) can be obtained from the INS measurements, though caution is needed in the case of multicomponent systems
due to different scattering lengths $\bar{b}_\ell$ for different atomic species.
\subsubsection{Inelastic neutron scattering experiments for on-center systems} 
\label{subsub_INS_On}
INS measurements on polycrystalline powder samples of type-I $\mathcal{R}_{8}$Ga$_{16}$Ge$_{30}$ ($\mathcal{R}$=Ba and Sr) have been performed by 
\textcite{Hermann:2005a} and \textcite{Christensen:2006aa}. 
\textcite{Hermann:2005a} have investigated the phonon density of states (PDOS) in the THz frequency range associated with $\mathcal{R}$(2) guest atoms in both $\mathcal{R}$=Ba and Sr. 
The case $\mathcal{R}$=Ba belongs to the category of quasi-on-center system showing the crystalline Umklapp process of phonon thermal conductivity $\kappa_{\rm ph}(T)$,
while the $\mathcal{R}$=Sr belongs to off-center system yielding the glass-like plateau thermal conductivity at around 10\,[K], as shown in Figs.\,\ref{fig:takaba_fig14} and \ref{fig:takaba_fig16}.
\textcite{Christensen:2006aa} have observed the PDOS of three powder-samples of $\mathcal{R}$=Sr, and $n$- and $p$-type $\mathcal{R}$=Ba.
See Fig.\,\ref{fig:takaba_fig28}.
These results\,\cite{Hermann:2005a,Christensen:2006aa} involve information about both quasi-on-center and off-center systems, so we will discuss in detail these results in the next subsection
\ref{subsub_INS_Off} by focusing on the essential difference between the dynamics of on-center and off-center systems.

\begin{figure}[t]
\begin{center}
\includegraphics[width = 0.9\linewidth]{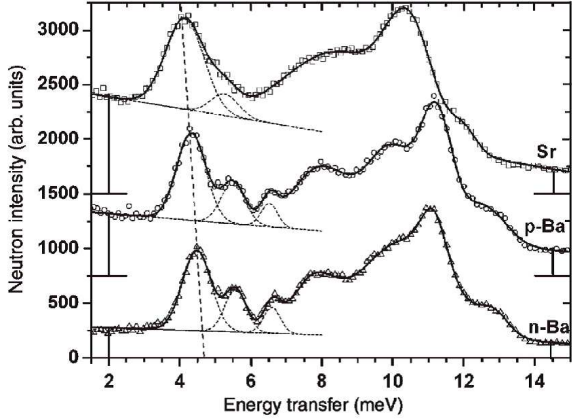}
\caption{Neutron scattering intensity corresponding to phonon densities of states\,(PDOS) 
for type-I $n$-Sr$_{8}$Ga$_{16}$Ge$_{30}$, $p$-Ba$_{8}$Ga$_{16}$Ge$_{30}$ and $n$-Ba$_{8}$Ga$_{16}$Ge$_{30}$.
Gaussian fits have been shown for the first curves.
After \textcite{Christensen:2006aa}.
}
\label{fig:takaba_fig28}
\end{center}
\end{figure}

\textcite{Lee:2007a} have carried out coherent INS experiments on a
single crystal of Ba$_{8}$Ga$_{16}$Ge$_{30}$ belonging to a quasi-on-center system using triple-axis spectrometer.  
They obtained the phonon dispersion relations along the [100] direction in the THz frequency range. 
The results close to the $\Gamma$-point have proved the  optical mode with $T_{\rm u}$ symmetry associated with the
Ba(2) guest atom at $E$=4.5\,[meV].
It is remarkable that the anti-crossing between the optic mode and the acoustic mode in the THz region was observed at around {$\qq$}=(0.45,0,0) in the Brillouin zone, which indicates that the Ba(2) guest atoms play a crucial role for splitting out the acoustic modes attributable to the vibrations of networked cages. 

\textcite{Christensen:2008a} have observed the avoided crossing in Ba$_{8}$Ga$_{16}$Ge$_{30}$ containing quasi-on-center Ba(2) guest atom by using a 13\,[g] single crystal.
The dispersion curve of a (330) reflection along [110] direction exhibits the avoided crossing between longitudinal acoustic phonon and the optic mode attributing to the Ba(2) guest atom at $h$=0.225 of (3,3,0)+[$h$,$h$,0].
See Fig.\,\ref{fig:takaba_fig29}.
They have estimated the phonon life-time $\tau$ from the data of the scattered intensity of incident neutron.
The $\tau$ of the phonons around the avoided crossing area is about 2\,[ps], which is much longer than 0.2\,[ps] obtained from the simple relation $\kappa_{\rm ph}(T)$=$C_{\rm ph}(T)v^2\tau/3$ using known values of the $\kappa_{\rm ph}$, the specific heat $C_{\rm ph}(T)$ of acoustic phonons, and the average phonon velocity $v$.
This discrepancy indicates that this simple relation for $\kappa_{\rm ph}(T)$ is not applicable in the region of  flattened dispersion relations of acoustic phonons.

\begin{figure}[t]
\begin{center}
\includegraphics[width = 0.9\linewidth]{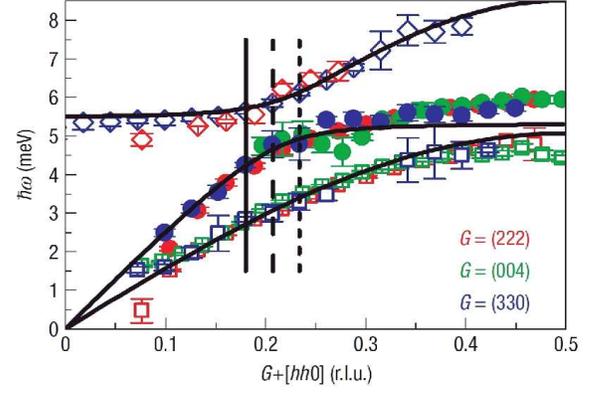}
\caption{(Color online)~
Phonon dispersion curves for quasi-on-center Ba$_{8}$Ga$_{16}$Ge$_{30}$
along [hh0] around the positions G= (222) (red), G= (004) (green) and G= (330) (blue). 
The error bars represent the standard deviation of the fitted energies.
After \textcite{Christensen:2008a}.
}
\label{fig:takaba_fig29}
\end{center}
\end{figure}

\textcite{Christensen:2009a} have carried out INS experiments for powder samples of type-I Ba$_{8}\mathcal{Y}_{x}$Ge$_{46-x}$ ($\mathcal{Y}_x$=Ni$_6$, Cu$_6$, Zn$_8$, Ga$_{16}$) under pressure of 9\,[kbars].
The phonon thermal conductivities of these samples show the crystalline Umklapp peak at around 10\,[K] except $p$-type Ba$_{8}$Ga$_{16}$Ge$_{30}$\,\cite{Bentien:2006a}.
\textcite{Christensen:2009a} have imposed chemical pressure by atomic substitution, and the physical pressure of 9\,[Kbars] was applied using a clamp cell. 
The volume reduction induced by the physical pressure increases the energy of the modes  associated with the guest atom.
A softening of the mode energies was observed upon cooling the sample.
Both $p$-type and $n$-type BGG showed a similar temperature dependence.

\begin{figure}[t]
\begin{center}
\includegraphics[width = 0.8\linewidth]{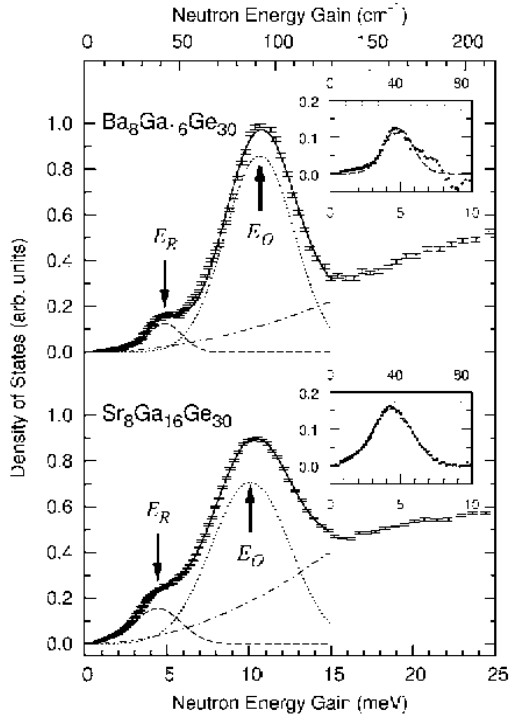}
\caption{The phonon densities of states\,(PDOS) of type-I Ba$_{8}$Ga$_{16}$Ge$_{30}$ and Sr$_{8}$Ga$_{16}$Ge$_{30}$ obtained from INS measurements. 
The dashed line is the contribution of the modes associated with Ba(2) guest atoms and Sr(2), the dotted-dashed line is the parabolic Debye contribution, and the dotted line is a Gaussian fit to the optical phonon density of states below 15\,[meV]. Because both the neutron scattering lengths of Ba and Sr are different, the measured PDOS have been arbitrarily scaled. 
Insets: The Gaussian peaks at $E_{\rm R}$ associated with Ba(2) and Sr(2).
Deviations in the inset for type-I Ba$_{8}$Ga$_{16}$Ge$_{30}$ are related to the Ba(1) mode.
After \textcite{Hermann:2005a}.
}
\label{fig:takaba_fig30}
\end{center}
\end{figure}


\begin{figure}[t]
\begin{center}
\includegraphics[width = 0.8\linewidth]{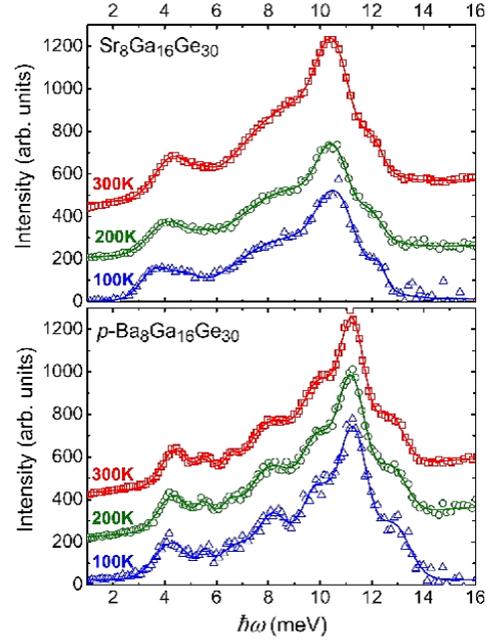}
\caption{(Color online)~Inelastic neutron scattering (INS) data as a function of temperature 300, 200, and 100 \,[K] for type-I
Sr$_{8}$Ga$_{16}$Ge$_{30}$ and $p$-type Ba$_{8}$Ga$_{16}$Ge$_{30}$. 
These are given by the squares, circles, and triangles, respectively. 
After \textcite{Christensen:2009a}.
}
\label{fig:takaba_fig31}
\end{center}
\end{figure}

\textcite{Koza:2010a} have performed INS experiments for Ba$_{8}$Zn$_{x}$Ge$_{46-x-y}\O_y$
with $x$=0, 2, 4, 6, 8 and $y$=3-3$x$/8, where $\O$ denotes a vacancy.
They have studied the modulation of
the eigenfrequency distribution of these compounds by progressively substituting Ge by
Zn in addition to probing its dependence on temperature between 2 and 300\,[K]. 
A number of peaks were shifted toward higher
energies by about 1$-$2\,[meV] as a result of the substitution.
Measurements on temperature-dependence  have demonstrated a hardening of the overall frequency distribution upon cooling. 
They observed a softening upon cooling of the lowest peak energy at 4.5$-$4.8\,[meV]  with a relative shift of 5\% from 300\,[K] too 2\,[K], whose peak is attributable to the Ba(2) guest atom. 

\textcite{Euchner:2012a} have made a high-resolution INS study of the PDOS and the phonon dispersion relation on Ba$_{8}$Ni$_{6-x}$Ge$_{40+x}$.
They have obtained evidence of spectral weight transfer between acoustic and optical phonons due to the strong hybridization.
These data exclude an interpretation in terms
of independent oscillators, e.g., the Einstein model or the soft potential model, since the relevance of hybridized modes between the Ba(2) guest atom and network cages are evident.  
The phonon life times are at least an order of magnitude larger than those estimated from  $\kappa_{\rm ph}(T)$. 

\subsubsection{Inelastic neutron scattering experiments on off-center systems}
\label{subsub_INS_Off}
\textcite{Hermann:2005a}  have  carried
out INS measurements on the PDOS at room temperature on 2.81 and 1.34\,[g] samples of polycrystalline samples of type-I Ba$_{8}$Ga$_{16}$Ge$_{30}$ (BGG) and Sr$_{8}$Ga$_{16}$Ge$_{30}$ (SGG).
Figure \ref{fig:takaba_fig30} shows weighted PDOS for BGG and SGG, which were extracted from the INS data by integrating the scattering over a 2$\theta$ range of 30$^\circ$-130$^\circ$ and
with a subsequent subtraction of the background\,\cite{Hermann:2005a}. 
The observation has confirmed 
that the local mode due to the $\mathcal{R}$(2) guest atom is independent of the integration range.
The local modes of both Ba(2) and
Sr(2) guest atoms have exhibited a linewidth that exceeds the instrumental resolution.
These results have indicated a damping of these local modes due to the interactions with network cages. 
The observed linewidth for Sr(2) local modes is larger than that for the Ba(2) local mode, which indicates a more pronounced interaction of the Sr guests with network cages.

Type-I Eu$_{8}$Ga$_{16}$Ge$_{30}$ ($\beta$-EGG) belongs to the  category of off-center systems showing a plateau in  $\kappa_{\rm ph}(T)$ as in the case of SGG.
However, INS experiments on $\beta$-EGG was not possible because Eu(2) has a very large neutron absorption cross section\,\cite{Hermann:2005a}.
But nuclear inelastic scattering (NIS) measurements are available to obtain the PDOS associated with the Eu(2) guest atom.
This technique utilizes the high brilliance of synchrotron radiation to obtain an element specific PDOS similar to Moessbauer absorption spectroscopy\,\cite{Hermann:2006a}.
The NIS has the advantage, different from INS experiments, 
that the nuclear fluorescence yields an ideal averaging over the entire Brillouin zone.
Because of the resonant nature of the NIS technique, the determination of the PDOS in $\beta$-EGG provides clear evidence that the Eu(2) guest atom neither interacts nor participates in any high energy vibrational
modes.
Furthermore, the microscopic determination of the PDOS in terms of NIS measurements suggests the presence of low-lying local modes associated with Eu(2) guest atoms in $\beta$-EGG\,\cite{Hermann:2005a}.

\begin{figure}[t]
\begin{center}
\epsfysize=4.5in 
\epsfbox{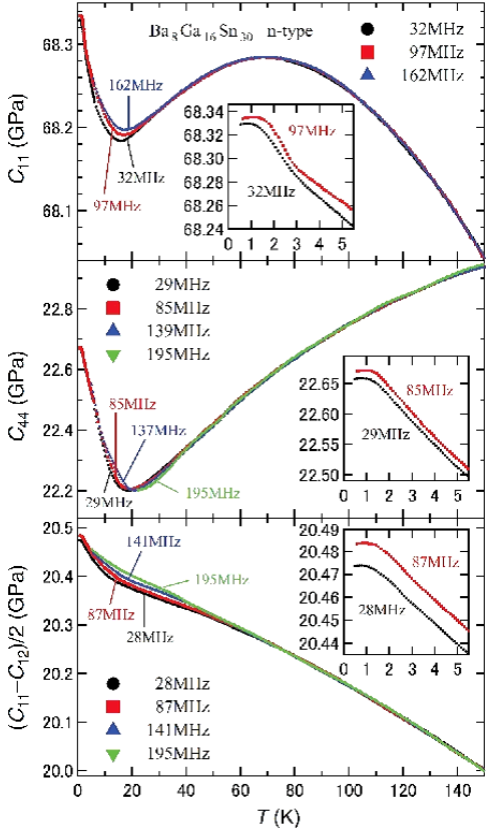}
\caption{(Color online)~
Temperature dependences of elastic stiffness constants $C_{11},C_{44},$ and $(C_{11}-C_{12})/2$
in $n$-type type-I Ba$_{8}$Ga$_{16}$Sn$_{30}$.
The insets represent the data in an expanded scale below 5\,[K].
After \textcite{Ishii:2012a}.
} 
\label{fig:takaba_fig32} 
\end{center}
\end{figure}
\begin{figure}
\epsfysize=4.5in 
\epsfbox{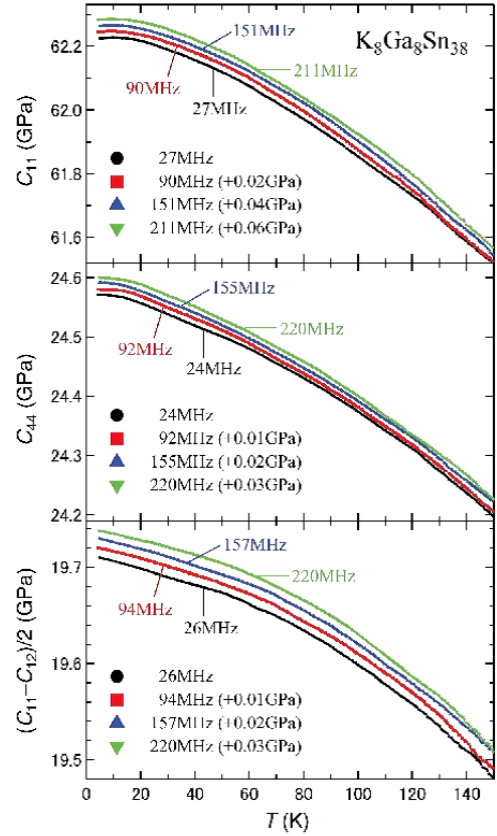}
\caption{\label{fig:takaba_fig33} (Color online)
Temperature dependences of elastic stiffness constants $C_{11},C_{44},$ and $(C_{11}-C_{12})/2$
in type-I K$_{8}$Ga$_{8}$Sn$_{38}$.
After \textcite{Ishii:2012a}.} 
\end{figure}

\textcite{Lee:2008a} have performed coherent INS experiments to investigate phonon  dispersion relations for  SGG
belonging to off-center systems.
They have observed that the optic $T_{\rm u}$ mode of Sr(2) appears around 4.0\,[meV].
Compared with the results of the low-lying modes in
BGG containing on-center Ba(2) guest atoms by \textcite{Lee:2007a}, the avoided-crossing due to the Sr(2) guest atom is less clear\,\cite{Lee:2008a}.

\textcite{Christensen:2009a} have carried out INS experiments on powder samples of SGG and $p$-type BGG at temperatures of 100\,[K], 200\,[K], 300\,[K], which are shown  with squares, circles, and triangles, respectively in Fig.\,\ref{fig:takaba_fig31}.
It should be noted that, with decreasing temperature the {\it width} of the peak around 5\,[meV] increases.
These behaviors coincide with the results of infra-red absorption by \textcite{Mori:2009a, Mori:2011a}.
The physical origin of this anomalous broadening will be discussed in the next section \ref{Dispersion_1}.

Here we should mention interesting works on clathrate hydrates in terms of incoherent INS\,\cite{Tse:2001a,Gutt:2002a,Baumert:2003a} and nuclear resonant IXS\,\cite{Klug:2011a} on the avoided crossing  between low-lying optical modes and acoustic phonons.
This was clearly demonstrated as the strong coupling between the local vibrations of guest molecules and water-framework vibrations.

\textcite{Ishii:2012a} have performed ultrasonic measurements on type-I Ba$_{8}$Ga$_{16}$Sn$_{30}$ ($\beta$-BGS) and K$_{8}$Ga$_{8}$Sn$_{38}$ (KGS) single crystals.
They have investigated the elastic softening of the elastic stiffness constant $C_{44}$ relevant to the strain component $e_{xy}$, 
which show the softening of $C_{44}$ up to 20\,[K] for $\beta$-BGS.
No charge-carrier dependence is observed
between $n$-type and $p$-type $\beta$-BGS.
For KGS, however, such tendencies have not been found, as shown in Figs.\,\ref{fig:takaba_fig32} and \ref{fig:takaba_fig33}\,\,\cite{Ishii:2006a, Ishii:2012a}.
These are the evidence that local vibrations of Ba(2) atom in $\beta$-BGS strongly couple with transverse ultrasonic waves. 
The significant softening on the bulk modulus in $\beta$-BGS contrasts
to the continuous hardening in KGS.

\section{PHONON DISPERSION RELATIONS}
\label{Dispersion_1}
\subsection{Theoretical aspects of THz frequency dynamics of type-I clathrate compounds}
\label{sec:theoretical}
\subsubsection{Molecular dynamics calculations}
Molecular dynamics (MD) calculations have been done for clathrate hydrates with encaged guest atoms or molecules\,\cite{Inoue:1996a, Tse:1997a, Baumert:2003a, English:2009a, English:2011a}.
It has been pointed out\,\cite{Tse:1997a} that the avoided crossing arises from the coupling between local vibrations of guest atoms and acoustic phonons of network cages.
\textcite{Myles:2007a} have performed MD calculations on the dispersion relations of acoustic and optic phonons of type-I Ba$_{8}$Ga$_{16}$Ge$_{30}$ and Ba$_{8}$Ga$_{16}$Si$_{5}$Ge$_{25}$. 
Their calculation is based on first-principle density functional theory (DFT) using a plane-wave basis and the pseudo-potential method. 
Acoustic modes of both
materials lie below 30$-$35\,[cm$^{-1}$]\,(=3.7$-$4.3\,[meV]), and most of optic modes show flat dispersion.
However, the optic modes of type-I 
Ba$_{8}$Ga$_{16}$Ge$_{30}$ in the range 100$-$190\,[cm$^{-1}$]\,(=12.4$-$23.6\,[meV]) and of type-I Ba$_{8}$Ga$_{16}$Si$_{5}$Ge$_{25}$ in the range 110$-$175\,[cm$^{-1}$]\,(=13.6$-$21.7\,[meV])
display significant dispersion.
The DFT calculations were made on the total energy and electric structures for type-I Ba$_{8}$Ga$_{16}$Sn$_{30}$ under hydrodynamic pressure\,\cite{Li:2012a}. 

\textcite{Koza:2010a} carried out the lattice-dynamical calculations in order to interpret the PDOS obtained from INS measurements for Ba$_{8}$Zn$_{6}$Ge$_{40}$ and Ba$_{8}$Ge$_{43}$ belonging to quasi-on-center systems.
In both compounds, eigenmodes relevant to the Ba guest are 
primarily located in the energy regime below 14\,[meV].
The low-energy modes at 3.10\,[meV] are mainly due to Ba(2) cations and 
Ba(1) cations contribute to the modes at 8.14\,[meV].
The partial contributions of the host lattice
constituents Ge and Zn indicate larger amplitudes in
the energy range of Ba(1) than in that of Ba(2) eigenfrequencies.
They have pointed out that
the suppression of the acoustic bandwidth
is accomplished not only by the hybridization of Ba
with Ge dynamics, but also by the lowest energy eigenstates at the Brillouin-zone boundary. 



\textcite{Johnsen:2010a} have carried out DFT calculations for Ba$_{8}\mathcal{T}_{6}$Ge$_{40}$ ($\mathcal{T}$=Cu, Ag, and Au) to explain the atomic dynamics probed by their INS experiments. 
The DFT calculations were in good agreement with the  PDOS data of INS concerning low-energy phonon-modes of $n$-type Cu compounds, but showed discrepancies for $p$-type Ag and Au compounds with glass-like thermal conductivities.

\textcite{Euchner:2012a} have performed first-principle DFT calculations on the phonon dispersion relations of the following three cases: type-I Ba$_{8}$Ge$_{40}$Ni$_{6}$, Ge$_{46}$ framework without guest atoms, and Ge diamond structure.
They have made a rescaling of the energy axis to achieve an agreement with experimental data for Ba$_{8}$Ge$_{40}$Ni$_{6}$.
The rescaling parameter of 1.4 is relatively large, indicating an underestimated Ba-Ge interaction within their DFT approach.
They have calculated the participation ratio (PR) characterizing  the local or extended natures of eigenmodes, in which the value of PR close to unity means extended modes with simultaneous displacements of relevant atoms.
Those of PR close to zero indicate the local modes associated with local vibrations of the Ba(2) guest atom.
The flat branches at 6$-$8\,[meV] in Ba$_{8}$Ge$_{40}$Ni$_{6}$ were identified
as the modes associated with the Ba(2) guest atom. 

The constraints of the limited size in \textit{ab~initio} calculations make it difficult to treat the effect of disorder of the atomic configuration, whereas the system of quasi-on-center Ba-Ge-Ni is possible to create structures with different Ni content.
These are Ba$_{8}$Ge$_{40}$Ni$_{6}$ and Ba$_{8}$Ge$_{42}$Ni$_{4}$, which have different occupation of Ni at the 6$c$ position.
This is the only site that exhibits disorder in Ba-Ge-Ni clathrate structure. 
Since the 6$c$ position is
located in the large 24-atom trapezohedron, the Ni content yields distortions of the cages. 
\textcite{Euchner:2012a} have claimed that the disorder distribution of Ni and Ge yields a broadening of the guest modes.

\textcite{Matsumoto:2009a} have theoretically investigated the effect of a quadratic anharmonic potential to explain the temperature dependence of the optical conductivities\,\cite{Mori:2009a,Mori:2011a}.
They have calculated unequally spaced energy-levels of a one-dimensional single well and of double-well potentials.
The dipole interaction of the guest atoms with electric fields was assumed to 
induce transitions among vibrational states with unequally spaced energies.
They have calculated the natural line broadening and the shift of the peak frequency. 
In the case of a single-well
potential, a softening of the peak frequency and an asymmetric narrowing of the line width with decreasing
temperature were explained as a shift in the spectral weight to lower-level transitions. 
However, the calculated results for one-dimensional double wells showed that the spectral width of the lowest mode decrease  with deceasing temperature.
This is in conflict with the experimental results showing the reverse effect by 
\textcite{Mori:2011a},
This discrepancy indicates that this type of isolated potential for guest atoms is unsatisfactory for describing the experimental results.

\textcite{Safarik:2012a} have theoretically investigated the error arising from the use of the harmonic Debye-Waller factor to explain strong anharmonic vibrations of off-center Eu(2) guest atoms in type-I Eu$_{8}$Ga$_{16}$Ge$_{30}$ 
by assuming a one-dimensional anharmonic potential. 
They have assessed the
error in the values and temperature dependence of the thermal average square displacement for Eu(2) guest atoms.
The harmonic approximation led at most to a $\sim$ 25\% error.

To summarize, MD calculations are powerful for gaining insight on the characteristics of vibrational properties of on-center systems with translational-invariance symmetry.
It is obvious that MD calculations are not efficient for off-center systems without translational symmetry. 
However, the results of MD calculations definitely exclude an interpretation of the Boson-peak like hump of the PDOS at around 0.5\,[THz] observed for off-center systems from adiabatic isolated-oscillator pictures such as the Einstein model
or the soft potential model, since the hybridized modes between $\mathcal{R}$(2) guest atoms and network cages are crucial. 

\subsubsection{Anharmonic potential expressed in terms of the relative coordinate}
\label{Anharmonic_potential_Relative}
In this subsection, 
we describe the THz frequency dynamics by constructing  a model from simple and general points of views.
This type of the model enables us grasp the essential points of physics involved.
The atomic configuration of type-I clathrate compounds with off-center guest atoms has been given by means of diffraction measurements
\,\cite{Nolas:2000a, Bentien:2005a, Christensen:2006a, Christensen:2009a} and extended x-ray absorption fine-structure (EXAFS) studies\,\cite{Baumbach:2005a, Jiang:2008a}.

As described in Sec.\,\ref{sec:type_I_structures},  $\mathcal{R}$(2) guest atoms in these compounds take  on- or off-center position
in 14 hedrons depending on the ratio of cage size and atomic radius of guest atom.
For example, the guest atoms in 14 hedron in type-I Ba$_8$Ga$_{16}$Sn$_{30}$ ($\beta$-BGS)
take the off-center position at $|\UU_\ell|=0.43$\,[{\AA}] from the center
of the tetradecahedral cage\,\cite{Suekuni:2008a}.
Note that the amplitude of $\mathcal{R}$(2) guest atoms are of the order of 0.05\,[{\AA}], which is sufficiently small compared with the $|\UU_\ell|$.
The deviation $|\UU_\ell|=0.43$\,[{\AA}] is 7.4\% of
the nearest-neighbor distance $d=5.84$\,[{\AA}] between Ba$^{2+}$ atoms.
Hence, the potential function for off-center guest atoms is shaped like the sheared  bottom of a wine bottle.
For example, four sites of 24$k$ off-center positions are schematically illustrated in
Fig.\,\ref{fig:takaba_fig34}\,\cite{Avila:2008a}.

\begin{figure}[t]
\begin{center}
\includegraphics[width = 0.6\linewidth]{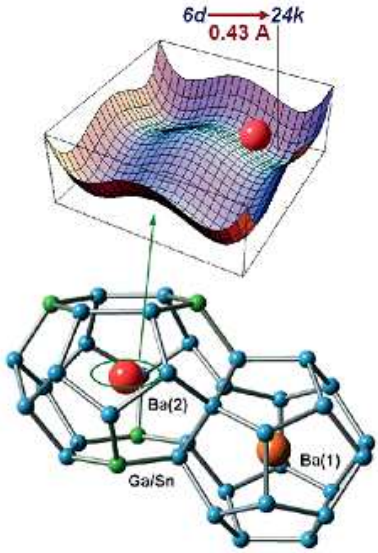}
\caption{(Color online)~
Schematic illustration of the anharmonic potential for the $\mathcal{R}$(2) guest atom in over-sized cages in type-I clathrate compounds.
After \textcite{Avila:2008a}.
}
\label{fig:takaba_fig34}
\end{center}
\end{figure}

The THz frequency dynamics is the most important part for elucidating peculiar thermal and dynamic properties of off-center systems.
This frequency range is mainly concerned with the  acoustic and the lowest optic bands.
For this situation, we focus on the dynamics of cages and $\mathcal{R}$(2) guest atoms.
Cages are treated to possess total mass $M$ with effective charge $-e_C^*$, and guest atoms do mass $m$ with charge $e_G^*$.
The position vector of the $\ell$-th cage at time $t$ was defined in  section \ref{INS_On_center} as $\RR_\ell+\rr_\ell(t)$,
where $\RR_\ell$ is the equilibrium position of the $\ell$-th cage center.
The  vector $\rr_\ell(t)$ represents a small deviation from  $\RR_\ell$.
The position of an $\mathcal{R}$(2) guest atom is defined by the
vector $\UU_\ell+\uu_\ell(t)$, where $\UU_\ell$ is
the position of an $\mathcal{R}$(2) guest atom from $\RR_\ell$, and $\uu_\ell(t)$ is a small deviation from  $\UU_\ell$ at time $t$.
Note that $\UU_\ell\neq 0$ corresponds to the case of an off-center guest atom,
whereas the on-center case becomes $\UU_\ell=0$.
This clarifies the difference
between on-center and off-center systems on the same theoretical basis.
The definition of coordinates is given in Fig.\,\ref{fig:takaba_fig27}}.

\textcite{Nakayama:2011a} have employed the following form of the anharmonic potential, which is  applicable to both the off-center and the on-center systems.
This is expressed in terms of ``relative displacement" $\ww_\ell(t)=\uu_\ell(t)-\rr_\ell(t)$ as
\begin{equation}
\label{eq:anharmonic}
V_{\rm anh}=\sum_\ell\left[\frac{\xi}{2}
\vert\UU_{\ell}+\ww_{\ell}(t)\vert^2
+
\frac{\eta}{4}\vert\UU_{\ell}+\ww_{\ell}(t)\vert^4\right],
\end{equation}
where the parameter $\xi$ takes a positive or negative value, and $\eta>0$.
This type of potential expressed by the relative coordinate $\ww_\ell(t)$ allows us to legitimately treat the THz frequency dynamics of hybridized modes  consisting of guest atoms and network cages beyond the adiabatic approximation.

The potential $V_{\rm anh}$ of Eq.\,(\ref{eq:anharmonic}) involves two types of anharmonic potentials depending on positive or negative $\xi$, namely wine-glass or wine-bottle type potential.
Thus, Eq.\,(\ref{eq:anharmonic}) makes it possible to treat in a unified way
both on-center and off-center systems.

\subsubsection{Spontaneous symmetry breaking of off-center systems}
For the case of off-center systems,
the parameter set
$\xi<0$ and $\eta>0$ in Eq. (\ref{eq:anharmonic}) yields a potential like the bottom of wine-bottle,
in which the minimum becomes
$V_{\rm anh,min}=-\xi^2/(4\eta)$ at $\vert\UU_\ell\vert^2=-\xi/\eta$.
The non-zero $\UU_\ell$ enables to rewrite the potential of 
Eq.\,(\ref{eq:anharmonic}) as
\begin{eqnarray}
\label{eq:wine}
V_{\rm anh}=
\frac{\xi}{2}\sum_{\ell}\vert\UU_{\ell}+\ww_{\ell}(t)\vert^2\left[ 1-\frac{\vert\UU_{\ell}+\ww_{\ell}(t)\vert^2}{2\vert\UU_{\ell}\vert^2}\right] .
\end{eqnarray}

\begin{figure}[t]
\begin{center}
\includegraphics[width = 0.5\linewidth]{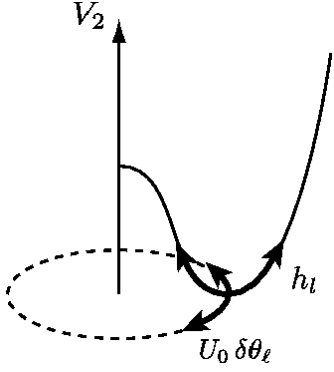}
\caption{Illustration showing the emergence of two degrees of freedom of motion; stretching 
$h_\ell(t)$ and libration $U_0\delta\theta_\ell(t)$.
}
\label{fig:takaba_fig35}
\end{center}
\end{figure}

We describe the vibrations around the equilibrium position of off-center
guest atoms by the following complex number,
\begin{eqnarray}
\label{eq:rotation}
W_\ell=\e^{i(\theta_\ell+\delta\theta_\ell(t)/\sqrt{2})}\left[ U_0+\frac{h_\ell(t)}{\sqrt{2}}\right] ,
\end{eqnarray}
where  $W_\ell$ represents the vector $\UU_\ell+\ww_\ell$.
The equilibrium position of the $\mathcal{R}$(2) guest atom from the center of the cage
is given by $U_0\e^{i\theta_\ell}$, which indicates that
the potential of Eq. (\ref{eq:wine}) preserves the {\it local} gauge symmetry.
The phase factor can be considered as a dynamical variable representing the guest position trapped in one of our hindering potentials caused by local symmetry breaking. 
The variable $h_\ell(t)$ in Eq. (\ref{eq:rotation}) represents a small fluctuation
along the radial direction from the randomly oriented angle $\theta_\ell$.
Figure~\ref{fig:takaba_fig35} depicts two parameters representing these two modes.
It is straightforward to include another degree of freedom $\delta\varphi_\ell(t)$ perpendicular to the plane introduced in Eq.\,(\ref{eq:rotation}).

The reformed potential 
function  with respect to small quantities $h_\ell(t)$ and $\delta\theta_\ell(t)$ is given from Eq.\,(\ref{eq:wine}) for $|\ww_\ell|<|\UU_\ell|$, as
\begin{eqnarray}
\label{eq:wine2}
   V_{\rm anh}=\frac{1}{2}\sum_{\ell}(\tilde{\xi}_s h_\ell^2+\tilde{\xi}_\theta
         U_0^2\delta\theta_\ell^2),
\end{eqnarray}
where the effective force constants $\tilde{\xi}_s$ and $\tilde{\xi}_\theta$ in Eq.\,(\ref{eq:wine2}) are introduced to redefine $\xi$ in Eq.\,(\ref{eq:wine}) in order to express the differences between angular
motion and stretching motion.
In addition, these force constants involve the effect of the random orientation of $\left\lbrace U_\ell \right\rbrace$
and anharmonicity through the following definitions,
\begin{equation}
\label{Effective}
\tilde{\xi}_s=-\xi_s\left(\frac{1+\left\langle  h_\ell^2\right\rangle/U_{0}^2}{2}\right),~~
\tilde{\xi}_\theta=-\xi_\theta
\left(\frac{1+\left\langle\delta\theta_\ell^2\right\rangle}{2}\right),
\end{equation}
where the angular brackets in the parentheses represent a thermal average on the anharmonic term in Eq.\,(\ref{eq:wine}), which should be proportional to $T$ at higher temperatures.

\subsection{Equations of motion}
\label{Equation_motion}
\subsubsection{Effect of charge fluctuation}
The relative displacements $\lbrace\ww_\ell\rbrace$
from the center of the cage induce
electric dipoles for both of the on-center and the off-center systems as in the case of ionic crystals\,\cite{Fano:1960a}.
The total potential energy due to the inter-site dipole-dipole interaction is
expressed as
\begin{eqnarray}
\label{eq:dipoles2}
V_{\rm dip}=\sum_{\ell,m}\frac{e_G^{*2}}{2R_{\ell m}^3}[\ww_\ell\cdot\ww_{m}-3(\ww_{\ell}\cdot\hat{\RR}_{\ell m})(\ww_{m}\cdot\hat{\RR}_{\ell m})],
\end{eqnarray}
where the dipole $\pp_\ell$ is defined by
$\pp_\ell=e_G^*\ww_\ell$, and $\hat{\RR}_{\ell m}$ is
a unit vector in the direction of the vector $\rr_{\ell m}=\rr_\ell-\rr_{m}$.
The Fourier transformed expression of Eq.(\ref{eq:dipoles2}) under the random phase approximation
is obtained as\,\cite{Cohen:1955a}
\begin{eqnarray}
\label{eq:Fourier_Dipoles2}
\nonumber
V_{\rm dip}&=
\frac{(e^*)^2}{4}\sum_{\ell,\kk,\mu}\vert \qq_{\kk\mu}\vert^2\mathrm{e}^{-i\kk\cdot\rr_{0\ell}}
\frac{1-3(\hat{\qq}_{\kk\mu}\cdot\hat{\RR}_{0\ell})^2}{R_{0\ell}^3}\\
&+ \mathrm{c.c.},
\end{eqnarray}
where $\qq_{\kk\mu}$ means the Fourier transform of $\ww_\ell$ relevant to the mode $\mu$,
and $\textrm{c.c.}$ indicates the complex conjugate of the first term.

Equation~(\ref{eq:Fourier_Dipoles2}) can be expressed, with using the Lorentz sums, as
\begin{eqnarray}
\label{eq:Fourier_Dipoles4}
V_{\rm dip}=\frac{m\omega_{\rm p}^2}{2}\sum_{\kk\mu}\vert \qq_{\kk\mu}\vert^2L_\mu,
\end{eqnarray}
where $L_\mu=(\hat{\qq}_{\kk\mu}\cdot\hat{\kk})^2-1/3$
for small $\kk$ for a cubic lattice, 
$\omega_{\rm p}^2=4\pi n_{\rm C}(e_{\rm G}^{*})^2/m$ is the squared plasma frequency of guest atoms with mass $m$, and charge $e_{\rm G}^*$ in cgs esu units and
$n_{\rm C}$=$1/a^3$ is the number density of charges.
We should note that $L_\|$=2/3 for longitudinal modes
and $L_\perp$=$-1/3$ for transverse modes, respectively\,\cite{Cohen:1955a}.

\subsubsection{Fourier-transformed representation of equations of motion}
Each cage is elastically connected with nearest neighbor cages.
We denote the coupling strength by the harmonic force constants $f_{\|}$ and $f_{\perp}$,
which represent longitudinal (dilation) and transverse (shear) modes.
With these quantities, the potential energy is expressed by
\begin{eqnarray}
\label{eq:harmonic}
V_{\rm lat}=\sum_{\ell,\mu}\frac{f_{\mu}}{2}\vert\rr_{\ell,\mu}-\rr_{\ell+1,\mu}\vert^2,
\end{eqnarray}
where $\mu$ denotes the species of the three modes: $\|$, $\perp$, and $\perp'$.

Network cages compose a cubic lattice\,\cite{Nolas:2000a, Bentien:2005a, Baumbach:2005a,Christensen:2006a},
which is invariant under translation by any lattice vector. 
This enables us make the Fourier transformation of the potential function given by
\begin{equation}
\label{eq:harmonic1}
V_{\rm lat}=2\sum_{\kk,\mu}f_{\mu}\vert \QQ_{\kk\mu}\vert^2\sin^2\left( \frac{\kk\cdot\aa}{2}\right),
\end{equation}
where $\aa$ is a lattice vector.
The Fourier transformation of the anharmonic potential expressed by the relative coordinate $\ww_\ell$ is made as
\begin{eqnarray}
\label{eq:Fourier_anharmonic5}
\bar{V}_{\rm anh}=\frac{1}{2}\sum_{\kk,\mu}\bar{\xi}_\mu\vert \qq_{\kk\mu}\vert^2,
\end{eqnarray}
where $\bar{\xi}$ represents the force constants of the both on-center and off-center systems.

The nature of the force constants $\bar{\xi}_\mu$ in Eq.\,(\ref{eq:Fourier_anharmonic5}) 
are quite different for on-center and off-center systems.
The crucial difference is that $\tilde{\xi}_\mu$ for off-center systems involves both the librational ($\tilde{\xi}_\theta$)
and the radial ($\tilde{\xi}_s$) degrees of freedom, while $\hat{\xi}$ for on-center systems {\it is  isotropic}, and independent of the mode $\mu$, whether longitudinal or transverse.
The equations of motion for two variables $\qq_{\kk\mu}, \QQ_{\kk\mu}$ are obtained by replacing $\bar{\xi}_\mu\to\hat{\xi}$ in Eq. (\ref{eq:motion3}).
On-center systems impose the condition
$\xi>0$, $\eta>0$ in Eq. (\ref{eq:anharmonic}).
This yields the equilibrium position of on-center guest atoms $\UU_\ell (t)=0$.
In this case, the force constant $\bar{\xi}$ becomes $\hat{\xi}=
\xi+\eta\langle\vert\ww_{\ell}(t)\vert^2\rangle/2$
by taking account of the thermal average of the anharmonic terms.

The equations of motion for two variables $\qq_{\kk\mu}, \QQ_{\kk\mu}$
with $\mu=\|,\perp$ are obtained from the Euler-Lagrange equation as
\begin{eqnarray}
\label{eq:motion3}
&m\omega_{\kk\mu}^2(\qq_{\kk\mu}+\QQ_{\kk\mu})
=(\bar{\xi}_\mu+m\omega_{\rm p}^2L_\mu)\qq_{\kk\mu},\\
\label{eq:motion33}
&(m+M)\omega_{\kk\mu}^2 \QQ_{\kk\mu}=4f_{\mu}\sin^2\left( \frac{\kk\cdot\aa}{2}\right) \QQ_{\kk\mu}-m\omega_{\kk\mu}^2\qq_{\kk\mu}.
\end{eqnarray}
Here, the above equations of motion are applicable to both cases of on-center or off-center system.
It is straightforward to introduce the effect of random orientation of off-center guest atoms by taking $\bar{\xi}_\mu\to\bar{\xi}$ ($\tilde{\xi}_\theta\leq\bar{\xi}_\mu\leq\tilde{\xi}_s$).

\begin{figure}[t]
\begin{center}
\includegraphics[width = 0.95\linewidth]{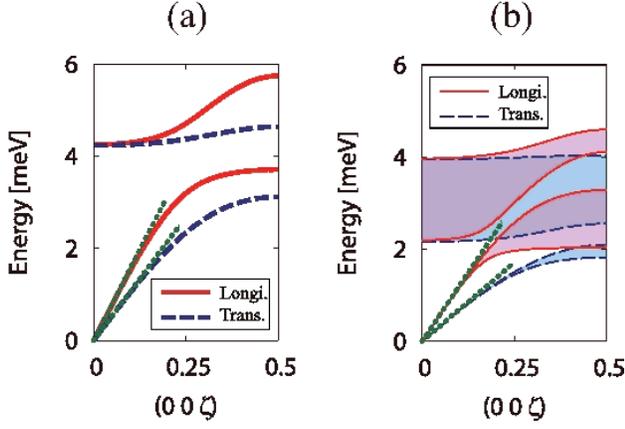}
\caption{
(Color online)~(a) Phonon dispersion curves along the [001] direction for BGG on-center systems.
Dispersion relations for longitudinal modes are plotted with solid lines and for transverse modes with dashed lines.
Dotted linear lines from the origin represent the long-wave limit of the acoustic dispersion relations.
The degeneracy of optic modes at the $\Gamma$-point is observed at 4.2\,[meV] arising from the on-centered symmetric potential.
(b) Phonon dispersion curves along the [001] direction for $\beta$-BGS  off-center systems.
Dispersion relations for longitudinal (transverse) modes are
illustrated by the region bounded by solid (dashed) lines.
Dotted linear lines from the origin again represent the long-wave limit of acoustic dispersion relations.
Note that optic modes observed at the $\Gamma$-point are not separated at 2.2\,[meV] and 4.1\,[meV].
Adopted from \textcite{Nakayama:2011a}.
}
\label{fig:takaba_fig36}
\end{center}
\end{figure}

\subsection{Phonon dispersion relations for off-center and on-center systems in the THz frequency region}
\label{subsubsec:phonon_dispersion}
\subsubsection{On-center System}
\label{subsubsec:on_center}
The force constants $f_\|,f_\perp$ in Eq.\,(\ref{eq:harmonic}) for Ba$_8$Ga$_{16}$Ge$_{30}$  (BGG) for an on-center system are obtained using the sound velocities
$v_\|=4,096$\,[m/sec] and $v_\perp=2,795$\,[m/sec]\,\cite{Christensen:2008a}.
These yield $f_\|=26.4$\,[N/m] and $f_\perp=12.3$\,[N/m] with using $M=7.04m$ and $a=10.78$\,[\AA].
The force constant $\hat{\xi}$ for $n$-type BGG on-center systems is obtained  at 2\,[K] in the same manner as for the data of Raman scattering $\omega_{0}/2\pi=32$\,[cm$^{-1}$] ($=0.96$\,[THz])
\,\cite{Takasu:2006a, Suekuni:2010a} as $\hat{\xi}=8.23$\,[N/m].
Equation (\ref{eq:motion3}) involves the squared plasma frequency $\omega_{\rm p}^2$ arising from the fluctuation of charged guest atoms defined in Eq. (\ref{eq:Fourier_Dipoles4}).
The magnitude of the plasma frequency should be $\omega_{\rm p}/2\pi$\,$\cong 0.09$\,[THz] using the mass ($m$=137\,[u]) and the charge of Ba$^{2+}$ ($e_G^*=2e$)
by taking the number density $n_{\rm C}$=0.628$\times10^{27}$\,[m$^{-3}$] and the relative electric susceptibility $\epsilon_r/\epsilon_0$\,$\cong 10$, the same as that of Si crystal.

\textcite{Nakayama:2011a} have calculated analytically the dispersion relations for BGG  on-center systems from Eqs.~(\ref{eq:motion3}) and (\ref{eq:motion33}) using the force constants mentioned above, as is shown in
Fig.\,\ref{fig:takaba_fig36}(a).
The acoustic phonon dispersions in Fig.\,\ref{fig:takaba_fig36}(a) for BGG
are flattened below $\omega_{0\mu}$ for both transverse and longitudinal modes.
The eigenfrequencies of optic modes at the $\Gamma$-point are degenerate, reflecting that the isotropic force constant $\hat{\xi}$ is independent of the mode $\mu$.

The experimental data on the dispersion relations in terms of coherent inelastic neutron scattering are available for $n$-type BGG on-center systems\,\cite{Lee:2007a, Christensen:2008a}.
The calculated results shown in Fig.\,\ref{fig:takaba_fig36}(a) using the smallest number of parameters of $f_\mu$ and $\hat{\xi}$ recover well
the inelastic neutron scattering  data for BGG\,\cite{Lee:2007a, Christensen:2008a}.

\subsubsection{Off-center system}
\label{subsubsec:off_center}
\begin{figure}[t]
\begin{center}
\includegraphics[width = 0.9\linewidth]{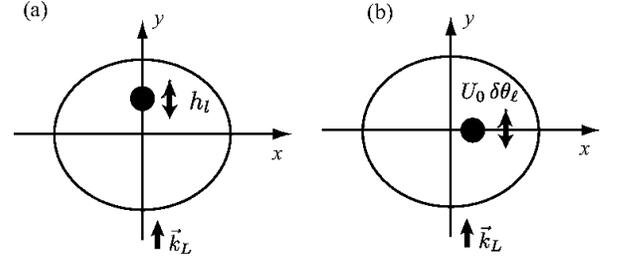}
\caption{Illustration showing the coupling mechanism
between acoustic phonons and $\mathcal{R}$(2) guest atoms. The figure (a) is the case that the off-center guest atom  takes the far-side position for incident longitudinal ($\|$) acoustic phonons with the wave vector $\kk_{\rm ph}$, and (b) shows $\mathcal{R}$(2) guest atom sits perpendicular position to the same longitudinal acoustic phonons.
The coupling becomes effective to parallel components of relevant displacements regardless of the transverse or longitudinal modes.
}
\label{fig:takaba_fig37}
\end{center}
\end{figure}

Equations (\ref{eq:motion3}) and (\ref{eq:motion33}) describe the hybridization between the vibrations of cages and the off-center $\mathcal{R}$(2) guest atoms in addition to the effect of the random orientation of off-center guest atoms on the spectral width of the phonon dispersion relations.
The hybridization occurs as a result of the coupling between {\it parallel} components of displacements regardless of the transverse ($\perp$) or longitudinal ($\|$) modes.
Figures\,\ref{fig:takaba_fig37}(a) and (b) illustrate the physical meaning of this coupling mechanism. 
For example, consider the case in which longitudinal acoustic phonons with polarization vector parallel to the $y$-axis are incident along $y$-axis when the $\mathcal{R}$(2) guest atom takes the far-side position as depicted in Fig.\,\ref{fig:takaba_fig37}(a). 
These longitudinal acoustic phonons dominantly couple with the stretching component $h_\ell(t)$
parallel to the polarization vector. 
Provided that the $\mathcal{R}$(2) guest atom sits on the right side along the $x$-axis as shown in Fig.\,\ref{fig:takaba_fig37}(b), the longitudinal acoustic phonons incident along the $y$-axis couple with the libration component $U_0\delta\theta_\ell$, which is parallel to the polarization vector of the acoustic phonons.
Thus, the coupling constant continuously spans from the librational one $\tilde{\xi}_\theta$ to the stretching one $\tilde{\xi}_s$. 
The same arguments hold for the coupling of transverse acoustic phonons.
These cause the broadening of the spectral width
through the distribution of force constants
$\tilde{\xi}_\theta\leq\tilde{\xi}_\mu\leq\tilde{\xi}_s$
with the definition $\mu=\perp$ or $\|$ \,\cite{Nakayama:2011a}.

The force constants $f_\|,f_\perp$ in Eq.\,(\ref{eq:motion33}) can be
obtained from the sound velocities  $v_\|$=3,369\,[m/sec] and $v_\perp$=1,936\,[m/sec] for type-I Ba$_8$Ga$_{16}$Sn$_{30}$ ($\beta$-BGS) of off-center systems measured by \textcite{Suekuni:2008a} using the relation $v_\mu=a\sqrt{f_\mu/(m+M)}$, which provide $f_\|$=18.1\,[N/m] and $f_\perp$=5.77\,[N/m] by employing masses $m$=136\,[u] for the Ba$^{2+}$ atom and $M$=8.55$m$ for the cage, and the lattice spacing between unit cell $a$=11.68\,[\AA].

The force constants $\tilde{\xi}_\mu$ in Eq. (\ref{eq:wine2}) can be determined from the data of optical spectroscopy since the spectra for optic modes provide the information at $\kk$=0, namely, at the $\Gamma$-point.
Taking $\kk\to 0$ in Eq.\,(\ref{eq:motion33}) and combining it with 
Eq.\,(\ref{eq:motion3}), the squared eigenfrequency for optic modes is given by
\begin{eqnarray}
\label{eq:optic_eigenfrequency}
\omega_{0\mu}^2=\frac{\tilde{\xi}_\mu+m\omega_{\rm p}^2L_\mu}{m}\left( 1+\frac{m}{M}\right),
\end{eqnarray}
where $\mu$ denotes the longitudinal or transverse polarization of optic modes.
The electrostatic interaction $V_{\rm dip}$ for off-center systems becomes the same as
in the case of on-center BGG.

The force constant $\tilde{\xi}_\mu$ in Eq.\,(\ref{eq:optic_eigenfrequency}) for $\beta$-BGS off-center systems is obtained from
the data of Raman scattering\,\cite{Takasu:2006a, Takasu:2010a} and far-infrared spectroscopy\,\cite{Mori:2009a, Mori:2011a}.
\textcite{Mori:2009a, Mori:2011a} have observed infrared active spectra at 7\,[K], with the lowest-lying peak at
$0.71$\,[THz], and with a line-width broadening of 0.57\,[THz]
for $\beta$-BGS off-center systems by means of THz time-domain spectroscopy.
This spectrum should be assigned to the librational mode of off-center rattling guest-ions, which couples with the transverse acoustic mode.

The spectrum of the $A_{\rm 1g}$ stretching mode coupled with the  longitudinal mode is not observed for $\beta$-BGS off-center systems owing to technical reasons \,\cite{Takasu:2006a, Suekuni:2010a}.
For other type of off-center systems, e.g., SGG, the spectra at 2\,[K] are obtained at 48\,[cm$^{-1}$],
and the spectrum at 36\,[cm$^{-1}$] for $\beta$-EGG off-center systems for the $A_{\rm 1g}$ mode \,\cite{Takasu:2006a, Suekuni:2010a}.
By extrapolating these data, the eigenfrequency of the A$_{1g}$ mode
of $\beta$-BGS is estimated as
$\omega_{0}/2\pi=30$\,[cm$^{-1}$] ($=0.9$\,[THz]), from which the force constant $\tilde{\xi}_\|$ associated with longitudinal optic mode can be evaluated.
These assignments lead to $\tilde{\xi}_\|$=7.32(1$\pm$ 0.25)\,[N/m] and $\tilde{\xi}_\perp$=2.21(1$\pm$0.25)\,[N/m],
taking the plasma frequency $\omega_{\rm p}/2\pi$=0.09\,[THz] from Eq.\,(\ref{eq:optic_eigenfrequency}).
The contribution from this plasma frequency is only 10\% at the $\Gamma$-point.

The calculated dispersion relations for $\beta$-BGS off-center systems are given in Fig.\,\ref{fig:takaba_fig36}(b).
It is remarkable, see in Fig.\,\ref{fig:takaba_fig36}(b), that the spectral width becomes broader and gapless in the region of avoided crossing.
This is attributable to the random orientation of guest atoms.

Experimental far-infrared spectroscopy data
at low temperatures\,\cite{Mori:2009a, Mori:2011a} for off-center systems provide interesting line-width broadening features.
The spectral width of about 0.57\,[THz] of the lowest-lying infrared active optic modes 0.71\,[THz] at 7\,[K] decreases with \textit{increasing} temperature.
This feature contradicts the assumption that the anharmonicity of
the potential Eq.\,(\ref{eq:anharmonic}) is a key element for interpreting the origin of the line-width broadening\,\cite{Matsumoto:2009a}.
These experimental results\,\cite{Mori:2009a, Mori:2011a} suggest that the random configuration of  $\UU_\ell$ yields the broadening of optical spectra at low temperatures.

Thermal-averaged squared-displacements, $\langle h_\ell^2\rangle$ and  $\langle\delta\theta_\ell^2\rangle$, are proportional to temperature $T$.
Raman scattering\,\cite{Takasu:2006a, Takasu:2010a, Suekuni:2010a} and far infrared spectroscopy\,\cite{Mori:2009a, Mori:2011a} experiments have shown  that the spectral energies belonging to the lowest band monotonically decrease with decreasing temperature $T$.
These observations can be interpreted by means of thermal-averaged anharmonic terms given by $\langle h_\ell^2\rangle$ and  $\langle\delta\theta_\ell^2\rangle$ given in Eqs. (\ref{eq:wine2}) and (\ref{Effective}), which are proportional to temperature $T$ under a self-consistent phonon approximation. 

Equations (\ref{eq:motion3}) and (\ref{eq:motion33}) yield the frequency range of the avoided crossing given by
\begin{eqnarray}
\label{eq:anti_eigenfrequency}
\delta\omega_{c\mu}\cong\left[\frac{m}{M}\left(\frac{\bar{\xi}_\mu}{m}+\omega_{\rm p}^2L_\mu\right)\right]^{1/2}
\cong\omega_{0\mu}\sqrt{\frac{m}{M}},
\end{eqnarray}
where the last relation gives a small contribution arising from the plasma frequency compared with the first term in the parenthesis.
This relation is valid since the contribution from the plasma frequency is only 10\% of $\omega_{0{\perp}}$ for $\beta$-BGS.
Equation (\ref{eq:anti_eigenfrequency}) indicates that the
frequency $\delta\omega_{c\mu}$ at avoided crossing is governed
by the quantities $\omega_{0\mu}$ and the square root of mass ratio $\sqrt{m/M}$.
The frequency  $\delta\omega_{c\mu}$ for off-center systems is much smaller than the case of on-center systems due to the inequality $\bar{\xi}_\mu<\hat{\xi}$.


\begin{figure}[t]
\begin{center}
\includegraphics[width = 0.8\linewidth]
{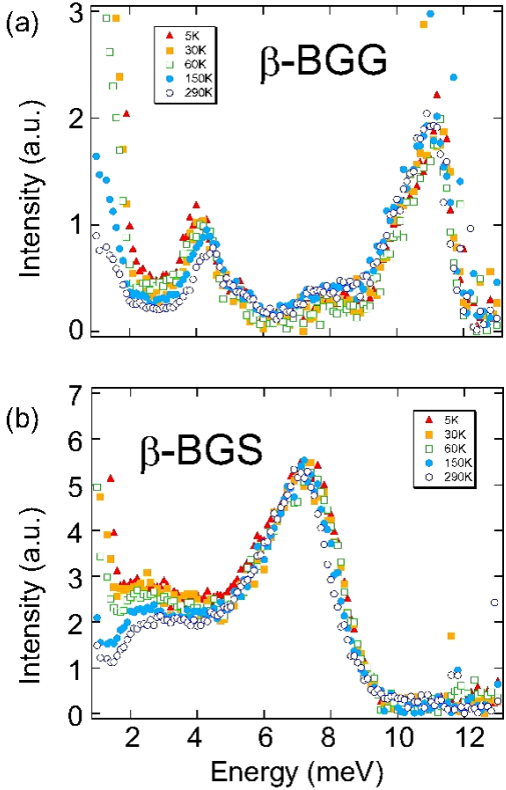}
\caption{(Color online)~
Phonon densities of states for both of type-I Ba$_8$Ga$_{16}$Ge$_{30}$ (BGG) (a) and Ba$_8$Ga$_{16}$Sn$_{30}$  ($\beta$-BGS) (b) 
obtained by means of coherent INS experiments in the temperature range from 5\,[K] to 290\,[K]. 
The data provide integrated densities of states from 0.4 to 0.7\,[$\AA^{-1}$] in $\kk$-space.
See text for the assignments of the spectra.
After \textcite{Nakamura:2010a}.
}
\label{fig:takaba_fig38}
\end{center}
\end{figure}


\subsubsection{Comparison of calculated dispersion relations to INS experiments}
\label{sec:comparison}
Coherent INS measurements for both BGG and $\beta$-BGS were performed in the temperature range from 5\,[K] to 290\,[K] using the cold neutron disk-chopper spectrometer  AMATERAS installed in Materials and Life Science Experimental Facility (MLF), Japan Proton Accelerator Research Complex (J-PARC)~\cite{Nakamura:2010a}.
The results integrated in the range from 0.4\,[$\rm{\AA}^{-1}$] to 0.7\,[$\rm{\AA}^{-1}$] are given in Fig.\,\ref{fig:takaba_fig38}\,(a) for BGG and Fig.\,\ref{fig:takaba_fig38}\,(b) for $\beta$-BGS.
These data provide important information on the THz frequency dynamics of clathrate compounds. 

First, for the data of quasi-on-center BGG, Fig.\,\ref{fig:takaba_fig38}(a) shows three sharp peaks at 4.0, 7.5 and 11\,[meV].
The assignments of these spectra can be made in comparison with optical spectroscopies described in Sec.\,\ref{optic_spectroscopy}.
The lower peaks at 4.0\,[meV] and 7.5\,[meV] have been observed in both Raman scattering\,\cite{Takasu:2006a}
and infra-red spectroscopies\,\cite{Mori:2009a}.
These have been assigned as $T_{\rm 2g}$ and $E_{\rm g}$ modes attributable to Ba(2) guest atoms encapsulated in 14 hedrons. 
The higher spectrum at 11\,[meV] corresponds to Ba(1) guest atoms in 12 hedrons.
The lower peak at around 4.0\,[meV] varies considerably with $T$ from 5\,[K] to 290\,[K], while two higher peaks do not show such a strong temperature dependence. 
The strong dependence on temperature manifests the relevance of the ``wine-glass" type anharmonic potential employed in Eq.\,(\ref{eq:anharmonic}) for Ba(2) guest atoms in BGG. 
Calculated results for dispersion relations for BGG are given in Fig.\,\ref{fig:takaba_fig36}\,(a).

Figure \ref{fig:takaba_fig38}\,(b) on off-center $\beta$-BGS shows two peaks at 2$-$4\,[meV] and 7\,[meV].
The lower hump arises from the vibrations of Ba(2) guest atoms in the 14 hedron, and the peak at 7\,[meV] is attributable to Ba(1) guest atoms in the 12 hedrons.
There is no clear gap between 2 and 4\,[meV], which is distinctly different from the result for BGG given in Fig.\,\ref{fig:takaba_fig38}\,(a).
The data in Fig.\,\ref{fig:takaba_fig38}\,(b) indicate that the lower-energy side of the hump at around 2\,[meV] is appreciably affected by temperature compared with the higher-energy side at around 4\,[meV]. 

The calculations given in Fig.\,\ref{fig:takaba_fig36}\,(b) have clarified that the lower side is concerned with the hybridization between libration vibrations of guest atoms and the transverse acoustic modes from networked cages. 
At the same time, the higher-energy side of the hump results from the hybridization between the stretching vibrations of guest atoms and the longitudinal acoustic phonons.
Thus, the INS data indicate that the libration motion of Ba(2) guest atoms in $\beta$-BGS experience much stronger anharmonicity compared with that of the stretching motion.
This is consistent with the view that the thermal-averaged anharmonic term $\langle\delta\theta_\ell^2\rangle$ in Eq.\,(\ref{Effective}) is large enough compared with the stretching term $\langle h_\ell^2\rangle$.
This is a reasonable interpretation for the observations on INS experiments by \textcite{Nakamura:2010a} for BGG and $\beta$-BGS.

To conclude this subsection, we emphasize that the observed densities of states given in Fig.\,\ref{fig:takaba_fig38} recover the calculations on BGG and $\beta$-BGS shown in Fig.\,\ref{fig:takaba_fig36}, very well.
In particular, the gapless density of states given in  Fig.\,\ref{fig:takaba_fig38}\,(b) for $\beta$-BGS is well reproduced in Fig.\,\ref{fig:takaba_fig36}(b).

\subsubsection{The origin of Boson-peak like excess density of states}
Since the specific heats of type-I clathrate compounds containing off-center guest atoms, for example, $\beta$-BGS or SGG, exhibit a hump at around $T\simeq$\,4\,[K], the corresponding modes have been observed by Raman scattering\,\cite{Takasu:2006a, Takasu:2010a, Suekuni:2010a, Kume:2010a} and infra-red measurements\,\cite{Mori:2009a, Mori:2011a}.

The relation between the phonon specific heat $C_{\rm ph}(T)$ and the phonon density of states $D(\omega)$ is expressed by the formula,\begin{eqnarray}\label{eq:Specific_1a}   
C_{\rm ph}(T)=\frac{-1}{k_{\rm B}T^2}\sum_\mu \left[\int_0^\infty \frac{\partial n_\mathrm{B}(\beta\hbar\omega)}{\partial\beta}\hbar\omega_\mu D\left( \omega_\mu\right) d\omega_\mu\right],
\end{eqnarray}
where $n_{\rm B}$ is the Bose-Einstein  distribution function and the inverse temperature is defined  by $\beta = 1/k_{\rm B}T$. 
The density of states $D(\omega)$ can be obtained in principle by inverting Eq.(\ref{eq:Specific_1a}) using the observed data of $C_{\rm ph}(T)$, though this procedure involves an uncertainty.
However, specific heat measurements are not affected by the mode-selection rule on excited modes as in the case of optical spectroscopies, in which active
modes are distinguished by infrared absorption, Raman scattering, and hyper-Raman scattering experiments. 
Optical spectroscopies provide information close to the $\Gamma$-point.
These are different from humps 
observed in specific heats that represent averaging over a wide sector of $\kk$-space.

The characteristic frequency of guest atoms in cages is much smaller than the Debye frequency $\omega_{\rm D}$ expressing the upper limit of propagating acoustic phonons of the network cages without guest atoms.
The coupling of guest atoms and cages yields the hybridization of dispersion curves and creates the flat and broad band.
This is the origin of the Boson-peak like mode observed in type-I clathrae compounds containing off-center guest atoms.
Thus, the flat band with broad spectra at around 0.5\,[THz] shown in Fig.\,\ref{fig:takaba_fig36}(b) is the origin of the Boson-peak like excess density of states.

\section{GLASS-LIKE SPECIFIC HEATS BELOW 1\,[K]}
\label{sec:thermal_theor}
\subsection{Low temperature specific heats and two-level tunneling states}
\label{sec:twolevels}

\subsubsection{Two-level tunneling model}
\label{two-level model}
\begin{figure}[t]
\begin{center}
\includegraphics[width = 0.6\linewidth]{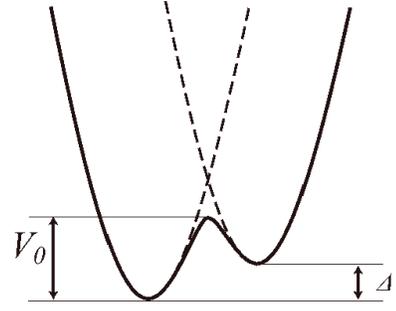}
\caption{Schematic illustration of double well potential representing the two-level tunneling state.
}
\label{fig:takaba_fig39}
\end{center}
\end{figure}

In off-center type-I clathrate compounds, the disorder is introduced by randomly oriented guest atoms.
This is the origin of glass-like specific heats observed for off-center systems at low temperatures.
At $T$\,$\lesssim$\,1\,[K], a quantum
mechanical description is needed to describe the states contributing to $T$-linear specific heats.
For structural glasses, it has been postulated that an atom or a group of atoms can occupy one of two potential minima\,\cite{Anderson:1972a, Phillips:1972a}. 
See the reviews by
\textcite{Hunklinger:1986r} and \textcite{Phillips:1987r}.
Each tunneling state can be simply represented by assuming  a double-well potential shown in Fig.\,\ref{fig:takaba_fig39}, 
where the abscissa gives the position of the tunneling element in multi-dimensional configuration space. 
This tunneling model provides a good phenomenological basis on which various observations can be consistently explained.
It should be emphasized that the idea of double-well potential is based on the adiabatic approximation.
This is acceptable since the wavelengths of excited acoustic phonons below 1\,[K] are sufficiently larger than the scale of tunneling elements.
The THz frequency dynamics attributable to the hybridization between local vibrations of guest atoms and acoustic phonons due to network cages should take into account the non-adiabatic aspect, as been described in Sec.\,\ref{Dispersion_1}.

We choose a basis set of ($\varphi_1, \varphi_2$)  of the appropriate potential $V_1$ and $V_2$ belonging to ground energies ($E_1, E_2$). 
By taking the zero of energy as the mean of $E_1$ and $E_2$, the Hamiltonian matrix becomes
\begin{equation}
 \label{eq:TLS_matrix2}
   \mathcal{H}=\frac{1}{2}
   \begin{pmatrix}
   -\epsilon &\Delta\\ \Delta&\epsilon
    \end{pmatrix}
    ,
\end{equation}
where the tunnel splitting $\Delta$ due to the overlap of the wavefunctions is given by
\begin{equation}
 \label{TLS_Splitting0}
   \Delta=2\left\langle \varphi_1|H|\varphi_2\right\rangle. 
\end{equation}

The overlap integral $\Delta$ generally involves an exponential decay given by
\begin{equation}
 \label{TLS_Splitting1}
   \Delta=\hbar\Omega\mathrm{e}^{-\lambda}=
   \hbar\Omega\mathrm{e}^{-\ell\sqrt{
    2mV_0}/\hbar},
\end{equation}
where $\hbar\Omega$ is approximately equal to $( E_1 + E_2)/2$, while $\ell$ and $V_0$ are the separation and the barrier height between the two wells, and $m$ is the mass of the tunneling element.
Typical values of the tunneling parameter $\lambda$ can be estimated from the requirement
that $\Delta$ must be approximately equal to $k_{\rm B}T$ if the tunneling states dominantly contribute to
thermal properties at a temperature $T$. 
This requires, for example,  $\Delta$\,$\cong$1\,[K] at 1\,[K], which gives approximately 5 for $\lambda$ with $\hbar\Omega$ 
equal to 100\,[K]. 

For off-center guest atoms, the overlapping integral  $\Delta$ is due to the angular rotation of off-center  guest atoms given by 
\begin{equation}
\label{Delta_0}
   \Delta=\hbar\Omega\, 
   \e^{-\sqrt{2IV_0}\delta\theta/\hbar},
\end{equation}
where the moment of inertia $I=25.4$\,[u\AA$^2$]  for $\beta$-BGS.
The zero-point energy scale $\hbar\Omega$ of the guest atom trapped in one of  four local-potential minima is estimated, from the uncertainty principle, to be $\hbar\Omega\approx 4.6$\,[K] for an actual spatial size $\approx(0.3\,[\mathrm{\AA}])^3$ for $\beta$-BGS.
This energy scale $\hbar\Omega\approx 4.6$\,[K](=4\,[meV]) is comparable with that of the Boson-peak like hump in $\beta$-BGS observed from specific heats and spectroscopic measurements.

The Hamiltonian matrix of Eq. (\ref{eq:TLS_matrix2}) provides  eigenvalues
\begin{equation}
 \label{eq:Diag_1}
   E=\pm\frac{1}{2}\left(\epsilon^2+\Delta^2 \right)^{1/2}.
\end{equation}
The distribution function of $\epsilon$ must be
an even function because both the positive and negative values of $\epsilon$ are equally likely. 
The range of energy variation of $\epsilon$ is determined by the thermal energy where atoms or molecules become free from the constraint of neighboring atoms.
In structural glasses, this corresponds to the
glass-transition temperature $T_\mathrm{g}$.
The glass-transition temperature obeys the Tammann's rule of the form
$T_\mathrm{g}$\,$\cong 2T_{\mathrm{melt}}/3$\,[K].
Thus, the distribution function $f \left(\epsilon,\Delta\right)$ is a reasonably slowly varying function on $\epsilon$ in the range of interest 10\,mK$<\epsilon/k_\mathrm{B}<1$\,[K], so that $f(\epsilon, \Delta)$ can be taken as independent of $\epsilon$.
Because of the exponential dependence of $\Delta$ on $\lambda$, 
only a relatively small range of $\lambda$ is sampled for a large range of $\Delta$.

We should remark here that $\epsilon$
expresses the asymmetry energy of \textit{two-level} systems, i.e., the thermal activation energy needed for the rearrangement of local microscopic structures in the regime $\epsilon/k_\mathrm{B}\gg$ 1\,[K].
Thus, $\epsilon$ does not manifest the energies of the excess density of states related to the Boson peak
observed at 3$-$10\,[K] for structural glasses.

\subsubsection{Specific heats below 1\,[K]}
By introducing the density of states $n(E)$ per volume per energy, the specific heat at low temperatures is given by
\begin{eqnarray}
 \label{C}
   C_{\rm tun}(T) = \frac{1}{4k_{\rm B}T^2}\int_{0}^{\Delta_0/2}
    n(E)E^2\sech^2\(\frac{\beta E}{2}\)dE,
\end{eqnarray} 
where $\beta=(k_\mathrm{B}T)^{-1}$.
The form of the distribution function $n(E)$ is not known $\textit{a~priori}$, but $n(E)$ should be a continuous function in the temperature regime $T\ll\Delta_0/k_\mathrm{B}$ where $\Delta_0$ is the upper bound of $E$. 

If we employ a simpler view that the states are distributed uniformly from -$\Delta_0/2$ to $\Delta_0/2$ such as $n(E)=\bar{N}/\Delta_0$, where $\bar{N}$ is the number density of tunneling sates per volume, Eq.\,(\ref{C}) yields at $T\ll\Delta_0/k_{\mathrm{B}}$,
\begin{eqnarray} 
 \label{C_1} 
   C_{\rm tun}\cong\frac{\pi^2\bar{N}}{3\Delta_0}k_{\rm B}^2T.
\end{eqnarray}
Above 1\,[K], the part of the asymmetry energy $\epsilon$ in Eq.\,(\ref{eq:Diag_1}) is a dominant term
in Eq.\,(\ref{C_1}).
In this case, using observed values of specific heats of silica glass
of $C_{\rm tun}$\,$\cong$\,2.5$\times$10$^{-4}\,[\mathrm{mJ\, cm^{-3}K^{-1}}]$ at $T$\,=0.1\,\,[K] and the upper bound of  $\Delta_0$\,$\cong$\,$T_{\mathrm{g}}$\,$\cong$\,500\,[K] in Eq.\,(\ref{C_1}), the number density of  tunneling states is estimated as $\bar{N}$\,$\cong 10^{20}$\,$[\mathrm{cm}^{-3}]$.

We can estimate the number density of tunneling states
$\bar{N}$ by taking the upper bound as  $\Delta_0$\,$\cong$\,500\,[K] in Eq.\,(\ref{C_1}) as $\bar{N}$\,$\cong$\,$1.1\times 10^{17}$\,$[\mathrm{cm}^{-3}]$.
Therefore, in silica glass only 10$^{-5}$ states per SiO$_2$ element contribute to tunneling states.

The structural glasses are in non-equilibrium states,
so that observed specific heats should vary  logarithmically with the measuring time,
implying the relaxation in a multi-valley potential in configuration space\,\cite{Anderson:1972a, Phillips:1972a}.
Here, we do not enter into the details of this interesting phenomenon. The readers can find the expression of specific heats involving relaxation time $\tau$ in articles by \textcite{Hunklinger:1986r} and \textcite{Phillips:1987r}.

The two-level tunneling model provides a good phenomenological picture of observed universal phenomena.
However, it is difficult to identify the tunneling entity in glasses due to its structural complexity.
Compared with structural glasses, type-I clathrate compounds containing off-center guest atoms are microscopically  well defined, and it is possible to understand the tunneling entity on an atomic scale.
This is the major goal of the next sections.

\subsubsection{Failure of the noninteracting picture for tunneling elements}
\label{subsec:failure}
One of the important features of type-I clathrate compounds found in experiments
\,\cite{Nolas:1995a,Nolas:1996a,Nolas:1998a,Nolas:1998aa,
Nolas:1998aaa,Nolas:1998aa,Nolas:2000a,Nolas:2001r,
Cohn:1999a,Sales:1996a,Sales:1998a,Sales:2001a,
Bentien:2004a,Bentien:2005a, Avila:2006a,Avila:2006aa,Suekuni:2007a,
Suekuni:2008a,Suekuni:2008aa,Suekuni:2010a, Xu:2010a}
is that guest atoms take either the on-center or off-center position depending on the size of cages or, equivalently, the ionic radii of guest atoms.

\begin{figure}[t]
\begin{center}
\includegraphics[width = 0.9\linewidth]{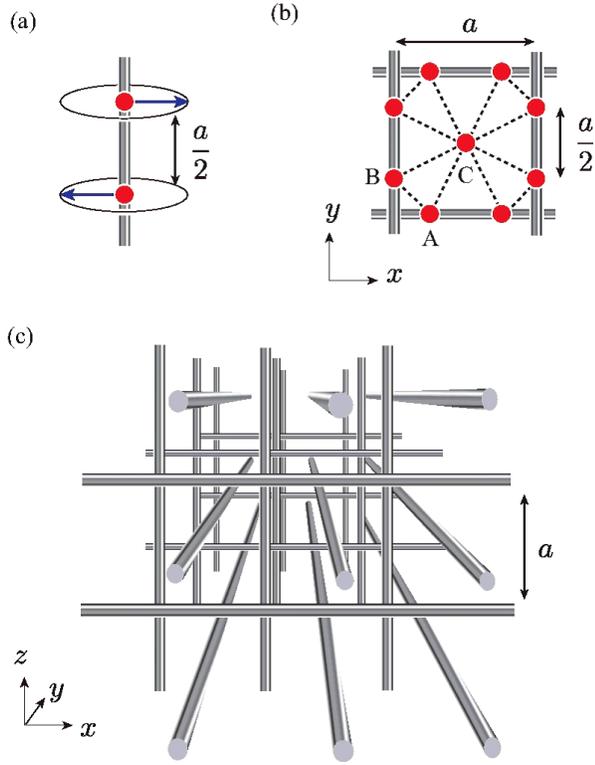}
\caption{(Color online) Schematic illustration of the configuration of the guest atoms in the over-sized cages in $\beta$-BGS.
The four-fold inversion axes are directed along $x$, $y$, $z$.
(a) The deviation from the on-center position (filled circles) induces the electric dipoles (arrows).  Here two nearest dipoles are depicted.  The electric dipoles rotate in the plane perpendicular to the axis linking the nearest dipoles.
(b) The configuration of the on-center positions of the guest atoms.  The filled circles represent the positions of off-center guest atoms, around which the electric dipoles are induced.  The sites A, B, and C in (a) are seated on the chains parallel to $x$, $y$, and $z$, respectively: $A=(a/4,0,a/4)$, $B=(0,a/4,3a/4)$, $C=(a/2,a/2,a/2)$.  The distance between the next-nearest neighbors (dashed lines) is $\sqrt{3/8}\,a$.
Note that these three dipoles constitute an \emph{equilateral triangle} and easily generate a frustrated situation.
(c) The 3D configuration of the dipole chains are illustrated.
} \label{fig:takaba_fig40}
\end{center}
\end{figure}

A typical deviation $r_0$=0.43\,[{\AA}] in $\beta$-BGS has been obtained from diffraction experiments\,\cite{Avila:2006aa,Avila:2008a}.
$p$-type BGG is slightly off-center by 0.15\,\AA\,\cite{Christensen:2006a, Jiang:2008a}. 
In this connection, it should be noted that $n$-type BGG shows crystal-like thermal conductivity, while $p$-type behaves with glass-like thermal properties\,\cite{Avila:2006a, Avila:2006a, Avila:2008a}.
This is one piece of evidence that the long-range dipole-interaction plays a crucial role in type-I clathrates, since the Coulomb interaction between the cages and the guest cation Ba$^{2+}$ in BGG is shielded in $n$-type electron-rich BGG, while the dipole interaction in $p$-type becomes relevant even for small dipole moments.
The network configuration consisting of off-center guest atoms is schematically illustrated in Fig.\,\ref{fig:takaba_fig40}.

At first glance, individual off-center guest atoms
seem to independently contribute to glass-like thermal properties at low temperatures.
This naive view misleads in that it suggests an isolated \textit{non-interacting} picture could explain the observed glass-like behaviors.
Such a \textit{non-interacting} picture 
does \textit{not} reproduce the observed temperature dependence as well as the magnitudes of the specific heats for type-I clathrate compounds with off-center guest atoms  at 1\,[K] and below\,\cite{Nakayama:2008a}.
In fact, as will be discussed in the next section, the tunneling between the nearby potential minima in configuration space generated by 
a combination of interacting dipoles is crucial for interpreting  glass-like thermal properties at low temperatures.

In an isolated non-interacting picture, the problem is reduced to the state of a single cage where $\mathcal{R}$(2) guest atoms experience a \textit{hindering} potential $V_{\rm h}(\theta)$ along the azimuthal direction\,\cite{Nolas:2001r,Sales:2001a}.
The barrier height of the hindering potential 
$V_{\rm h}$ between nearby potential wells is estimated to be of the order of 10\,[K].
Actually, first principles calculations have shown $V_{\rm h}$ to be $\cong$20\,[K] for Sr$^{2+}$(2) guest atoms in SGG\,\cite{Madsen:2005a}.

Since the separation of neighboring wells is small, for example, $\pi U_0/2$=0.67\,[{\AA}] in $\beta$-BGS,
$\mathcal{R}$(2) guest atom to tunnel to a nearby potential minimum at lower temperatures $T\ll V_{\rm h}$, where the off-center guest atom executes zero-point motion at one of the four wells with an energy $\hbar \Omega$.

The tunneling splitting $\Delta$ due to the angular rotation of off-center  guest atoms is given by 
\begin{eqnarray}
\label{Delta_1}
\Delta=\hbar\Omega\,\e^{-\sqrt{2IV_{\rm h}}\delta\theta/\hbar},
\end{eqnarray}
where $\delta\theta\leq\pi/2$ is the angle between two nearby minima.
Since the zero-point energy $\hbar\Omega$ of the guest atom is estimated to be $\hbar\Omega\approx 4.6$ \,[K] for $\beta$-BGS, we can estimate the most probable lower-bound as $\Delta_{\rm min}=0.03$ \,[K] by using  the hindering potential height $V_{\rm h}(\theta)\approx 10$\,[K] and the  moment of the inertia of dipoles $I=25.4$\,[u\AA$^2$] for $\beta$-BGS into 
Eq.\,(\ref{Delta_1}).

The lower bound of the integral in Eq.\,(\ref{C}) should be $\Delta_{\rm min}/2$.
Introducing the dimensionless variable $x$ defined by $x=E/(2k_\mathrm{B}T)$,
the function $x^2\sech^2(x)$ in the integrand of Eq.\,(\ref{C}) has a maximum at around $x=1$.
The contribution to the integral of the product of $x^2\sech^2(x)$ 
and $n(2xk_{\rm B}T)$ should become sharply smaller at temperatures below 
$T$\,$\lesssim\Delta_{\rm min}/k_{\rm B}\approx 0.03$\,[K],
but the experimental data of specific heats for type-I clathrate compounds  do not show this tendency.

Furthermore, the isolated \textit{non-interacting} picture is based on  the idea that \textit{every} off-center guest atom contributes to the specific heat.
Taking the distribution function $n(E)=\bar{N}/\Delta_0$ with $\Delta_0\approx V_{\mathrm{h}}\approx 20$\,[K] and $\bar{N}$
=6, where 6 is the number of guest atoms in a unit cell of $\beta$-BGS, Eq.\,(\ref{C_1}) becomes 
\begin{equation}
   C_{\rm tun}\cong 1.0\times10^4\,T \,\,\,[\mathrm{mJ\,mol^{-1}\,K^{-1}}],
\end{equation}
which is two orders of magnitude larger than the observed value for $\beta$-BGS $C_{\rm tun}$\,$\cong 30\,T$\,[mJ\,mol$^{-1}$\,K$^{-1}$]\,\cite{Suekuni:2008a, Suekuni:2008aa}.
Thus, the \emph{non-interacting} picture based on the assumption that every off-center guest-ion would independently contribute to the tunneling states yields the specific heat $C_\mathrm{tun}(T)$ of two orders of magnitude \emph{larger} than the observed values at 1\,[K], in conflict with observations.
It is therefore not reasonable to employ the non-interacting picture to explain glass-like properties
of type-I clathrate compounds observed at $T$\,$\lesssim 1$\,[K]\,\cite{Nakayama:2008a}.

\subsection{Interacting-dipoles in type-I clathrate compounds}
\label{subsec:interacting}
\subsubsection{Multi-valley potentials in configuration space}

\begin{figure}[t]
\begin{center}
\includegraphics[width = 0.9\linewidth]{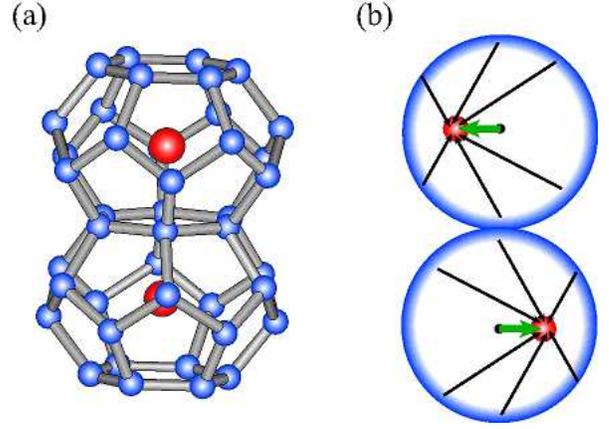}
\caption{(Color online) (a) Illustration of two 14-hedrons cages along $z$-direction consisting of anions and guest cations.
(b) The cage (outer circle) and the symmetry-broken off-center guest atom compose an effective electric dipole moment (thick arrow), which is the vector sum of each dipole (thin arrow).
} \label{fig:takaba_fig41}
\end{center}
\end{figure}

In type-I clathrate compounds with divalent guest ions, the deviation of the guest ion from the center of the cage  
induces an \textit{electric dipole moment} due to the difference of the charges between the guest atom with the charge +2e and the ion with the charge -e constituting cages.
The strength of the electric dipole moment can be estimated to be  $p$=4.1 Debye in the case of $\beta$-BGS with the deviation $U_0$=0.43\,[{\AA}] from the center.
The deviation is about 7.2\% of the distance between the neighboring 14-hedrons ($a/2$=5.84\,[{\AA}], where $a$ is the lattice constant).
Thus, the cages with off-center guest atoms intrinsically possess electric dipoles, and that it is crucial to take into account these characteristics.
See Fig.\,\ref{fig:takaba_fig41}.

\textcite{Bentien:2005a} have pointed out, from crystallographic studies\,\cite{Bentien:2000a, Bentien:2002a} and thermal conductivity measurements\,\cite{Bentien:2004a}, that the Ba(2) guest can be off-center in $p$-type Ba$_{8}$Ga$_{16}$Ge$_{30}$.
\textcite{Avila:2006aa} have found that $p$-type BGG shows crystal-like thermal conductivity, while $n$-type behaves as glass-like.
This indicates that the long-range interaction between divalent guest atoms becomes relevant owing to anti-shielding by doped holes in $p$-type, while $n$-type BGG
takes the opposite tendency. 
$p$-type BGG are slightly off-center by 0.15\,[\AA]\,\cite{Christensen:2006a, Jiang:2008a} with much smaller dipole moments compared with the case of $\beta$-BGS. 
This is one piece of evidence that the long-range dipole-interaction is crucial for interpreting the thermal properties of type-I clathrates since the Coulomb interaction between cation guest-ion Ba$^{++}$ in BGG is shielded in $n$-type electron-rich BGG, while the dipole-dipole interaction in $p$-type becomes relevant even for small dipole moments.

In type-I clathrate compounds, off-center guest atoms are in 14-hedron cages, and its nearest-neighbor off-center guest atoms are located in the next two  14-hedron cages, which share the \textit{same} four-fold inversion axis.
The four-fold inversion axes are directed along the $x, y, z$ axes due to the cubic symmetry of $\beta$-BGS as illustrated in Fig.\,\ref{fig:takaba_fig40}(a).
The key for yielding glass-like behavior is that these dipoles constitute an equilateral-triangle structure among next-nearest neighbor dipoles, as depicted in Fig.\,~\ref{fig:takaba_fig40}(b).
This configuration generates a frustrated situation necessary to the emergence of glass-like behavior below 1\,[K].

To make our argument clear,
let us consider two electric dipoles $\pp_\ell$ and $\pp_{m}$ separated by a distance  $|\RR_{\ell m}|$.
The dipolar interaction is given by the following form,
\begin{eqnarray}
 V_{\ell m}=\frac{1}{4\pi\ve_r|\RR_{\ell m}|^3}
\left[ \pp_\ell\cdot\pp_{m} - 3(\pp_\ell\cdot\hat{\RR}_{\ell m}) (\pp_{m}\cdot\hat{\RR}_{\ell m})\right] ,
\label{Vij}
\end{eqnarray}
where $\varepsilon_r$ is the dielectric constant of the clathrate, and $\hat{\RR}_{\ell m}=\RR_{\ell m}/|\RR_{\ell m}|$ is a unit vector.
The potential function for two coupled dipoles $\pp_1$ and $\pp_2$ along an axis
becomes $V_{12}=V_{\rm h}(\theta_1)+V_{\rm h}(\theta_2)+W_{12}(\theta_1, \theta_2)$
with $W_{12}=p^2\cos(\theta_1-\theta_2)/(4\pi\varepsilon_r R_{12}^3)$,
where two global minima (maxima) in $(\theta_1,\theta_2)$ configuration space appear at  $|\theta_1-\theta_2|=\pi$~$( 2\pi)$
since the dominant term for a nearest-neighbor pair is the first term in Eq.\,(\ref{Vij}).
This configuration acts as a new hindering potential in addition to the four-fold inversion symmetric potential $V_{\rm h}$.
This argument can be straightforwardly extended to the case of multiple pairs providing many potential minima in configuration space $\mathcal{P}=(\theta_1,\theta_2,\theta_3, \cdots)$
, where the potential function is $V_{123\cdots}= \sum V_{\rm h}(\theta_\ell)+\sum V_{\ell m}$.

\begin{figure}[t]
\begin{center}
\includegraphics[width = 0.6\linewidth]{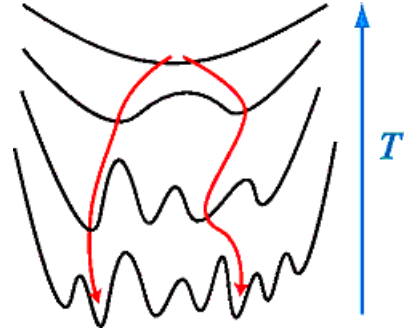}
\caption{(Color online) Schematic illustration of 
free-energy landscape representing non-equilibrium states.
} \label{fig:takaba_fig42}
\end{center}
\end{figure}

The energy scale of the dipole-dipole interactions between nearest neighbors is given by its maximum value,
\begin{eqnarray}
J_1=\frac{p^2}{4\pi\varepsilon_rR_1^3},
\label{J}
\end{eqnarray}
where $R_1$ is the distance between the nearest neighbors.
The actual distance between nearest-neighbor guest atoms in $\beta$-BGS is $R_{1}$=$a$/2=5.84\,[{\AA}].
The characteristic energy scale for nearest-neighbor coupling is then  estimated as $J_1$\,$\cong$\,$6\varepsilon_r/\varepsilon_0$\,[K].
Taking into account the dielectric constant for semiconductors in the range $5$\,$\lesssim\varepsilon_r/\varepsilon_0\lesssim20$, it turns out that $\mathcal{R}$(2) guest atoms experience strong electric fields.
The energy-scale of the dipole-dipole interaction is estimated to be of the order of a few 10\,[K].
They are no longer regarded as isolated dipoles.
Combining four-fold inversion symmetry of dipoles with the frustrated situation due to the equilateral triangle (See Fig.\,~\ref{fig:takaba_fig40}(b)), many local minima are created in a hierarchical potential map in configuration space $\mathcal{P}$, where tunneling should occur in a multi-valley potential in configuration space
with simultaneous local structural rearrangements of an appropriate number of guest atoms.
See the illutration of multi-valley potential depicted in Fig.\,\ref{fig:takaba_fig42}.

\subsubsection{Explicit form of specific heats below 1\,K}
\label{Explicit_form}
\label{subsubsec:explicitform}
The range of the energy distribution $n(E)$ can be determined from the energy at which the dipoles become free from the constraint of neighboring dipoles.
The maximum coupling strength from  neighbor dipoles $\Delta_0$\,$\cong z_1J_1+z_2J_2\cong 6J_1$ with configuration numbers $z_1=2$ and $z_2=8$ ($J_3\ll J_1, J_2$).
We introduce the ratio between the number of tunneling states $\bar{N}$ and the total number of off-center guest atoms $N$ given by $\eta=\bar{N}/N<1$.
Since $\bar{N}/J_1=3\pi\varepsilon_r\eta/p^2$ using $R_1=a/2$ into Eq.\,(\ref{J}), Eq.\,(\ref{C_1}) becomes the simple form
\begin{eqnarray}
\label{C3}
   C_\mathrm{tun}\cong\frac{6\pi^3\varepsilon_r\eta}{p^2}k_{\rm B}^2T.
\end{eqnarray}
Equation~(\ref{C3}) predicts larger specific heats for  smaller dipole moments $p$, but $\eta$ vanishes with $p\rightarrow0$\,\cite{Nakayama:2008a}.

A carrier-type dependence on $C_{\rm tun}(T)$ has been observed by \textcite{Suekuni:2008a} for $n$- and $p$-type $\beta$-BGS with  deviations $U_0=0.434$ and 0.439\,[\AA], respectively.
The observed $C_{\rm tun}(T)$ below 1\,[K] for $n$-type $\beta$-BGS is a few percent larger than that of the $p$-type, as shown in Fig.\,\ref{fig:takaba_fig18}.
This accords with the prediction of Eq.\,(\ref{C3}).

The comparison of Eq.\,(\ref{C3}) with the observed magnitude of a specific heat $C_{\rm tun}(T)$\,$\cong$30$T$\,[$\mathrm{mJ\,mol^{-1}\,K^{-1}}$] below 1\,[K] for $\beta$-BGS\,\cite{Suekuni:2008a, Suekuni:2008aa}
provides important information on the tunneling states. 
Taking $\varepsilon_r$\,$\cong$\,$10\varepsilon_0$, $\eta$ becomes $\cong$0.06 from $C_{\rm tun}(T)$\,$\cong 50(\varepsilon_r/\varepsilon_0)\eta T$\,[$\mathrm{mJ\,mol^{-1}\,K^{-1}}]$.
This indicates that only 6\% of the off-center guest ions contribute to the specific heats \textit{on average}.
This implies that the averaged number of dipoles simultaneously rearranged by tunneling is $1/\eta$\,$\cong 20$.

The formulation of the specific heat from Eq.\,(\ref{C3}) is based on the tunneling states caused by the dipole-dipole interaction; then  $C_{\rm tun}(T)$ vanishes  when the dipole-dipole interaction becomes irrelevant.
This is a reason why the specific heats of on-center  clathrate compounds show the Debye specific heat at low temperatures.

\subsubsection{Tunneling states in alkali-halide crystals containing CN$^-$ ions}
The previous subsection has highlighted the role of  off-center dipoles in type-I clathrate compounds.
In this connection, we should mention low temperature thermal properties of alkali-halide crystals containing CN$^{-}$ ions which provide good examples showing the relevance of long-range dipole-dipole interactions.
The alkali-halide mixed crystals have served as 
a model for the exploration of the low-energy excitations characteristic to structural glasses
\,\cite{Narayanamurti:1970r, DeYoreo:1986a, Loidl:1988a, Winterlich:1995a, Topp:2002a}.
The low temperature thermal properties of 
(KBr)$_{1-x}$(KCN)$_x$ single crystals in the range of 
$0.25\leq x\leq 0.70$ have investigated by 
\textcite{DeYoreo:1986a} and \textcite{Topp:2002a}.

Although (KBr)$_{1-x}$(KCN)$_x$ is crystalline, it is disordered in two ways.
The CN$^-$ ions and the Br$^-$ ions are randomly distributed over the anion sub-lattice. 
This disorder by itself is unlikely to produce emergence of a glass-like behavior since mixed crystals such as KBr:KI involves the same disorder, but does not show glass-like behavior\,\cite{Nathan:1976a}.
In the case of KBr:KCN, the CN$^-$molecules are disordered with respect to their orientations in the cubic lattice.

The observed specific heats are well described by the relation given in Eq.\,(\ref{Two-level_Specific}).
For $x$=0.25, 0.50, and 0.70, the values of the small factor $\delta$ giving the best fit for the long-time data are $\delta$=0.03, 0.00, and 0.08, respectively.
The observed values of specific heats below 1\,[K] for the case of $x$=0.25 yields the number density of $\bar{N}$\,$\cong 10^{17}$\,[cm$^{-3}$].
This is the evidence that all of CN$^-$ elements do not contribute to tunneling states, similar to the case in off-center clathrate compounds described in the previous section.
The role of long-range interactions between elastic dipoles randomly distributed in crystals
have been theoretically investigated \,\cite{Randeria:1988a, Grannan:1990a, Grannan:1990b}.

We should mention the INS experiments performed for the KCN:CN system by \textcite{Walton:1974a}.
They have observed at 5\,[K]
a clear anti-crossing of the acoustic phonon dispersion relation between 0.4 and 0.5\,[THz] corresponding to 
$T_{\rm 1u}$ and $T_{\rm 2u}$ modes of a CN$^-$ molecule in a KCl crystal doped with 6$\times$10$^{19}$ CN$^-$\,[cm$^{-3}$].
However, the anti-crossing was not observed at room temperature.
The strong dependence of INS intensities on temperature indicates that the coupling between acoustic phonons
and  libration modes of CN$^{-}$ depend on the state of the CN$^{-}$ molecule, namely, the libration modes are well characterized at low temperatures compared with those at room temperature.

\section{UNDERLYING MECHANISMS OF GLASS-LIKE  THERMAL CONDUCTIVITIES}
\label{subsec:origin_kappa11}
The reduction of phonon thermal conductivities $\kappa_{\rm ph}(T)$ is the key element to increase the dimensionless figure of merit $ZT$ according to the ``phonon-glass electron-crystal" concept\,\cite{Slack:1995r}.
\textcite{Berman:1949a} discovered in structural glasses the unexpected anomaly of $\kappa_{\rm ph}(T)$.
He measured $\kappa_{\rm ph}(T)$ of several samples of silica glass in the temperature range 2.2$-$90\,[K] and found the plateau behavior in the temperature range 5$-$10\,[K] with a magnitude several orders of magnitude smaller than that of crystal silica.
However, these properties attracted little attention until the work of \textcite{Zeller:1971a}, who discovered three characteristic temperature regions;
(i) $\kappa_{\rm ph}(T)\propto T^{2-\delta}$ below 1\,[K],
(ii) the plateau region of $\kappa_{\rm ph}(T)$ at around 5$-$10\,[K], and 
(iii) $\kappa_{\rm ph}(T)$ show the $T$-linear rise subsequent to the plateau region above 10\,[K].
The same features of $\kappa_{\rm ph}(T)$ have been found for a variety of glasses and amorphous materials\,\cite{Zeller:1971a, Freeman:1986a}. 

\textcite{Nolas:1998aa} have pointed out that the $\kappa_{\rm ph}(T)$ of off-center SGG polycrystalline samples behaves like those of structural glasses.
The same characteristics have been observed for other type-I clathrate compounds containing off-center guest atoms\,\cite{Nolas:1995a,Nolas:1996a,Nolas:1998a,Nolas:1998aa,
Nolas:1998aaa,Nolas:1998aa,Nolas:2000a,Nolas:2001r,
Cohn:1999a,Sales:1996a,Sales:1998a,Sales:2001a,
Bentien:2004a,Bentien:2005a, Avila:2006a,Avila:2006aa,Suekuni:2007a,
Suekuni:2008a,Suekuni:2008aa,Suekuni:2010a, Xu:2010a},
where the temperature dependence and the magnitude of 
$\kappa_{\rm ph}(T)$ are almost identical to those of 
structural glasses; these are the power law of $\kappa_{\rm ph}(T)\propto T^{2-\delta}$ in the temperature below $T$\,$\simeq$\,1\,[K], the plateau-temperature region at $T$=1$-$10\,[K], and the subsequent
rise of $\kappa_{\rm ph}(T)$ proportional to $T$ at the high-temperature end of the plateau.

This section describes theoretical interpretations taking into account the quantum mechanical aspects of these unique behaviors observed in type-I clathrate compounds belonging to off-center systems.

\subsection{Thermal conductivities below 1\,K}
\label{subsec:tunneling_kappa}
$\kappa_{\rm ph}(T)$ can be evaluated on the assumption that heat is carried by acoustic phonons obeying linear-dispersion relations.
This is expressed by
\begin{equation}
\kappa_{\rm ph}\left(T\right) =\frac{1}{3}\sum_{\rm \mu}\int_0^{\omega_{\rm c\mu}} \hbar\omega_\mu\frac{\partial n_\mathrm B}{\partial T}v_{\rm \mu}^2\tau_{\rm \mu}\left(T, \omega\right)D(\omega_\mu)\mathrm{d}\omega_\mu,
\label{eq:kappa}
\end{equation}
where $n_\mathrm{B}$ is the Bose-Einstein distribution function, and
$D(\omega_\mu)$, $v_{\mu}=\partial\omega_\mu(k)/\partial k$ and $\tau_{\mu}$  are the Debye density of states, the group velocity, and the life-time of acoustic phonons of the mode $\mu$, respectively.
Note that the Matthiessen's rule for the life-time $1/\tau_{\mu}=\sum_i 1/\tau_i$ should hold for independent scattering process $i$.

The cut-off frequency $\omega_{\rm c\mu}$ means the cross-over frequency from linear dispersion to flat dispersion of acoustic phonons of the mode $\mu$, at which the group velocity $v_\mu$ of acoustic phonons of the mode $\mu$ almost vanishes. 
At temperatures below 1\,[K], this cut-off frequency becomes irrelevant, whereas this frequency becomes crucial above 1\,[K].

For type-I clathrates with off-center guest atoms at $T$\,$\lesssim1$\,[K], dominant scattering arises from the interaction between tunneling states and thermally excited acoustic phonons.
The scattering rate of a phonon of the mode $\mu$
due to the interaction with a tunneling state 
is given by\,\cite{Phillips:1987r}
\begin{equation}
\label{scatter_rate}
\frac{1}{\tau_{\mu}(\omega)}=\sum_E\frac{\pi g_\mu^2 \omega_\mu}{\rho v_{\mu}^2}\(\frac{\Delta}{E}\)^2
\tanh\(\frac{\beta\hbar\omega}{2}\)\delta\left(\hbar\omega-E\right),
\end{equation}
where $g_\mu$ is the deformation coupling constant and $\rho$ the mass density ($\rho=6.01\times10^3$~$[\mathrm{kg\,m^{-3}}$] for $\beta$-BGS).
The averaging procedure by the distribution function $f(\ve,\lambda)$ defined in \ref{two-level model} can be simplified by replacing $\sum_E\delta(\hbar\omega-E)\to\int n(E)\mathrm{d}E$ as in the case of Eq.\,(\ref{C}).
The dominant tunneling process occurs at $\Delta$\,$\cong E$, then Eq.\,(\ref{scatter_rate}) provides 
\begin{equation}
\label{scatter_rate2}
\frac{1}{\tau_{\mu}(T, \omega)}=\frac{\pi^2 g_\mu^2 \omega_\mu}{2\rho v_{\mu}^2}\(\frac{\ve_r\eta}{p^2}\)
\tanh\(\frac{\beta\hbar\omega}{2}\),
\end{equation}
where the density of states is taken as $\bar{N}/\Delta_0$\,$\cong\pi\ve_r\eta/2p^2$
due to the interacting-dipole picture given in Sec.\,\ref{Explicit_form}.
By combining the above scattering rate with Eq.\,(\ref{eq:kappa}), $\kappa_{\rm tun}(T)$ becomes
\begin{eqnarray}
\kappa_{\rm tun}(T) \cong \frac{\rho k_\mathrm{B}^3 v_\mathrm{s} }{2\pi^{2}\hbar^2g^2 (\varepsilon_r\eta/p^2)}T^2,
\label{eq:kappa-T2}
\end{eqnarray}
where the velocity $v_{\rm s}$ and the deformation coupling constant $g$  refer to the average values for the three modes of acoustic phonons.
A more convenient form involving the specific heat $C_{\rm tun}$ is obtained as
\begin{eqnarray}
\kappa_{\rm tun}(T)
\cong\(\frac{3\pi\rho v_\mathrm{s}\eta k_\mathrm{B}^3}{\hbar^2C_{\rm tun}/T}\)\(\frac{T}{g/k_\mathrm{B}}\)^2.
\label{eq:kappa-C}
\end{eqnarray}
This relation suggests that thermal conductivities 
$\kappa_{\rm tun}(T)\propto T^{2-\delta}$ with small factor $\delta$ reflecting $C_{\rm tun}(T)\propto T^{1+\delta}$\,\cite{Kaneshita:2009a}.
The small factor $\delta$ originates from the distribution function $f(\epsilon, \lambda)$ as iscussed in Sec.\,\ref{Explicit_form}.
Equation (\ref{eq:kappa-C}) predicts that the magnitude of $\kappa_{\rm tun}(T)$ should be proportional to the inverse of $C_{\rm tun}(T)$, namely, $\kappa_{\rm tun}(T)\to$ small when $C_{\rm tun}(T)\to$ large.
This relation holds for  $\kappa_{\rm tun}(T)$ of $n$- and $p$-type $\beta$-BGS as observed in Figs.\,\ref{fig:takaba_fig07} and \ref{fig:takaba_fig18}.

We employ the following physical parameters of $\beta$-BGS.
The average value of $v_{\mu}$ is $v_{\rm s}$=2.3$\times 10^3$\,$\left[\mathrm{m\,s^{-1}}\right]$ from the actual values of velocities, $v_{C_{11}}$=3369\,$\left[\mathrm{m\,s^{-1}}\right]$, $v_{C_{11}-C_{12}}$=1969\,$\left[\mathrm{m\,s^{-1}}\right]$, and $v_{C_{44}}$=1844\,$\left[\mathrm{m\,s^{-1}}\right]$.
These yield
\begin{eqnarray}
\label{eq:kappa_C2}
   \kappa_{\rm tun}(T)=2.0\times 10^5\(\frac{T}{g/k_\mathrm B}\)^2~\left[\mathrm{W\,K^{-1}\,m^{-1}}\right].
\end{eqnarray}
Using the experimental data of $\kappa_{\rm tun}(T)$\,$\cong$0.02$T^2\,\left[\mathrm{W\,K^{-1}\,m^{-1}}\right]$ for $\beta$-BGS\,\cite{Suekuni:2008a}, the deformation coupling $g$ is estimated as $g\simeq 0.3$\,[eV].
This is a reasonable value because the deformation coupling constants $g$ in glasses are in the range of 0.1\,$-$1\,[eV]\,\cite{Anderson:1972a,Phillips:1972a}.

\subsection{Plateau region of thermal conductivities at around 5\,K}
For off-center type-I clathrate compounds, some mechanisms have been proposed to explain $\kappa_{\rm ph}(T)$ 
in the plateau temperature region\,\cite{Dong:2001a, Bridges:2004a, Hermann:2005aa, English:2009a}.
\textcite{Bridges:2004a} argued a mechanism for plateau thermal conductivities of off-center clathrate compounds.
\textcite{Hermann:2005aa} have considered the role of Einstein oscillators in filled skutterudites in view of explaining their lattice dynamics.
\textcite{Dong:2001a} have performed molecular dynamics (MD) calculations for $\kappa_{\rm ph}(T)$ on clathrate compounds with and without guest atoms in cages.
The addition of guest atoms in cages produced a reduction of phonon thermal conductivity. 
\textcite{English:2009a} have investigated the mechanism for $\kappa_{\rm ph}(T)$ in methane hydrate, which shows the same plateau temperature region as the case of off-center clathrate compounds.
These works have employed MD calculations combined with the linear response formula for $\kappa_{\rm ph}(T)$,  not taking into account quantum aspects such as the annihilation or creation of acoustic phonons $via$ anharmonic interactions.
The importance of the three-phonon process, Umklapp process, was first pointed by \textcite{Peierls:1929a} to explain the peak of $\kappa_{\rm ph}$ observed at about 10\,[K] for crystals.
MD calculations at the present stage cannot reproduce these \textit{quantum} processes.
This is the most difficult task in MD calculations on heat transport simulations. 

\subsubsection{Umklapp process for on-center systems}
Type-I clathrate compounds belonging to on-center systems possess the symmetry of translational invariance.
This implies that the wave vector $\kk$ of acoustic phonons carrying heat are well defined in the whole Brillouin zone.
Acoustic phonons are scattered mainly by two mechanisms:
Rayleigh elastic scattering due to imperfections and the anharmonic phonon-phonon inelastic scattering.
The Rayleigh scattering due to the static imperfections proportional to $1/\tau_{\rm R}(\omega)\propto\omega^4$  does not become the dominant scattering mechanism in type-I clathrate compounds.
This is because the wavelength of excited phonons is much larger than the scale of imperfections.
So, we should have the situation $v_{\mu}\tau_{\mu}(T, \omega)> L$ at low temperatures, where $L$ is the size of the crystalline part.
This leads to $\kappa_{\rm ph}(T)\propto C_{\rm ph}\propto T^3$ for on-center type-I clathrate compounds.

With increasing temperature above 1\,[K], the wave vectors of thermally excited phonons approaches the middle of the Brillouin zone, and the Umklapp process starts to contribute to the decrease of  $\kappa_{\rm ph}(T)$ at around $T\sim$10\,[K]. 
See Figs.\,\ref{fig:takaba_fig14}
and \ref{fig:takaba_fig16}.

Coherent INS data for on-center BGG  given in Fig.\, \ref{fig:takaba_fig29} show that the avoided crossings for transverse acoustic phonons occur at $\vert\kk\vert\gtrsim\vert\GG\vert/4$ with the reciprocal lattice vector $\GG$.
This allows the Umklapp process $\kk_1+\kk_2=\kk_3+\GG$ for acoustic phonons which simultaneously satisfy energy conservation law for anharmonic three-phonon processes.
These crystalline features of $\kappa_{\rm ph}(T)$ are clearly manifested in experimental data of on-center clathrate compounds shown in Figs.\,\ref{fig:takaba_fig14}
and \ref{fig:takaba_fig16}.

At high temperatures $T\geq\hbar\omega_{c\mu}/k_{\rm B}$, the number of excited phonons are proportional to $T$.
The scattering probability $1/\tau_\mu$ due to ahnharmonic interactions of acoustic phonons of the mode $\mu$ is proportional to the number of excited phonons $\propto T$ resulting in  the mean free path  $\ell_\mu=v_{\mu}\tau_\mu\propto 1/T$. 
Since the average velocity $v_{\mu}^2$ in Eq.\,(\ref{eq:kappa}) should be independent of $T$,
$\kappa_{\rm ph}(T)\propto T^{-1}$ as observed in Figs.\,\ref{fig:takaba_fig14}
and \ref{fig:takaba_fig16}.
\subsubsection{Plateau temperature region of off-center systems}
$\kappa_{\rm ph}(T)$ of type-I clathrate compounds containing off-center guest atoms take the universal form as described in Sec.\,\ref{subsec:kappa}.
The energy range of the plateau region in $\kappa_{\rm ph}(T)$ overlaps with that of the Boson-peak like excess density of states in off-center type-I clathrate compounds in the THz frequency region.
The physical origin of the plateau temperature region
should be interpreted in reference to this excess density of states. 

The upper cut-off frequency $\omega_{c\mu}$ of the integral in the general formula for $\kappa_{\rm ph}(T)$ in Eq.\,(\ref{eq:kappa}) represents the cross-over frequency from the linear dispersion relation to the flat dispersion relation of acoustic phonons, where the group velocity of acoustic phonons of the mode $\mu$ vanishes.
The frequency $\omega_{c\mu}$ behaves as a mobility edge for acoustic phonons.
Thus, the cross-over frequency $\omega_{c\mu}$ is crucial to reveal the plateau behavior of $\kappa_{\rm ph}(T)$.

Transverse acoustic phonons with two degrees of freedom of the modes mainly contribute to phonon thermal transport.
The dispersion relations of $\beta$-BGS theoretically calculated in Fig.\,\ref{fig:takaba_fig36}(b)
show a wide flat region for the acoustic branch of transverse phonons arising from the coupling with the low-lying librational modes\,\cite{Nakayama:2011a}.
The calculated results given in Fig.\,\ref{fig:takaba_fig36}(b) definitely show that the avoided crossing starts at the region below  $\vert\kk\vert\cong\vert\GG\vert/4$.
This result indicates that acoustic phonons carrying heat are limited to those with the wave number $\vert\kk\vert\lesssim\vert\GG\vert/4$.
This leads to the plateau temperature region of $\kappa_{\rm ph}(T)$ similar to the case of the Dulong-Petit limit of the Debye specific heats, i.e., $\kappa_{\rm ph}(T)$=constant.
Namely, conventional heat transport occurs via already excited phonons yielding a saturation in $\kappa_{\rm ph}(T)$, which is referred to as the plateau.

The cross-over frequency $\omega_{c\mu}$ is obtained by subtracting the frequency $\delta\omega_{c\mu}$ given by Eq. (\ref{eq:anti_eigenfrequency}) as $\omega_{c\mu}=\omega_{0\mu}-\delta\omega_{c\mu}$.
The use of Eq.\,(\ref{eq:motion33}) leads to
\begin{equation}
\label{eq:avoided_freq}
\omega_{c\mu}\cong
 \omega_{0\mu}\left( 1-\sqrt{\frac{m}{M}}\right).
\end{equation}
The peak frequency of the thermal distribution at the temperature $T$ becomes $\hbar\omega_{c\mu}$\,$\cong 3.83k_{\rm B} T$, taking into account the Stefan shift.
Then, the relation between the lowest optic eigenfrequency $\omega_{0\perp}$ and the onset temperature $T_{\rm p}$ of the plateau
becomes, from Eq. (\ref{eq:avoided_freq})
\begin{equation}
\label{eq:crit_freq}
T_{\rm p}\cong\hbar\omega_{0\perp}\frac{\left( 1-\sqrt{m/M}\right)}{3.83k_{\rm B}}.
\end{equation}
The above relation between $T_{\rm p}$ of $\kappa_{\rm ph}(T)$ and $\hbar\omega_{c\mu}$ of the phonon dispersion relation holds for type-I clathrates with off-center rattling guest atoms.

\subsection{$T$-linear rise above the plateau temperature region}
\label{subsec:T_linear_kappa}
\subsubsection{Thermal transport due to the hopping of local modes associated with guest atoms}
At the temperature regime $T$\,$\gtrsim$10\,[K] above the plateau, the $\kappa_{\rm ph}(T)$ of off-center type-I clathrate compounds exhibit the empirical law $\kappa_{\rm ph}(T)=\kappa_{\rm plateau}+\kappa_{\rm add}$, where $\kappa_{\rm add}\,=\,\alpha T$\,\cite{Cohn:1999a, Nolas:2000a, Nolas:2001r, Sales:2001a, Bentien:2005a, Avila:2006a, Suekuni:2007a, Suekuni:2008a, Avila:2008a}.
The prefactors $\alpha$ become $\alpha=0.009$\,[$\mathrm{W\,K^{-2}\,m^{-1}}$] for $n$-type and $\alpha=0.007$\,[$\mathrm{W\,K^{-2}\,m^{-1}}$] for $p$-type $\beta$-BGS\,\cite{Avila:2006a, Suekuni:2007a, Suekuni:2008a, Avila:2008a}.
Of importance is that $\kappa_{\rm ph}(T)$  vanishes when extrapolating the temperature $T\rightarrow0$, as seen from Fig.\,\ref{fig:takaba_fig14}.
It implies that a new \textit{additional} heat-transport channel opens up above the plateau temperature region.
This feature, termed $\kappa_{\rm add}(T)$, at high temperatures is the same as those of structural glasses or amorphous materials\,\cite{Zeller:1971a, Stephens:1930a, Freeman:1986a}.
This is remarkably different from the cases of on-center type-I clathrate compounds, which behave as $\kappa_{\rm ph}(T)\propto 1/T$ above $T\gtrsim$10\,[K]. 

The plateau temperature regions of $\kappa_{\rm ph}(T)$  are also observed for silica aerogels\,\cite{Bernasconi:1992a}.
In silica aerogels, there are two types of phonons, low-energy extended acoustic phonons and high-energy localized phonons separated at the cross-over frequency $\omega_{\rm c}$. 
\textcite{Alexander:1986a} have proposed a hopping mechanism of localized modes
to explain the $T$-linear dependence of $\kappa_{\rm add}(T)$ of silica aerogels. 
\textcite{Jagannathan:1989a} and \textcite{Nakayama:1999a, Nakayama:1999b} have introduced this mechanism to explain the $T$-linear rise of $\kappa_{\rm add}(T)$ above the plateau in structural glasses.
\textcite{Hashimoto:2011a} have demonstrated for silica glass the hopping of local modes associated with the Boson peak.
They have employed transient saturation spectroscopy for $^{167}$Er$^{3+}$ ions doped in a silica glass fiber in the range 2.5$-$30\,[K], a temperature range taken above the plateau region.
The result is consistent with the idea that local modes carry heat by their hopping assisted by acoustic phonons.

\begin{figure}[t]
\begin{center}
\includegraphics[width = 0.5\linewidth]{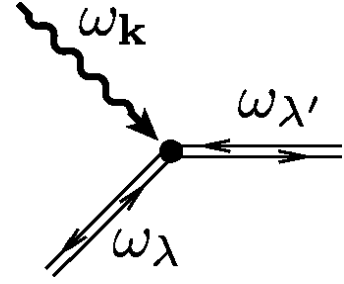}
\caption{
Diagram illustrating a first-order process of local-mode hopping.  
The local mode (double lines) interacts with the acoustic phonon (wavy line) with the eigenfrequency $\omega_\kk$, hopping from the initial state with $\omega_\lambda$ to the final with $\omega_{\lambda'}$, where $\omega_{\lambda'}>\omega_\lambda>\omega_\kk$.  
The set of arrowheads in the opposite directions on the double lines indicate that the local modes consist of the superposition of plane waves with opposite directions.
}
\label{fig:takaba_fig43}
\end{center}
\end{figure}

Here we present a simple explanation for the hopping mechanism contributing to $\kappa_{\rm ph}(T)$ in the temperature regime above the plateau.
The formula of Eq.\,(\ref{eq:kappa}) for the thermal conductivity based on the propagation of extended acoustic phonons is not applicable to the hopping mechanism.
At such high temperatures, the additional channel of  heat transfer is opened up via the diffusion of local modes associated with guest atoms, where the hopping of local modes occurs via the anharmonic interaction with extended acoustic phonons as shown in Fig.\,\ref{fig:takaba_fig43}.

The thermal conductivity due to the phonon-assisted hopping is given by\,\cite{Alexander:1986a} 
\begin{eqnarray}
   \kappa_{\rm hop}(T)=\sum_{\lambda}C_\lambda(T)\frac{R_{\lambda}^2}{3\tau_{\lambda}(T)},
\label{eq:Hopping_Mechanism_0}
\end{eqnarray}
where $R_{\lambda}$
is the hopping distance of the local mode $\lambda$ and $R_{\lambda}^2/\tau_\lambda$ is the thermal diffusivity due to the local mode hopping.
The decay rate $1/\tau_{\lambda}(T)$ is caused by the interaction with acoustic phonons, and $C_\lambda$ the specific heat associated with the local mode $\lambda$.
In the higher temperature regime above the plateau region $T\geq\hbar\omega_{c\mu}/3.83k_{\rm B}$, the specific heat $C_\lambda$ follows the Dulong-Petit relation in the form $C_\lambda=k_{\rm B}/V$, where $V$ is the volume of the system.
Substituting this relation into Eq.\,(\ref{eq:Hopping_Mechanism_0}), we have the thermal conductivity formula given by
\begin{eqnarray}
   \kappa_{\rm hop}\cong\frac{k_{\rm B}}{3V}\sum_{\lambda}\frac{R_{\lambda}^2}{\tau_{\lambda}(T)},
\label{eq:Hopping_Mechanism}
\end{eqnarray}
where the sum over $\lambda$ means the integral on $\omega_\lambda$ is defined by 
\begin{eqnarray}
   \sum_{\lambda}\to \sum_\mu\int_{\omega_{c\mu}}^{\omega_{c\mu}+
   \Delta\omega_\lambda}d\omega_\lambda N_{\rm ex}(\omega_\lambda).
\label{eq:Def_Integral}
\end{eqnarray}
Here  $N_{\rm ex}$ is the excess density of states and $\Delta\omega_\lambda$ is its bandwidth.
The decay rate $1/\tau_\lambda(T)$ is proportional to the Bose-Einstein distribution function $n_{\rm B}(T)$ of acoustic phonons of the form $n_{\rm B}$\,$\cong k_{\rm B}T/\hbar\omega_{\rm c\mu}$ in the high temperature regime $T\gtrsim\hbar\omega_{c\mu}/3.83k_{\rm B}$. 
These procedures give rise to the following simple relation for $\kappa_{\rm hop}(T)$,
\begin{eqnarray}
   \kappa_{\rm hop}(T)\simeq\frac{Nk_{\rm B}^2}{V}\frac{R(\omega_{\rm c\mu})^2T}{\hbar\omega_{\rm c\mu}\tau(\omega_{\rm c\mu})},
\label{eq:Hopping_Mechanism_1}
\end{eqnarray}
where the factor 3 is canceled out by taking into account the contribution from the three acoustic phonon modes.
We can estimate from Eq.\,(\ref{eq:Hopping_Mechanism_1}) the actual value of the  hopping contribution to $\kappa_{\rm hop}(T)$ above the plateau region by the following arguments.

With increasing temperature above $T\gtrsim\hbar\omega_{\rm c\mu}/3.83k_{\rm B}$, acoustic phonons are hybridized with vibrations of guest atoms
manifesting the Ioffe-Regel limit $kR(\omega_{\rm c\mu})$\,$\cong 1$, or equivalently, $\omega_{\rm c\mu}\tau(\omega_{\rm c\mu})$\,$\cong 1$\,\cite{Ioffe:1960a}.
This is the limit in which a quantum mechanical treatment is applicable. 
The average hopping distance $R(\omega_{\rm c\mu})$ is surely the distance between the nearest neighbor off-center guest atoms of the order of 1\,[nm].
Thus, the hybridized local mode $\lambda$ associated with the guest atoms in cages should satisfy the condition $\omega_{\rm c\mu}\tau(\omega_{\rm c\mu})$\,$\cong 1$.
Substituting these relations into Eq.\,(\ref{eq:Hopping_Mechanism_1}) together with the number density of the local mode $\lambda$ of $N/V=1/R(\omega_{\rm c\mu})^3$,
we have 
\begin{eqnarray}
   \kappa_{\rm hop}\simeq 0.002\,[{\rm WK}^{-1}{\rm m}^{-1}].
\label{eq:Hopping_Mechanism_2}
\end{eqnarray}
The estimated value agrees well with the observed prefactor $\alpha$ for off-center type-I clathrate compounds, reflecting that the $T$-linear rise above the plateau region is a universal phenomena.

The quantitative microscopic calculations on $\kappa_{\rm hop}$ should be made by taking into account the anharmonic interaction Hamiltonian between the acoustic phonon with the wave vector $\kk$ and the local modes $|\lambda\>$, $|\lambda'\>$, which is expressed as follows:
\begin{eqnarray}
H_{\rm hop}=C_{\rm{eff}} \sum_{\kk, \lambda, \lambda'}(A_{\mathbf{k}, \lambda, \lambda'}\, c_{\lambda}^{\dagger} c_{\lambda'} b_{\mathbf{k}}+\mathrm{h.c.}),
\label{H_3}
\end{eqnarray}
where the operators $b_\kk^{\dagger}$ and $c_\lambda^{\dagger}$ ( $b_\kk$ and $c_\lambda$) are the creation (annihilation) operators for an acoustic phonon and a local mode, respectively.
$H_{\rm hop}$ is proportional to the dilatation of cages due to incoming acoustic phonons. 
The explicit form of the coupling constant $A_{\mathbf{k}, \lambda, \lambda'}$ in Eq.\,(\ref{H_3}) is given by\,\cite{Nakayama:1999a, Nakayama:1999b, Kaneshita:2009a}
\begin{eqnarray}
A_{\mathbf{k}, \lambda,\lambda'}=\sqrt{\frac{1}{V}}\(\frac{\hbar}{2\rho}\)^{3/2}
\sqrt{\frac{\omega_{\mathbf{k}}}{\omega_{\lambda}\omega_{\lambda'}}} \(\frac{I}{v_{\mu}\ell_{\lambda}\ell_{\lambda'}}\),
\label{eq:A}
\end{eqnarray}
where $\rho$ is the mass density. 
The localization length $\ell_{\lambda}$ is the size of cages containing guest atoms, which is comparable with the hopping distance of the local mode  $\lambda$.
The factor $I$ in Eq.\,(\ref{eq:A}) is the overlap integral between local modes represented by 
\begin{eqnarray}
I\cong\(\frac{16}{\pi}\)\e^{-R_{\lambda}/\ell_{\lambda} +\i\mathbf{k}\cdot\mathbf{R_{\lambda}}},
\label{eq:I}
\end{eqnarray}
where the local mode decays exponentially.
As seen above, the explicit form of the local mode is not necessary, but only the order of $R_{\lambda}$ and $\ell_{\lambda}$.

For the process $|\kk; \lambda\>\rightarrow |\lambda'\>$ illustrated in Fig.\,\ref{fig:takaba_fig43}, first-order perturbation theory provides the hopping life-time of the local mode $\lambda$ at temperatures $k_B T \gg \hbar \omega_{c\mu}$ of the form
\begin{eqnarray}
\nonumber
\frac{1}{\tau_{\lambda}(T)}&=&\frac{2\pi}{\hbar^2}\sum_{\lambda,\lambda'}\mid A_{\kk,\lambda,\lambda'}\mid^2\\
&\times& \left[1+n(\omega_\kk)+n(\omega_{\lambda'}\right]
 \delta\left(\omega_\kk-\omega_{\lambda'}+\omega_\lambda\right).\label{eq:T}
\end{eqnarray}
The substitution of Eq.\,(\ref{eq:T}) into Eq.\,(\ref{eq:Hopping_Mechanism}) yields the hopping contribution to the thermal conductivity given by
\begin{eqnarray}
\kappa_{\rm hop}(T)=\frac{12^2C_{\rm eff}^2k_{\rm B}^2T }{\pi^4\rho^3v_s^{5}\ell_\lambda^5},
\label{eq:HP}
\end{eqnarray}
where the mass density $\rho$=6.01$\times10^3$\,[$\mathrm{kg\,m^{-3}}$] is
for $\beta$-BGS, $\ell_\lambda\simeq 1$\,[nm] and $1/v_s^{5}$ is the average of $1/v_{\mu}^{5}$ taking $\cong$3.6\,$\times$10$^{-17}$\,[$\mathrm{m^{-5}\,s^{5}}]$ for $\beta$-BGS.
Since the magnitude and temperature dependence of the thermal conductivities above the plateau temperature region of off-center type-I clathrate compounds are almost identical to those of structural glasses, it is appropriate to take the coupling constant $C_{\rm{eff}}$ of structural glasses, which provides
$C_{\rm{eff}}$=$-2.0\times$10$^{12}$\,[$\mathrm{N\,m^{-2}}$]\,\cite{Nakayama:1999a, Nakayama:1999b, Kaneshita:2009a}.
As a result, the thermal conductivity for $\beta$-BGS becomes $\kappa_{\rm ph}(T)$\,$\cong 0.003T$\,[$\mathrm{W\,K^{-1}\,m^{-1}}$] above the plateau $T\geq10$\,[K].
This coincides well with the experimental values
of $\beta$-BGS\,\cite{Avila:2006a, Suekuni:2007a, Suekuni:2008a, Avila:2008a}.

\subsubsection{Overall interpretation of glass-like thermal conductivities}
\label{subsubsec:highT_limit}
The analyses described from Sec.\,\ref{sec:thermal_theor} to Sec.\,\ref{Dispersion_1} lead to the
following interpretation on $\kappa_{\rm ph}(T)$ of off-center type-I clathrate compounds.
These are summarized as follows;
(i) two-level tunneling states generated by  \textit{interacting} electric dipoles explain the observed $\kappa_{\rm ph}(T)\propto T^{2-\delta}$ with a small value of $\delta$ below 1\,[K],
(ii) the plateau region, $\kappa_{\rm ph}(T)=\mathrm{constant}$, is a direct consequence of the {\it flattening} of acoustic modes due to the hybridization with local vibrations of off-center guest atoms, and
(iii) the subsequent $T$-linear rise above the plateau originates from the hopping of local modes assisted by  acoustic modes of networked cages.

There is an additional complication at much higher temperatures, as shown in Fig.\,(\ref{fig:takaba_fig14}) and (\ref{fig:takaba_fig18}), where
the $T$-linear rise in $\kappa_{\rm ph}(T)$ does not continue, but
$\kappa_{\rm ph}(T)$ shows a curling over above $T$\,$\simeq$\,100[K].
This temperature regime is important for practical application of type-I clathrates to efficient  thermoelectric materials.
In this regime, quenching of phonon-assisted hopping occurs via the breakdown of three-phonon anharmonic scattering\,\cite{Simons:1964a}.
The hysteresis becomes larger in adjusting to an instantaneous thermal equilibrium distribution, which
leads to a curling over of $\kappa_{\rm ph}(T)$ attributable to the hydrodynamic or Akhieser limit\,\cite{Akhieser:1939a}.
Thus, the key aspects of thermal conductivities of type-I clathrate compounds containing off-center guest atoms can be completely understood over the entire temperature range in a consistent way.
\section{SUMMARY AND CONCLUSIONS}
This article has reviewed the development of research 
on phonon-glass electron-crystal\,(PGEC) thermoelectric materials, in particular, focusing on experimental and theoretical aspects of type-I clathrate compounds.
After the PGEC concept was coined by \textcite{Slack:1995r}, much efforts have been spent in exploring efficient thermoelectric materials. 
As a candidate for PGEC materials, 
\textcite{Slack:1995r} have promoted clathrate compounds encapsulating guest atoms in cages.
Type-I clathrate compounds manifest the PGEC concept when the size mismatch is large enough between the guest ion and the tetrakaidecahedoral cage. 
Thereby, the guest atoms take off-center positions in the tetrakaidecahedron. 

The microscopic structure of type-I clathrate compounds have been clarified by means of the following experimental techniques: x-ray diffraction, neutron diffraction, resonant x-ray diffraction, and extended x-ray absorption fine structure spectroscopy.
These have been described in detail in Sec.\,\ref{sec:type_I_structures}.   
Specific heat measurements on type-I clathrate compounds with off-center guest atoms have revealed the existence of the Boson-peak like excess density of states over the Debye phonons at around several Kelvin. 
The characteristic temperatures $\theta$ of guest atom $\mathcal{R}$(2) have been estimated from the peaks in the plot of 
$C_{\rm V}(T)/T^3$\,{\rm~vs}.\,$T$. 
These results have been summarized in Table\,\ref{tab:tab1}.
In addition, the $T$-linear specific heats identical to those of glasses have been experimentally verified below about 1\,[K], as described Sec.\,\ref{sec:thermal_experiments}.

The excess densities of states over the Debye phonons have been also found by means of spectroscopic measurements: infrared absorption, Raman scattering, M$\ddot{\rm o}$ssbauer spectroscopy, and inelastic neutron scattering.
Optical spectroscopies have made possible the identification of the relevant modes contributing to the excess densities of states.
Raman scattering and infrared absorption experiments for type-I clathrate compounds containing off-center guest atoms have revealed the relevance of the T$_{\rm 2g}$ mode and the E$_{\rm g}$ mode to the excess densities of states.
These modes do not exist in the case of type-I clathrate compounds with on-center guest atoms. 
It is remarkable that the line widths of these modes increase on cooling and the spectral energies shift lower, which provide opposite behaviors to those expected for usual crystalline materials. 
This key observation has been confirmed by INS experiments as well\,\cite{Nakamura:2010a}.
These experiments and theoretical interpretation have been described in detail in Sec.\,\ref{sec:THz} and Sec.\,\ref{Dispersion_1}.

Furthermore, thermal conductivity measurements display glass-like behaviors characterized as follows: $T^2$-dependence at temperatures below 1\,[K], the plateau at temperatures of several [K], and the $T$-linear dependence above the plateau at around 10\,[K]. 
Surprisingly enough, these temperature dependences as well as the magnitudes are identical to those of structural glasses in the whole temperature range. 
We have described relevant experiments in Sec.\,\ref{sec:ele_experiments}, emphasizing that off-center guest atoms play a key role for the emergence of glass-like thermal and dynamical properties.

Concerning the efficiency of the thermopower, type-I Ba$_{8}$Ga$_{16}$Sn$_{30}$ manifests low thermal conductivities in addition to high absolute values of the Seebeck coefficients $S$ of 300\,[$\mu$V/K] for both $p$- and $n$-type crystals at 300\,[K]. 
The electrical resistivity $\rho$ at room temperature is in the range 20$-$40$\times10^{-3}$\,[$\Omega$cm], which is several times larger than that expected for efficient thermoelectric materials. 
This large $\rho$ is caused by low-carrier densities of 10$^{19}$\,[cm$^{-3}$] and low carrier mobilities 20$-$40\,[cm$^2$/Vs]. 
As a result, the dimensionless figure of merit $ZT=S^2T/\rho\kappa_{\rm tot}$ does not reach the desired value of 1.0, but exhibit maxima of 0.58 and 0.50 at around 450\,[K] for the $p$- and $n$-type samples, respectively. 
It is necessary to decrease the resistivity by doping carriers in order to realize the ``electron-crystal" concept.

Type-V$\hspace{-.1em}$I$\hspace{-.1em}$I$\hspace{-.1em}$I Ba$_{8}$Ga$_{16}$Sn$_{30}$ is a possible candidate for an efficient thermoelectric material. 
The Ba guest atoms in distorted dodecahedron  vibrate with the atomic displacement parameter (ADP) three times larger than those of cage atoms. 
These anisotropic and anharmonic vibrations depress $\kappa_{\rm ph}$ at higher temperatures above 100\,[K]. 
The actual values of $\kappa_{\rm ph}$ take the value 0.7\,[W/(mK)] at above 100\,[K], as shown in Fig.\,\ref{fig:takaba_fig17}.
These values are higher than that $\kappa_{\rm ph}$=0.4\,[W/(mK)] of $\beta$-BGS, but they are still lower than those of thermoelectric materials based on Bi-Te. 
The carrier density can be tuned by substituting various types of elements for the cage atoms Ga and Sn. 
For Cu substitution, $ZT$ for $p$-type and $n$-type samples increase to 0.9 at 480\,[K] and 1.45 at 520\,[K], respectively. 
These values exceed those of the Pb-Te based materials at the same temperature. 
Therefore, type-V$\hspace{-.1em}$I$\hspace{-.1em}$I$\hspace{-.1em}$I Ba$_{8}$Ga$_{16}$Sn$_{30}$ without toxic elements is regarded as a human-friendly thermoelectric material with operation temperatures around 400$-$600\,[K]. 
At higher temperatures, Si- and Ge-based clathrates have better structural stability than Sn-based clathrates.
Thermoelectric modules composed of similar clathrates based on Sn, Ge, and Si may have a conversion efficiency higher than 10$\%$ in the wide temperature range from 300\,[K] to 1200\,[K]. 

Finally, we should emphasize that the theoretical understanding of glass-like behaviors of $\kappa_{\rm ph}$ of off-center clathrate compounds has  benefited very much from accumulated research on structural glasses. 
However, in structural glasses, it is difficult to identify relevant entities or elements due to their complex microscopic structures\,\cite{Nakayama:2002r}.
This is the main reason why the arguments on the origin of the Boson peak  have continued for decades.
This being said, it is clear that off-center guest atoms in type-I clathrate compounds are a key ingredient for emerging glass-like behavior. 
This has made it possible to provide clear theoretical interpretation of glass-like thermal and dynamic properties for these compounds.

To sum up, the subjects described in this review involve interesting phenomena not only for the physics itself but also for the exploration of renewable-energy materials. 
We hope that this review will serve as a basis for further development of clathrate-related efficient thermoelectric materials.

\section{ACKNOWLEDGMENTS}
We are most grateful to M. A. Avila, T. Kume, T. Mori, Y. Takasu, and N. Toyota for many valuable discussions and useful suggestions. 
We thank H. Fukuoka, T. Hasegawa, I. Ishii, C.H. Lee, T. Suzuki, K. Tanigaki, H. Tou, M. Udagawa, K. Umeo, J. Xu, and S. Yamanaka for fruitful discussions.
Special thanks are to M. Nakamura and M. Arai for supplying the INS data of type-I clathrates prior to publication.
We are grateful to O. B. Wright for a critical reading of the manuscript.
T. N. wishes to acknowledge with gratitude the support and hospitality of Max-Planck Institute for the Physics of Complex Systems  during his stay in 2012-13.
This work was partially supported by a NEDO Grant No.\,09002139-0 and Grant-in-Aid for Scientific Research from the Ministry of Education, Culture, Sports, Science and Technology (MEXT) of Japan, 
Grants Numbers 18204032, 19051011, 20102004, 22013018, 22540404, and 22740225.

\bibliographystyle{apsrmp}







\end{document}